\documentclass[iop,apjl]{emulateapj}
\usepackage{epsfig,amsfonts,amsmath,graphicx,natbib,apjfonts,longtable,cases,afterpage}

\def\sci#1#2{#1 $\times 10^{#2}$}
\def\sth#1{#1 $\times 10^{3}$}
\def\snth#1{#1 $\times 10^{-3}$}
\def\snfo#1{#1 $\times 10^{-4}$}

\def\sfo#1{#1 $\times 10^{4}$}
\def\sfi#1{#1 $\times 10^{5}$}

\def\snbsi#1#2#3{$#1_{#2}^{#3} \times 10^{-6}$}
\def\snbfi#1#2#3{$#1_{#2}^{#3} \times 10^{-5}$}
\def\snbfo#1#2#3{$#1_{#2}^{#3} \times 10^{-4}$}
\def\snbth#1#2#3{$#1_{#2}^{#3} \times 10^{-3}$}
\def\snbt#1#2#3{$#1_{#2}^{#3} \times 10^{-2}$}

\def\sbo#1#2#3{$#1_{#2}^{#3}$}

\def\nod{\nodata}
\def\swift{{\it Swift}}
\def\chandra{{\it Chandra}}

\def\ein{1}
\def\ari{2}
\def\har{3}

\shorttitle{Short GRB Afterglows}
\shortauthors{Fong et al.}

\begin{document}

\title{A Decade of Short-duration Gamma-ray Burst Broad-band Afterglows: \\ Energetics, Circumburst Densities, and Jet Opening Angles}

\author{
W.~Fong\altaffilmark{\ein}$^{,}$\altaffilmark{\ari},
E.~Berger\altaffilmark{\har},
R.~Margutti\altaffilmark{\har}, \&
B.~A.~Zauderer\altaffilmark{\har}
}

\altaffiltext{1}{Einstein Fellow}

\altaffiltext{2}{Steward Observatory, University of Arizona, 933 North Cherry Avenue, Tucson, AZ 85721}

\altaffiltext{3}{Harvard-Smithsonian Center for Astrophysics, 60
Garden Street, Cambridge, MA 02138, USA}

\begin{abstract}

We present a comprehensive catalog and analysis of broad-band afterglow observations for 103 short-duration gamma-ray bursts (GRBs), comprised of all short GRBs from November 2004 to March 2015 with prompt follow-up observations in the X-ray, optical, near-infrared and/or radio bands. These afterglow observations have uncovered 71 X-ray detections, 30 optical/NIR detections, and 4 radio detections. Employing the standard afterglow synchrotron model, we perform joint probability analyses for a subset of 38 short GRBs with well-sampled light curves to infer the burst isotropic-equivalent energies and circumburst densities. For this subset, we find median isotropic-equivalent $\gamma$-ray and kinetic energies of $E_{\rm \gamma, iso}\approx 2 \times 10^{51}$~erg, and $E_{K,{\rm iso}}\approx (1-3) \times 10^{51}$~erg, respectively, depending on the values of the model input parameters. We further find that short GRBs occur in low-density environments, with a median density of $n\approx (3-15) \times 10^{-3}$~cm$^{-3}$, and that $\approx 80-95\%$ of bursts have densities of $n\lesssim 1$~cm$^{-3}$. We investigate trends between the circumburst densities and host galaxy properties, and find that events located at large projected offsets of $\gtrsim 10$~effective radii from their hosts exhibit particularly low densities of $n\lesssim 10^{-4}$~cm$^{-3}$, consistent with an IGM-like environment. Using late-time afterglow data for 11 events, we find a median jet opening angle of $\theta_j=16 \pm 10^{\circ}$. We also calculate a median beaming factor of $f_b \approx 0.04$, leading to a beaming-corrected total energy release of $E_{\rm true} \approx 1.6 \times 10^{50}$~erg. Furthermore, we calculate a beaming-corrected event rate of $\Re_{\rm true} = 270^{+1580}_{-180}$~Gpc$^{-3}$~yr$^{-1}$, or $\approx 8^{+47}_{-5}$~yr$^{-1}$ within a 200~Mpc volume, the Advanced LIGO/Virgo typical detection distance for NS-NS binaries.

\end{abstract}

\section{Introduction}

The afterglow emission from gamma-ray bursts (GRBs) provides a unique way to probe their basic properties: the energy scale, circumburst densities, and jet opening angles. Nearly two decades of long-duration GRB afterglow observations established a median beaming-corrected kinetic energy release of $\approx 10^{51}$~erg \citep{fks+01,bkf03,fb05,gbb+08,nfp09,lbt+14}, circumburst densities of $\approx 0.1-100$~cm$^{-3}$ \citep{pk02,yhs+03}, and opening angles of $\approx 2-20^{\circ}$ \citep{bfk03,fks+01,fb05,kb08,rlb+09,rmz15}. In some cases, the radial density profiles of their circumburst environments reflect the wind environments expected for massive stars \citep{cl00,yhs+03}.

The afterglows of short-duration GRBs ($T_{90} \lesssim 2$~sec; \citealt{kmf+93}) are uniformly fainter \citep{ber07,nak07,gbb+08,nfp09,ber10,ber14}, and have thus far been primarily utilized for providing precise burst localization, and therefore robust associations to host galaxies (e.g., \citealt{fbc+13}). These host studies have revealed that at least some short GRBs originate from older stellar populations, and are distinct in their global properties from the hosts of long GRBs \citep{ber09,lb10,fbc+13,ber14}. The local and galactic environments of short GRBs \citep{fbf10,ber10,fb13}, together with the lack of associated supernovae (e.g., \citealt{ffp+05,sbk+06,ktr+10,fbm+14}), and the discovery of a near-IR ``kilonova'' from the short GRB\,130603B \citep{bfc13,tlf+13}, have provided strong observational support for a compact object binary merger progenitor involving two neutron stars or a neutron star and a stellar mass black hole (NS-NS/NS-BH; \citealt{elp+89,npp92}).

As studies of the host galaxies of short GRBs have progressed at a rapid pace, knowledge of their basic explosion properties has been limited by both the paucity of afterglow detections and the relatively low detection rate of well-localized ($\sim$few arcsec uncertainty) short GRBs from the \swift\ satellite. Furthermore, afterglow studies greatly benefit from observations across several orders of frequency, which serve to provide significantly tighter constraints on the energy scales and circumburst densities than single band observations. Thus far, short GRB afterglow studies have either focused on a small number of events in a single band (e.g., \citealt{ber07,nak07,kkz+08,nkg+12}), or on radio through X-ray observations of individual bursts \citep{bpc+05,sbk+06,fbm+14}. Comparative studies relative to long GRBs have only served to argue for lower energy scales and circumburst densities, but have not provided actual distributions (e.g., \citealt{ber14}). Thus, there have been no attempts to utilize the full set of broad-band afterglow observations of short GRBs. 

The relative difficulty of detecting short GRB afterglows is likely a direct reflection of a combination of low explosion energies and low circumburst densities. Predictions for NS-NS/NS-BH mergers span several orders of magnitude in density, depending on the precise distribution of natal kick velocities, merger timescales, and host galaxy type \citep{pb02,bpb+06}. Similarly, different mechanisms of energy extraction to power a relavistic blast-wave can produce energy scales which differ by three orders of magnitude (e.g., \citealt{rj99a,ros05}). Furthermore, the calculation of the true short GRB energy scale and event rate directly depends on the geometry of their jets. Constraints on the energetics, densities, and jet opening angles from short GRB afterglows thus offer a way to study these fundamental questions. These basic properties also serve as critical inputs for the detectability of other electromagnetic counterparts to compact object mergers, and will directly affect follow-up strategies to gravitational wave signals detected with Advanced LIGO/VIRGO (e.g., \citealt{mb12,nkg+13}).

Taking advantage of the dedicated searches for short GRB afterglows at all wavelengths, we are now in a position to explore these basic properties for a large population of events for the first time. Here, we present the first comprehensive broad-band catalog of short GRB afterglows, representing a decade of observations since the launch of {\it Swift} in 2004, and use this sample to constrain short GRB energies and circumburst densities. In Section~\ref{sec:obs}, we introduce the sample and data reduction methods for X-ray through radio observations. In Section~\ref{sec:bb}, we model the temporal and spectral behavior of the afterglows, and use the observations to infer constraints on the energetics and circumburst densities for each burst. In Section~\ref{sec:results}, we present the energetics and circumburst densities for 38 events with well-sampled afterglow data sets. In Section~\ref{sec:disc}, we discuss the observational afterglow properties, jet opening angles, and investigate trends between the bursts and their galactic environments. Finally, in the Appendix, we provide X-ray, optical, and radio afterglow catalogs. In an upcoming work, we will address the effect of the basic inferred properties on the detectability of electromagnetic counterparts to gravitational wave sources.

Throughout the paper, all magnitudes are in the AB system and are corrected for Galactic extinction in the direction of each burst \citep{sfd98,sf11}. Unless otherwise noted, uncertainties correspond to $1\sigma$ confidence. We employ a standard $\Lambda$CDM cosmology with $\Omega_M=0.27$, $\Omega_\Lambda=0.73$, and $H_0=71$ km s$^{-1}$ Mpc$^{-1}$.

\section{Observations}
\label{sec:obs}

\subsection{Sample}

\afterpage{
\LongTables
\tabletypesize{\small}
\begin{deluxetable}{lccccc}
\tablecolumns{6}
\tabcolsep0.05in\scriptsize
\tablewidth{0pc}
\tablecaption{Short GRB Basic Information
\label{tab:info}}
\tablehead {
\colhead {GRB}                     &
\colhead {$T_{90}$}                &
\colhead {$z$}                     &
\colhead {X-ray}               &
\colhead {Opt/NIR}             &
\colhead {Radio}               \\
\colhead {}                        &
\colhead {(s)}                     &
\colhead {}                        &
\colhead {}                        &
\colhead {}                        &
\colhead {}                        
}
\startdata
050202 & $0.27$  & \nod & \nod & N & N \\
050509B & $0.04$  & $0.225$    & Y & N & N \\
050709  & $0.07$  & $0.161$    & Y & Y & N \\
050724A & $3.0$   & $0.257$    & Y & Y & Y \\ 
050813  & $0.6$   & $0.72/1.8$ & Y & N & N \\
050906  & $0.26$  & \nod       & N & N & N \\
050925  & $0.07$  & \nod       & N & \nod & N \\
051105A & $0.09$  & \nod       & N$^{b}$ & \nod & N \\
051210  & $1.3$   & $>1.4$     & Y & N & \nod \\
051221A & $1.4$   & $0.546$    & Y & Y & Y \\
\hline
060121  & $2.0$   & $<4.1$     & Y & Y & \nod \\  
060313  & $0.7$   & $<1.7$     & Y & Y & N \\ 
060502B & $0.09$  & $0.287$    & Y & N & \nod \\
060801  & $0.5$   & $1.130$    & Y & N & N \\
061006  & $0.4$   & $0.438$    & Y & Y & \nod \\
061201  & $0.8$   & $0.111$   & Y & Y & \nod \\
061210  & $0.2$   & $0.41$   & Y & N & N \\  
061217  & $0.2$   & $0.827$    & Y & N & \nod \\
\hline
070209  & $0.09$  & \nod       & N & N & \nod \\
070406$^{a}$  & $1.20$  & \nod       & N & N & \nod \\ 
070429B & $0.5$   & $0.902$    & Y & N & N \\ 
070707$^{a}$  & $1.1$   & $<3.6$     & Y & Y & \nod \\ 
070714B & $2.0$   & $0.923$    & Y & Y & N \\
070724A & $0.4$   & $0.457$   & Y & Y & N \\
070729  & $0.9$   & $0.8$      & Y & N & N \\
070809  & $1.3$   & $0.473$   & Y & Y & \nod \\ 
070810B & $0.08$  & \nod       & N & N & \nod \\ 
070923  & $0.05$  & \nod & \nod & \nod & N \\ 
071017$^{a}$  & $0.5$   & \nod & N & \nod & \nod \\
071112B$^{a}$ & $0.30$  & \nod & N & N & N \\
071227  & $1.8$   & $0.381$    & Y & Y & \nod  \\ 
\hline
080121$^{a}$ & $0.7$  & \nod & N & N & \nod \\
080123 & $0.4$    & \nod       & Y & N & \nod \\ 
080426 & $1.7$    & \nod      & Y & N & \nod \\ 
080503 & $0.3$    & $<4.2$     & Y & Y & N \\ 
080702A & $0.5$   & \nod       & Y & N & N \\ 
080905A & $1.0$   & $0.122$   & Y & Y & \nod \\ 
080919 & $0.6$    & \nod       & Y & N & \nod \\ 
081024A & $1.8$   & \nod       & Y & N & N \\
081024B$^{a}$ & $0.4$   & \nod  & N$^{b}$ & N & N \\ 
081226A & $0.4$   & $<4.1$     & Y & Y & N \\
081226B$^{a}$ & $0.7$   & \nod       & N & N & N \\ 
\hline
090305A & $0.4$   & $<4.1$     & Y & Y & \nod \\ 
090417A & $0.07$  & \nod & \nod & \nod & N \\ 
090426  & $1.3$   & $2.609$    & Y & Y & \nod \\ 
090510  & $0.3$   & $0.903$    & Y & Y & N  \\
090515  & $0.04$  & $0.403$   & Y & Y & N \\  
090607  & $2.3^{c}$ & \nod & Y & N & \nod \\ 
090621B & $0.14$  & \nod       & Y & N & N \\
090715A & $0.5$   & \nod & \nod & N & N \\ 
090916$^{a}$  & $0.3$   & \nod & N & \nod & \nod \\ 
091109B & $0.3$   & $<4.4$     & Y & Y & \nod \\
091117$^{a}$  & $0.43$  & \nod & N$^{b}$ & N & N \\ 
\hline
100117A & $0.30$  & $0.915$    & Y & Y & \nod \\ 
100206A & $0.1$   & $0.407$    & Y & N & \nod \\ 
100213A & $2.4^{d}$ & \nod       & Y & \nod & \nod \\ 
100625A & $0.3$   & $0.452$    & Y & N & N \\ 
100628A & $0.04$  & \nod       & N & N & N \\ 
100702A & $0.16$  & \nod       & Y & N & \nod \\ 
101219A & $0.6$   & $0.718$    & Y & N &  \nod\\ 
101224A & $0.2$   & \nod       & Y & \nod & N \\ 
\hline
110112A & $0.5$   & $<5.3$     & Y & Y & N \\
110112B$^{a}$ & $0.5$ & \nod   & N & N & N \\
110420B$^{a}$ & $0.08$ & \nod  & N$^{b}$ & N & N \\
111020A & $0.4$ & \nod         & Y & N & N \\
111117A & $0.5$ & 1.3          & Y & N & N \\
111121A & $0.45$ & \nod  & Y & \nod & N \\
111222A$^{a}$ & $0.3$ & \nod         & Y & \nod & \nod \\
\hline
120229A & $0.22$ & \nod  & \nod & N & N \\
120305A & $0.1$ & \nod         & Y & N & N \\
120521A & $0.45$ & \nod  & Y & N & N \\
120630A$^{e}$ & $0.6$ & \nod   & Y & \nod & \nod \\
120804A & $0.81$ & 1.3         & Y & Y & N \\
120817B$^{a}$ & $0.19$ & \nod  & N & N & \nod \\
121226A & $1.0$ & \nod         & Y & N & N \\
\hline
130313A & $0.26$ & \nod        & Y & N & N \\
130515A & $0.29$ & \nod        & Y & N & \nod \\
130603B & $0.18$ & $0.356$    & Y & Y & Y \\
130626A & $0.16$ & \nod        & N & \nod & \nod \\
130716A & $0.8$  & \nod        & Y & N & N \\
130822A & $0.04$ & \nod        & Y & N & N \\
130912A & $0.28$ & $<4.1$        & Y & Y & N \\
131004A & $1.54$ & 0.717       & Y & Y & N \\
131125A$^{f}$ & 0.5 & \nod     & \nod & N & \nod \\
131126A$^{f}$ & 0.3 & \nod     & \nod & N & \nod \\
131224A$^{a}$ & 0.8    & \nod        & N & \nod & N \\
\hline
140129B & 1.36 & $<1.5$ & Y & Y & \nod \\
140320A$^{e}$ & 0.45 & \nod & Y & N & \nod \\
140402A$^{a}$ & 0.03 & \nod & N & N & \nod \\
140414A$^{a}$ & 0.7 & \nod & N & N & \nod \\
140516A & 0.19 & \nod & Y & N & N \\
140606A$^{a}$ & 0.34 & \nod & N & N & \nod \\
140619B$^{a}$ & 0.5 & \nod & N & N & N \\
140622A & 0.13 & 0.959 & Y & N & N \\
140903A & 0.30 & 0.351 & Y & Y & Y \\
140930B & 0.30 & $<4.1$ & Y & Y & N \\
141202A$^{a}$ & 1.3 & \nod & N & \nod & \nod \\
141205A$^{a}$ & 1.1 & \nod & N & N & \nod \\
141212A & 0.30 & 0.596 & Y & N & N \\
\hline
150101A & 0.06 & \nod & Y & \nod & N \\
150101B & 0.018 & 0.134 & Y & Y & N \\
150120A & 1.20 & 0.460 & Y & N & N \\
150301A$^{a}$ & 0.48 & \nod & Y & \nod & \nod
\enddata
\tablecomments{Short GRBs with X-ray, optical/NIR or radio follow-up observations. ``Y'' = detection, ``N'' = non-detection, and \nod means there is no follow-up in that band. \\
$^{a}$ Observing constraint or delayed \swift/XRT observations. \\
$^{b}$ There are XRT observations but no X-ray flux upper limit is reported. \\
$^{c}$ This burst has $T_{90}=2.3 \pm 0.1$~s but is spectrally hard \citep{gcnr220}. \\
$^{d}$ This burst has $T_{90}=2.4 \pm 0.4$~s, but the spectral lag of $5 \pm 15$~ms indicates this is a short-hard burst \citep{gcnr268}. \\
$^{e}$ Delayed reporting of burst position preventing immediate ground-based follow-up. \\
$^{f}$ IPN-localized burst with no \swift\ follow-up. \\
}
\end{deluxetable} }

We present afterglow observations for 103 short GRBs discovered by the \swift\ satellite \citep{ggg+04}, {\it Fermi} satellite \citep{aaa+09,mlb+09}, {\it Konus-Wind} \citep{afd+95}, High Energy Transient Explorer~2 (HETE-2; \citealt{rac+03}), or the Interplanetary Network (IPN; \citealt{hga+10}) between November 2004 and March 2015. We restrict our sample to all bursts with $T_{90} \lesssim 2$~s and follow-up observations in any of the X-ray, optical, near-infrared (NIR) or radio bands on timescales of $\delta t \lesssim$~few days (where $\delta t$ corresponds to the time after the $\gamma$-ray trigger). We also include three events (GRBs\,050724A, 090607 and 100213A) which have $T_{90} \approx 2.5-3$~s but which exhibit the spectral hardness and negligible spectral lags typical of short GRBs \citep{gcnr220,gcnr268}. For bursts with optical/NIR follow-up, we only include data from bursts with afterglow detections or meaningful limits of $\gtrsim 20$~mag at $\delta t \lesssim 1$~day. Basic information for the sample of 103 events, including durations, redshifts, and the available follow-up in each observing band is presented in Table~\ref{tab:info}.

Of the 103 bursts in our sample, 96 (93\%) have X-ray follow-up observations, 87 (84\%) have optical/NIR observations, and 60 (58\%) have radio observations. These observations have uncovered 71 X-ray afterglows, 30 optical/NIR afterglows, and 4 radio afterglows, leading to detection fractions of 74\%, 34\%, and 7\%, respectively (Table~\ref{tab:info}). The observations for these bursts are catalogued in the Appendix (Tables~A1-A3). 

Thirty-one bursts in this sample have redshifts determined from their host galaxies (Table~\ref{tab:info}), while two events have spectroscopic redshifts from absorption in their afterglows, GRB\,090426A: \citep{aap+09,lbb+10} and GRB\,130603B \citep{cpp+13,dtr+14}. Furthermore, we place upper limits on the redshift from the detection of the optical afterglow, and therefore the lack of suppression blueward of the Lyman limit ($\lambda_0=912$\,\AA) or Lyman-$\alpha$ line ($\lambda_0=1216$\,\AA), for 11 bursts (Table~\ref{tab:info}). We also place a lower limit of $z>1.4$ for GRB\,051210 from the lack of detection of emission or absorption features in the spectrum of the host galaxy \citep{bfp+07}. In addition to the broad-band afterglow observations that have been published thus far, we present new optical/NIR observations for 11 bursts (Table~A2), and new radio observations for 25 events (Table~A3).

\subsubsection{Observing Constraints}

Of the 25 bursts with X-ray observations and no detected X-ray afterglow, 18 events had a delayed \swift/XRT response due to an observing constraint or burst discovery from another satellite (indicated in Table~\ref{tab:info}); thus, the non-detection of X-ray afterglows for these events are due to factors unrelated to the bursts themselves. Likewise, 12 bursts with optical observations and no detected optical afterglow have an observing constraint (e.g., delayed precise localization from \swift, crowded field, high Galactic extinction sightline, position contaminated by a bright star), making the detection of an optical afterglow particularly challenging (Table~A2). Only one burst (GRB\,150101B) with radio observations is severely contaminated by a neighboring bright source, thus preventing a deep radio limit at the afterglow position. Taking these observing constraints into account, we find that the vast majority (91\%) of bursts that have X-ray follow-up and no observing constraints result in an X-ray afterglow detection. Similarly, the fraction of detected optical afterglows increases to 40\% after correcting for observing constraints. The radio detection fraction remains at $\approx 7\%$.

\begin{figure*}
\begin{minipage}[c]{\textwidth}
\tabcolsep0.0in
\includegraphics*[width=0.505\textwidth,clip=]{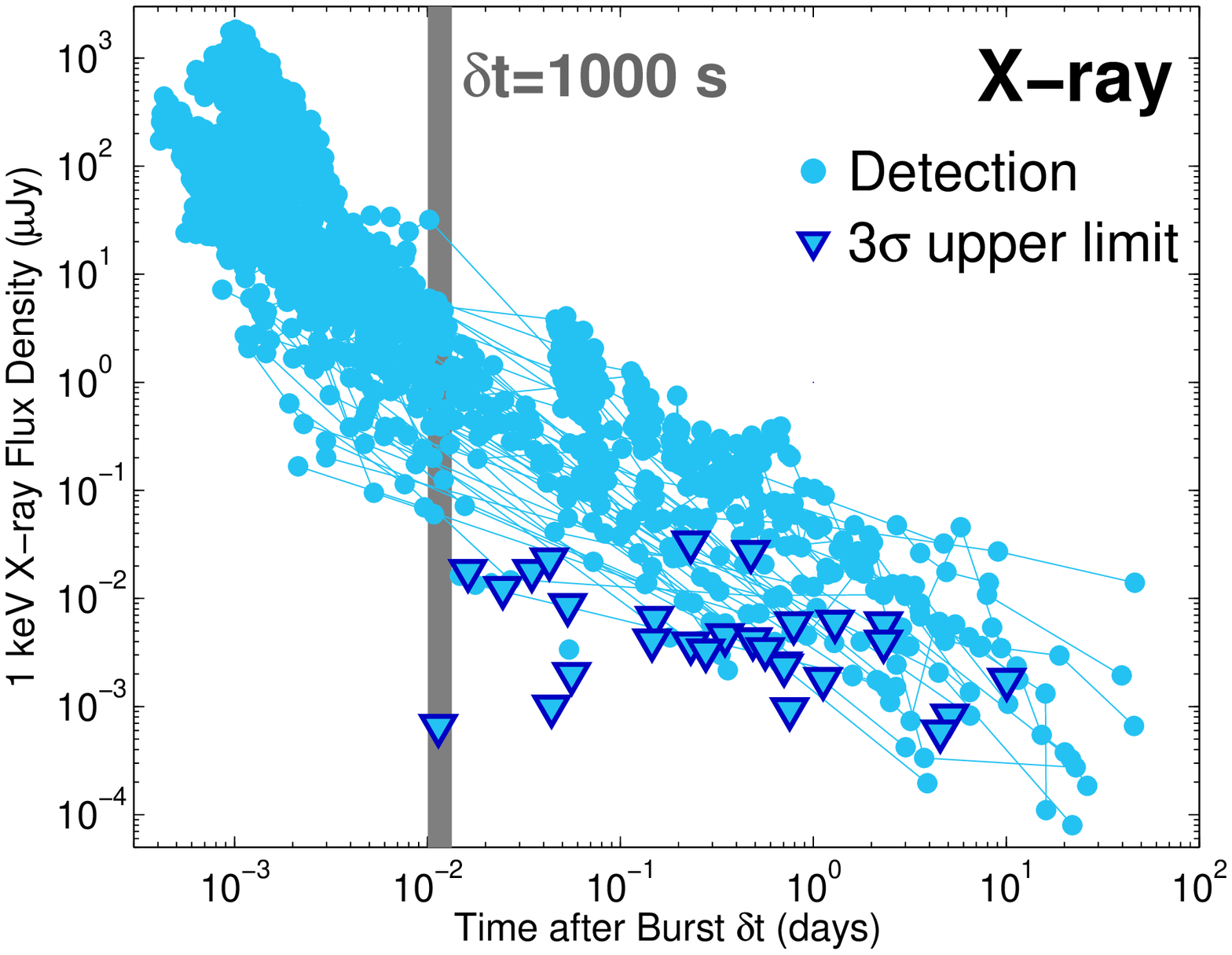}
\includegraphics*[width=0.495\textwidth,clip=]{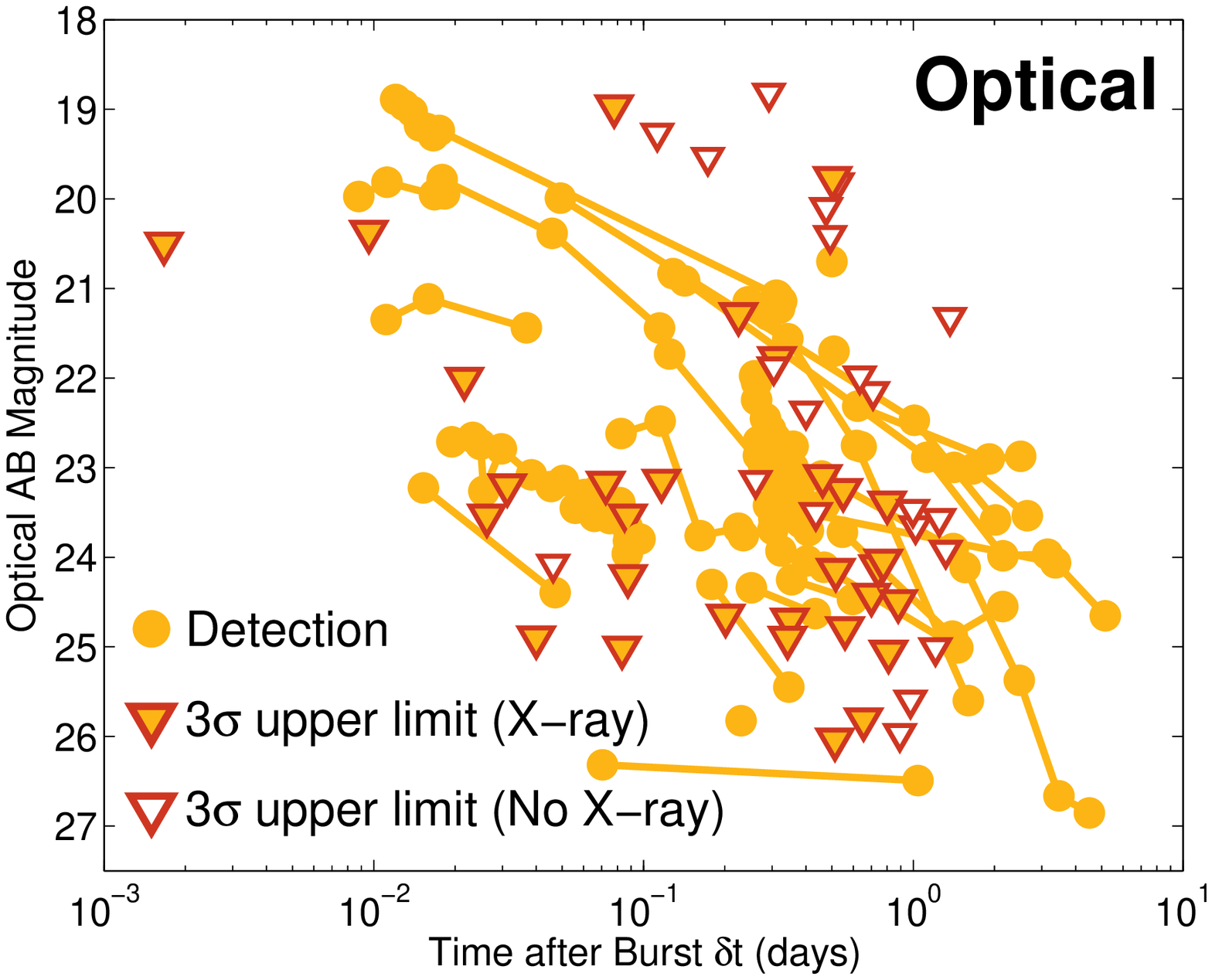} \\
\includegraphics*[width=0.51\textwidth,clip=]{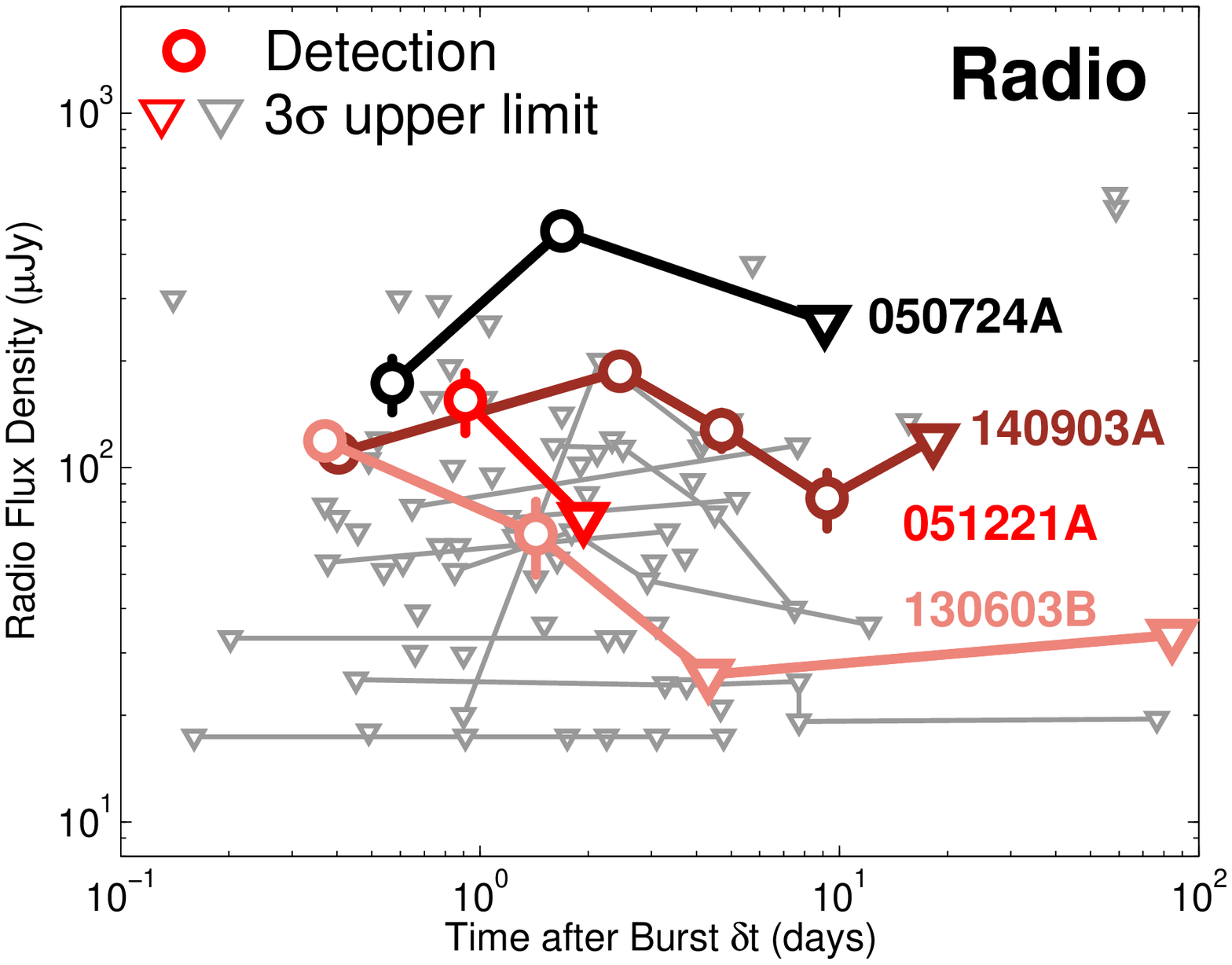}
\end{minipage}
\caption{Broad-band afterglow light curves of all short GRBs with follow-up observations between November 2004 and March 2015. In all panels, circles denote detections, triangles indicate $3\sigma$ upper limits, and solid lines connect observations for the same burst. {\it Top left:} $0.3-10$~keV X-ray afterglow light curves for 92 short GRBs with X-ray observations corresponding to the events in Table~A1. Included are $3\sigma$ upper limits for 29 bursts with no detected X-ray emission at $\delta t \gtrsim 1000$~sec (blue triangles). The data for $\delta t \gtrsim 1000$~sec (grey vertical band) are used in our subsequent afterglow analysis. {\it Top right:} Optical afterglow observations for 87 short GRBs corresponding to the events in Table~A2. The 30 bursts with detected optical afterglows are shown (orange circles), along with $3\sigma$ upper limits for bursts with an X-ray afterglow detection (filled triangles), and bursts with no X-ray afterglow detection (open triangles). {\it Bottom left:} Radio afterglow data for 60 short GRBs corresponding to the events in Table~A3. The light curves for four short GRBs with radio afterglow detections are shown: GRBs\,050724A (black), 051221A (red), 130603B (light red), and 140903A (maroon). Upper limits for the remaining 56 events with no detected radio afterglows are shown ($3\sigma$; grey triangles), including 25 bursts for which radio observations have not been published in the literature thus far.
\label{fig:lc}}
\end{figure*}

\subsection{X-rays}

We gather all available X-ray afterglow observations from the \swift\ light curve repository\footnote{\tt http://www.swift.ac.uk/xrt\_curves} \citep{ebp+07,ebp+09}, the GRB Coordinates Network (GCN) Circulars, and the literature (Table~\ref{tab:xcat}). The data were taken with the X-ray Telescope (XRT) on-board \swift, the {\it Chandra} X-ray Observatory, and the X-ray Multi-Mirror Mission ({\it XMM-Newton}). Ten bursts have {\it Chandra} observations, while three bursts have {\it XMM-Newton} observations (Table~\ref{tab:xcat}). We use unabsorbed fluxes and uncertainties in the $0.3-10$~keV energy band when they are available; otherwise, we use the count-rate light curves in the same energy range and convert to fluxes (described later in this section). For bursts with multiple upper limits, we only include those which help to constrain the temporal behavior of the X-ray light curve. Of the 96 short GRBs with X-ray observations, four events have no reported measurements or upper limits\footnote{These events are GRBs\,051105A, 081024B, 091117, and 110420B.}. Therefore, the resulting late-time X-ray afterglow catalog is comprised of 92 events (Table~\ref{tab:xcat}).

When applicable, we convert the count rate light curves to unabsorbed fluxes using the count-rate-to-unabsorbed-flux conversion factors provided by the \swift\ light curve repository. These factors are derived from the automatic spectral fitting routine \citep{ebp+09}. This routine fits the X-ray spectrum for each burst to an absorbed power law model characterized by photon index, $\Gamma$, and the intrinsic neutral hydrogen absorption column, $N_{\rm{H,int}}$, in excess of the Galactic column density in the direction of the burst \citep{kbh+05,wlb11,wsb+13}. We use spectral parameters extracted in the Photon Counting (PC) mode when available; otherwise, we use parameters from the Windowed Timing (WT) mode. In ten cases, the value of $N_{\rm{H,int}}$ is highly uncertain, but consistent with zero. Therefore, utilizing the median value of $N_{\rm{H,int}}$ may result in an over-estimate of the true unabsorbed flux. Instead of using the given conversion factors for these bursts, we calculate the unabsorbed fluxes using WebPIMMS\footnote{\tt https://heasarc.gsfc.nasa.gov/Tools/w3pimms.html}, setting $N_{\rm{H,int}}=0$. For 16 events, no count-rate-to-unabsorbed-flux conversion factor is available, so we employ a fiducial value of $1 \times 10^{-11}$~erg~cm$^{-2}$~s$^{-1}$~ct$^{-1}$ set by the median value for all of the events in our sample. Applying these conversion factors to each of the count-rate light curves from \swift/XRT, {\it Chandra} and {\it XMM-Newton}, we obtain the unabsorbed fluxes, $1\sigma$ uncertainties, and $3\sigma$ upper limits for each burst (Table~A1).

To enable comparison of the X-ray light curves to the optical and radio data, we convert the X-ray fluxes to flux densities, $F_{\nu,X}$, at a fiducial energy of 1~keV ($F_{\nu,X} \propto \nu^{\beta_X}$ where $\beta_X \equiv 1-\Gamma$). When no spectral information is available, we use a fiducial spectral index of $\beta_{X,{\rm med}}=-1.0$, set by the median value of the events in our sample. The flux densities, $1\sigma$ uncertainties, and $3\sigma$ upper limits are listed in Table~A1, and the resulting light curves are shown in Figure~\ref{fig:lc}. 

Finally, we present {\it Chandra} observations for three bursts that have not been published in the literature thus far: GRBs\,120804A (PI:~Burrows), 140930B (PI: Fong) and 150101B (PIs: Troja, Levan). We retrieve the pre-processed Level 2 data from the \chandra\ archive. We use the {\tt CIAO} data reduction package to extract a count-rate within a $2.5''$-radius source aperture centered on the X-ray afterglow position, and utilize source-free regions on the same chip to estimate the background. For GRB\,120804A, we obtain spectral parameters from earlier \chandra\ epochs of the same burst \citep{bzl+13}. For GRBs\,140930B and 150101B, we use {\tt CIAO/specextract} to extract a spectrum and obtain the spectral parameters. We then apply these parameters to the count rate to convert to flux density as described above (Table~A1).

\subsection{Optical/NIR}

For each burst we gather all available optical and NIR afterglow observations from the literature and GCN Circulars (see Table~A2 for references). When there are multiple upper limits for a given burst, we include only the deepest limits at $\delta t \lesssim 1$~day. We convert all magnitudes to the AB system using instrument-specific conversion factors when available, or the standard conversions following \citet{br07}. We correct all observations for Galactic extinction in the direction of each burst \citep{sfd98,sf11}, and convert AB magnitudes to flux densities, $F_{\nu,{\rm opt}}$. A log of observations for 87 events with optical/NIR follow-up is provided in Table~A2, and the light curves and upper limits are shown in Figure~\ref{fig:lc}. Bursts with no detected optical afterglow are further classified by the detection of an X-ray afterglow (Figure~\ref{fig:lc}).

We also present optical/NIR observations of 11 short GRBs that have not been published in the literature thus far: GRBs\,070724A, 100628A, 110420B, 120229A, 130716A, 140402A, 140606A, 140619B, 140930B, 150101B, and 150120A. These observations were enabled by target-of-opportunity programs on the twin 6.5-m Magellan telescopes (PI: Berger), the twin 8-m Gemini telescopes (PIs: Berger, Cucchiara, Fox, Tanvir), and the 6.5-m MMT (PIs: Berger, Fong). We use standard tasks in the IRAF/{\tt ccdred} package to process the Magellan and MMT data, and the IRAF/{\tt gemini} package to process the Gemini data. For each of these bursts, we obtained the first epoch of observations at $\delta t \approx 1-20$~hr and at least one additional set of observations at $\delta t \gtrsim 24$~hr to provide a template (Table~A2). To assess any fading between the two epochs, we perform digital image subtraction for each burst and filter using the ISIS software \citep{ala00}. With the exception of GRBs\,070724A, 140930B, and 150101B, we find no significant emission in any of the subtracted images. We therefore employ aperture photometry using standard tasks in IRAF to place $3\sigma$ upper limits on the optical/NIR afterglow brightness. To measure the afterglow brightness for GRBs\,070724A and 140903B, we employ aperture photometry on the detected point source in the subtracted image. For GRB\,150101B, the afterglow position in the subtracted image is contaminated by residual emission from its bright host galaxy; thus, we use PSF photometry to measure the afterglow brightness. Details of the photometry will be outlined in an upcoming paper (Fong~et~al., in prep.) The observational details, afterglow brightness and $3\sigma$ upper limits for these bursts are presented in Figure~\ref{fig:lc} and listed in Table~A2.

\subsection{Radio}

We gather all available radio afterglow data taken with the Karl G. Jansky Very Large Array (VLA), Westerbork Synthesis Radio Telescope (WSRT), Australia Telescope Compact Array (ATCA), and Combined Array for Research in Millimeter-wave Astronomy (CARMA). The resulting radio afterglow catalog is comprised of 60 short GRBs (Table~A3). The large majority of events, 53 ($88\%$), were observed with the VLA (Table~A3), 28 of which were observed with the upgraded VLA, which has a ten-fold improvement in sensitivity \citep{pcb+11}. We present new observations enabled by target-of-opportunity programs with the upgraded VLA (PI: Berger) and CARMA (PI: Zauderer) for 25 bursts (Table~A3). In 11 cases, we obtained multiple sets of observations to probe the radio emission on timescales spanning $\delta t \approx 1-10$~days.

For data calibration and analysis of VLA observations, we follow standard procedures in the Astronomical Image Processing System (AIPS; \citealt{gre03}). For CARMA observations, we use the Multichannel Image Reconstruction, Image Analysis and Display (MIRIAD) software package \citep{rtw+95}. For the majority of cases, we do not find any uncatalogued radio sources in or around the X-ray or $\gamma$-ray positions. To calculate the $3\sigma$ upper limits on the radio afterglow brightness, we measured the RMS noise in the map using a large source-free central region utilizing AIPS/{\tt IMSTAT} for VLA data and MIRIAD/{\tt IMSTAT} for CARMA data. The radio afterglow detections and upper limits for 60 short GRBs with radio observations are listed in Table~A3 and displayed in Figure~\ref{fig:lc}. Flux densities reported here supercede those reported in GCN Circulars.

Four bursts have detected radio afterglows, and all discoveries were made with the VLA (GRB\,050724A: \citealt{bpc+05,pan06}; GRB\,051221A: \citealt{sbk+06}; GRB\,130603B: \citealt{fbm+14}; GRB\,140903A: this work). In particular, we present the discovery of the radio afterglow of GRB\,140903A, which is detected at three frequencies: $1.4$, $6.0$ and $9.8$~GHz (Figure~\ref{fig:lc} and Table~A3), and will be further discussed in an upcoming work.

\subsection{Afterglow Brightness and TOO Response Times}

Thanks to \swift, X-ray and optical follow-up typically commences within $\approx 60$~sec after the GRB is detected. While \swift/XRT is responsible for nearly all of the X-ray afterglow detections, only three short GRBs have detected optical afterglows with \swift/UVOT (Table~A2). Thus, the detection of short GRB afterglows, or placement of meaningful upper limits, in the optical and radio bands relies on ground-based Target-of-Opportunity programs. For the 30 short GRBs with detected optical afterglows, the median optical afterglow brightness is $\approx 23.0$~mag at $\delta t \approx 7.0$~hr. The median $3\sigma$ limit placed on bursts with an X-ray afterglow and no detected optical afterglow is $\gtrsim 23.8$~mag with a median response time of $\delta t \approx 7.4$~hr, while the limit placed on bursts with no detected X-ray or optical afterglow is less constraining and more delayed, $\gtrsim 22.7$~mag at $\delta t \approx 12.2$~hr (Figure~\ref{fig:lc}). This is likely due to the more limited number of facilities with instruments that can cover the comparatively larger $\gamma$-ray positional error circles. However, since we only include bursts with optical limits of $\gtrsim 20$~mag, we are excluding a fraction of bursts with very shallow follow-up; thus the limits here for bursts with only $\gamma$-ray localizations are an optimistic representation of the entire population. In the radio band, the median $3\sigma$ upper limit for all observations is $\lesssim 74.1\,\mu$Jy, and the median response time to the first observation is $\delta t \approx 24.7$~hr. In more recent cases, we set unprecedented $3\sigma$ limits of $15-20\,\mu$Jy using the upgraded VLA on timescales of $\delta t \approx 1-10$~days.

\section{Broad-band Afterglow Analysis}
\label{sec:bb}

\afterpage{
\LongTables
\linespread{1.3}
\tabletypesize{\footnotesize}
\begin{deluxetable}{lcccc}
\centering
\tablecolumns{5}
\tablewidth{0pc}
\tablecaption{Short GRB Spectral and Temporal Power-law Indices
\label{tab:xrayspec}}
\tablehead {
\colhead {GRB} &
\colhead {$\alpha_X$} &
\colhead {$\beta_X$}            &
\colhead {$\alpha_{\rm opt}$}    &
\colhead {$\beta_{\rm opt}$}     \\  
}
\startdata
050509B & $-1.10 \pm 0.25$ & $-0.88 \pm 0.34$ & \nod & \nod \\
050709  & $-1.23 \pm 0.10$ & $-1.24 \pm 0.35$ & $-1.42 \pm 0.08$ & \nod  \\
050724A & $-0.93 \pm 0.08$ & $-0.81 \pm 0.15$ & $-1.74 \pm 0.11^{a}$ & $-0.82 \pm 0.03^{a}$ \\
050813  & $<-0.004$        & $-1.3^{+2.1}_{-1.3}$ & \nod & \nod  \\
051210  & \nod             & $-2.1 \pm 0.5$ & \nod & \nod \\
051221A & $-1.08 \pm 0.12$ & $-1.0 \pm 0.2$  & $-0.97 \pm 0.06$ & \nod  \\
060121  & $-1.23 \pm 0.20$ & $-1.07 \pm 0.16$ & $-0.60 \pm 0.24$ & \nod  \\
060313  & $-1.47 \pm 0.39$ & $-0.96 \pm 0.09$ & $-0.70 \pm 0.19$ & $-1.35 \pm 0.19$  \\
060502B & \nod                & $-2.07^{+1.50}_{-0.54}$ & \nod & \nod  \\ 
060801  & \nod & $-0.68 \pm 0.12$ & \nod & \nod \\
061006  & $-0.85 \pm 0.30$ & $-0.90 \pm 0.25$ & $-0.43 \pm 0.08$ & \nod \\
061201  & $-1.96 \pm 1.18$ & $-0.66 \pm 0.12$ & \nod & \nod \\
061210  & $-1.71 \pm 1.15$ & $-1.86^{+1.26}_{-0.61}$ & \nod & \nod \\
070429B & $<-1.12$ & $-2.0 \pm 0.63$ & \nod & \nod \\
070707  & $<-0.65$ & \nod & $-2.55 \pm 0.22$ & \nod \\
070714B & $-1.96 \pm 0.69$ & $-1.07 \pm 0.19$ & $-0.81 \pm 0.11$ & \nod \\
070724A & $-0.95 \pm 0.33$ & $-0.60 \pm 0.25$ & $<-0.15$ & $-0.58 \pm 0.02$ \\
070729  & \nod & $-0.5^{+0.44}_{-0.25}$ & \nod & \nod  \\
070809  & $-1.09 \pm 1.10$ & $-0.37^{+0.13}_{-0.07}$ & $-0.73 \pm 0.33$ & \nod \\
071227  & $-0.97 \pm 0.27$ & $-0.90 \pm 0.31$ & $<-0.24$ & \nod \\
080123  & $-0.77 \pm 0.19$ & $-1.6 \pm 0.4$ & \nod & \nod \\
080426  & $-1.54 \pm 0.33$ & $-1.03 \pm 0.16$ & \nod & \nod \\
080503  & \nod & $-1.53^{+0.26}_{-0.12}$ & \nod & \nod \\
080702A & $<-0.38$ & $-1.00 \pm 0.43$ & \nod & \nod \\
080905A & \nod & $-0.53 \pm 0.18$ & $-0.39 \pm 1.36$ & \nod \\
080919  & $-1.23 \pm 0.69$ & $-1.9^{+0.63}_{-0.38}$ & \nod & \nod \\
081024A & \nod & $-0.70 \pm 0.38$ & \nod & \nod \\
081226A & \nod & $-2.27^{+1.24}_{-0.32}$ & $-0.95 \pm 0.30$ & $-1.67 \pm 0.67^{b}$ \\
090305A  & \nod & \nod & $-0.74 \pm 0.09$ & $-0.71 \pm 0.27$ \\
090426A & $-1.15 \pm 0.16$ & $-1.04 \pm 0.09$ & $-0.58 \pm 0.16$ & $-0.94 \pm 0.06^{b}$ \\
090510  & $-1.79 \pm 0.63$ & $-0.75 \pm 0.08$ & $-2.37 \pm 0.29$ & $-0.85 \pm 0.05$ \\
090515  & \nod & $-1.53^{+0.78}_{-0.28}$ & $<-0.12$ & \nod \\
090607  & $<-0.72$ & $-1.2 \pm 0.4$ & \nod & \nod \\
090621B & $-1.48 \pm 0.54$ & $-2.7^{+0.68}_{-1.0}$ & \nod & \nod \\
091109B & $-0.83 \pm 0.28$ & $-1.1 \pm 0.3$ & $-0.49 \pm 0.45$ & \nod \\
100117A & \nod & $-1.6 \pm 0.3$ & $-1.60 \pm 0.33$ & \nod \\
100206A & \nod & $-2.05^{+1.07}_{-0.53}$ & \nod & \nod \\
100213  & \nod & $-2.0^{+1.2}_{-1.0}$ & \nod & \nod \\
100625A & $<0.37$ & $-1.5 \pm 0.2$ & \nod & \nod  \\
100702A & \nod & $-1.7 \pm 0.3$ & \nod & \nod  \\
101219A & $-1.37 \pm 0.13$ & $-0.8 \pm 0.1$ & \nod & \nod  \\
101224A & \nod & $-2.4^{+1.8}_{-0.6}$ & \nod & \nod \\
110112A & $-1.10 \pm 0.05$ & $-1.2 \pm 0.2$ & $<-0.32$ & \nod  \\
111020A & $-0.78 \pm 0.05$ & $-1.04 \pm 0.16$ & \nod & \nod  \\
111117A & $-1.21 \pm 0.05$ & $-1.0 \pm 0.2$ & \nod & \nod  \\
111121A & $-1.55 \pm 0.30$ & $-0.90 \pm 0.12$ & \nod & \nod \\
111222A & $<-0.30$ & $-0.7^{+1.4}_{-0.4}$ & \nod & \nod \\
120305A & \nod & $-2.2^{+0.6}_{-0.4}$ & \nod & \nod \\
120521A & \nod & $-0.8 \pm 0.25$ & \nod & \nod \\
120630A & \nod & $-0.8 \pm 0.3$ & \nod & \nod \\
120804A & $-1.02 \pm 0.10$ & $-1.1 \pm 0.1$ & \nod & \nod  \\
121226A & $-1.12 \pm 0.28$ & $-1.5 \pm 0.25$ & \nod & \nod \\
130313A & \nod & $-1.6^{+2.1}_{-2.5}$ & \nod & \nod \\
130515A & \nod & $-0.7 \pm 0.31$ & \nod & \nod \\
130603B & $-1.88 \pm 0.15$ & $-1.2 \pm 0.1$ & $-1.26 \pm 0.05$ & $-2.0 \pm 0.1^{b}$  \\
130716A & $<-0.52$ & $-1.1^{+0.56}_{-0.38}$ & \nod & \nod \\
130822A & \nod & $-0.89^{+1.1}_{-0.34}$ & \nod & \nod \\
130912A & $-1.33 \pm 0.18$ & $-0.57 \pm 0.13$ & $<-0.42$ & \nod \\
131004A & $-1.1 \pm 0.61$ & $-0.8 \pm 0.3$ & $-1.94 \pm 0.25$ & $-1.52 \pm 0.13^{b}$ \\
140129B & $-1.6 \pm 0.26$ & $-0.97 \pm 0.11$ & $-1.54^{c}$ & \nod \\
140320A & $<-0.23$ & $-3.8^{+2.6}_{-0.88}$ & \nod & \nod \\
140516A & $-0.42 \pm 0.14$ & $-0.70 \pm 0.44$ & \nod & \nod \\
140622A & $<-0.16$ & $-0.55^{+0.42}_{-0.18}$ & \nod & \nod \\
140903A & $-1.05 \pm 0.20$ & $-0.60 \pm 0.10$ & \nod & \nod \\
140930B & $-1.27 \pm 0.15$ & $-0.73 \pm 0.28$ & \nod & \nod \\
141212A & \nod & $-1.7^{+2.75}_{-0.81}$ & \nod & \nod \\
150101A & $<-0.12$ & $-0.40 \pm 0.56$ & \nod & \nod \\
150101B & $-1.07 \pm 0.15$ & $-0.67 \pm 0.17$ & $-1.01 \pm 0.62$ & \nod \\
150120A & \nod & $-1.0 \pm 0.31$ & \nod & \nod \\
150301A & \nod & $-0.7 \pm 0.13$ & \nod & \nod 
\enddata
\tablecomments{Error bars correspond to $1\sigma$ confidence. When applicable, $\alpha_X$ and $\alpha_{\rm opt}$ represent pre-jet break values. Values of $\beta_{\rm opt}$ are observed values and are uncorrected for intrinsic rest-frame extinction, $A_V$. \\
$^a$ These values are computed over the same time interval as the X-ray flare that is super-imposed on the underlying afterglow power-law decay. No optical detections exist for the underlying afterglow. \\
$^b$ This value corresponds to the spectral behavior after the jet break. \\
$^c$ No uncertainties are given for the optical light curve.
}
\end{deluxetable} }

We utilize the broad-band afterglow observations to constrain the explosion properties and circumburst environment of each burst. We adopt the standard synchrotron model for a relativistic blastwave in a constant density medium \citep{spn98,gs02}, as expected for a non-massive star progenitor. This model provides a mapping from the broad-band afterglow flux densities to the burst physical parameters: the isotropic-equivalent kinetic energy ($E_{\rm K,iso}$), circumburst density ($n$), fractions of post-shock energy in radiating electrons ($\epsilon_e$) and magnetic fields ($\epsilon_B$), and the electron power-law distribution index ($p$), with $N(\gamma)\propto \gamma^{-p}$ for $\gamma \gtrsim \gamma_{\rm min}$, where $\gamma_{\rm min}$ is the minimum Lorentz factor of the electron distribution.

The synchrotron spectrum is characterized by a flux normalization and three break frequencies: the self-absorption frequency ($\nu_a$), the peak frequency ($\nu_m$), and the cooling frequency ($\nu_c$). Constraints on the physical parameters require knowledge of where the synchrotron break frequencies are located with respect to the observing bands. In most cases, there is not enough information to constrain the locations of $\nu_a$ and $\nu_m$ with respect to the observing bands, so we make assumptions about their locations (detailed in the next sections). However, there are several cases in which we can use the available data to determine the location of $\nu_c$ with respect to the X-ray and optical bands. To determine the location of $\nu_c$, we first determine the temporal and spectral power-law indices ($\alpha$ and $\beta$, respectively, where $F_{\nu} \propto t^{\alpha}\nu^{\beta}$) from the X-ray and optical light curves and spectra. We then compare these indices to the standard relations given by the synchrotron model to determine whether $\nu_c$ is located above or below the X-ray band. This also allows us to calculate the value of $p$, and governs how the fluxes map to the burst physical properties \citep{gs02}.

\subsection{Temporal and Spectral Behavior}

\subsubsection{X-rays}
\label{sec:xfit}

To investigate the temporal behavior of the X-ray afterglows, we utilize $\chi^2$-minimization to fit a single power law model to each light curve in the form $F_{\nu,X} \propto t^{\alpha_X}$, with temporal index $\alpha_X$ as the single free parameter and the best-fit flux normalization $C_{0}$ given by

\begin{equation}
C_{0}=\frac {\sum_{i=1}^N \frac{F_{{\rm model},i} \times
F_{X,i}}{\sigma^2_{X,i}}}{\sum_{i=1}^N
\frac{F_{{\rm model},i}^2}{\sigma^2_{X,i}}},
\end{equation}

\noindent where $F_{{\rm model},i}$ are the un-normalized model fluxes, $F_{X,i}$ and $\sigma_{X,i}$ are the observed fluxes and uncertainties, respectively, and $N$ is the number of data points. Since early-time X-ray afterglow light curves are often subject to steep decays, plateaus, or flares which may contaminate the afterglow emission \citep{nkg+06,zfd+06,mcg+11,mzb+13}, we only utilize X-ray data at $\delta t \gtrsim 1000$~s, when bursts have typically settled into the power-law afterglow phase. For the X-ray light curves, we initially include all of the available data at $\delta t \gtrsim 1000$~s in the fit. In a few cases, there are light curve features beyond $\delta t \approx 1000$~s which significantly affect the fit: flares (GRBs\,050724A and 111121A), plateaus (GRB\,051221A), or steepenings (GRBs\,051221A and 111020A). For these bursts, we exclude the time intervals that contain such features in the fits. In 10 cases, there only exists a single detection and an upper limit at $\delta t \gtrsim 1000$~s, so we can only extract an upper limit for $\alpha_X$. The resulting best-fit values for $\alpha_X$, along with $1\sigma$ uncertainties, are listed in Table~\ref{tab:xrayspec}. Also listed are the X-ray spectral indices, $\beta_X$, from the relation $\beta_X \equiv 1-\Gamma$. The X-ray afterglows have weighted mean values and $1\sigma$ uncertainties of $\langle \alpha_X \rangle=-1.07^{+0.31}_{-0.37}$ and $\langle \beta_X \rangle=-0.89^{+0.38}_{-0.78}$, and median values of $\alpha_{X,{\rm med}} \approx -1.16$ and $\beta_{X,{\rm med}} \approx -1.06$.

\subsubsection{Optical}

We determine the temporal index of the optical observations ($\alpha_{\rm opt}$, where $F_{\rm \nu,opt} \propto t^{\alpha_{\rm opt}}$) in the same manner as described in Section~\ref{sec:xfit}, using the filter with the most well-sampled light curve for each burst. If there are multiple filters for which we can determine $\alpha_{\rm opt}$, we independently fit $\alpha_{\rm opt}$ for each filter and report the weighted mean (Table~\ref{tab:xrayspec}). A few short GRBs have well-measured optical spectral indices from contemporaneous multi-band data, but do not have a well-sampled light curve in a single filter, preventing a measurement of $\alpha_{\rm opt}$. For these events, we use the measured value of $\beta_{\rm opt}$ (see below) to extrapolate all of the available afterglow data to a single filter, and then determine the temporal decay index from these observations. In this manner, we are able to measure the optical temporal decay index for 19 short GRBs, and place upper limits in five cases for bursts with only a single detection and an upper limit. The best-fit optical temporal indices and $1\sigma$ uncertainties are listed in Table~\ref{tab:xrayspec}. The weighted mean for the 19 short GRBs with measured values is $\langle \alpha_{\rm opt} \rangle = -1.07^{+0.31}_{-0.61}$, and the median is $\alpha_{\rm opt, med} \approx -0.99$. 

If there are contemporaneous observations in multiple filters, we use these to determine the observed spectral slope, $\beta_{\rm opt}$ ($F_{\rm \nu,opt} \propto \nu^{\beta_{\rm opt}}$). For a few bursts, there are multi-band observations taken at different times, as well as a measurement of $\alpha_{\rm opt}$ from a well-sampled light curve in a single filter; in such cases, we use $\alpha_{\rm opt}$ to interpolate the data from multiple filters to a common time. To determine $\beta_{\rm opt}$, we then use $\chi^{2}$-minimization to fit the optical/NIR photometry to a power law model. The values of $\beta_{\rm opt}$ are listed in Table~\ref{tab:xrayspec}. The weighted mean for all short GRBs with determined spectral indices after incorporating rest-frame extinction (Section~\ref{sec:ext}) is $\langle \beta_{\rm opt} \rangle = -0.71^{+0.25}_{-0.51}$, and the median is $\beta_{\rm opt, med} \approx -0.88$. 

\subsubsection{Rest-Frame Extinction}
\label{sec:ext}

In four cases where there are contemporaneous observations in multiple optical/NIR filters, a single power law model provides a poor fit to the broad-band photometry ($\chi^2_{\nu} \gtrsim 3$). For these bursts, we include the line-of-sight rest-frame extinction from the host galaxy ($A_{V}^{\rm host}$) as a second free parameter in the fit. We constrain the extinction using the Milky Way extinction curve \citep{ccm89}, and employ the burst redshifts (Table~\ref{tab:info}) to obtain the rest-frame extinction. For bursts with no determined spectroscopic redshift, we assume $z=0.5$ set by the median of the short GRB population \citep{ber14}. We find non-zero extinction values for GRBs\,060121, 070724A, 081226A, and 130603B (Table~\ref{tab:prop}). The resulting values for the optical spectral indices, uncertainties, and $A_V^{\rm host}$ are listed in Table~\ref{tab:prop}.

For bursts where we do not have enough information from the optical/NIR bands alone to constrain the spectral behavior of the afterglow, we initially assume that there is no rest-frame extinction, $A_V^{\rm host}=0$. However, in eight cases, the difference in slope between the X-ray and optical bands is shallower than expected (e.g., $|\beta_{OX}| \lesssim |\beta_{X}|-0.5$, where $\beta_{OX}$ is the spectral index between the X-ray and optical bands), suggesting that either there is intrinsic extinction, or that the X-rays do not originate from the forward shock. Assuming the former explanation, we include the minimum amount of extinction required until the X-ray and optical solutions agree to within the $1\sigma$ uncertainties. In two cases, GRBs\,080919 and 101219A, there is only an upper limit on the optical afterglow brightness; thus we can only determine a lower limit on the rest-frame extinction. We note that GRB\,080919 has the highest value of rest-frame extinction, with $A_V^{\rm host}\gtrsim6$~mag. However, this burst has a sightline close to the Galactic plane and therefore has a highly uncertain Galactic extinction, which likely affects the inferred value for $A_V^{\rm host}$. For the 12 events with rest-frame extinction, we find values of $A_V^{\rm host} \approx 0.3-1.5$~mag. These $A_V^{\rm host}$ values are also listed in Table~\ref{tab:prop}.

\subsection{Determination of Electron Power Law Index and the Location of the Cooling Frequency}

\LongTables
\linespread{1.8}
\tabletypesize{\footnotesize}
\begin{deluxetable*}{lccccccccc}
\tablecolumns{9}
\tablewidth{0pc}
\tablecaption{Inferred Properties
\label{tab:prop}}
\tablehead {
\colhead {GRB}                     &
\colhead {$A_V^{\rm host}$}       		   &
\colhead {$\nu_c<\nu_X$?}          &
\colhead {$p$}                     &
\colhead {$\epsilon_B$}            &
\colhead {$E_{\rm \gamma,iso,52}$} &
\colhead {$<E_{\rm K,iso,52}>$}    &
\colhead {$\eta_\gamma$}           &
\colhead {$<n>$}           	   \\
\colhead {}                        &
\colhead {(mag)}                   &
\colhead {}                        &
\colhead {}                        &
\colhead {}                        & 
\colhead {($10^{52}$~erg)}         & 
\colhead {($10^{52}$~erg)}         & 
\colhead {}             	   &
\colhead {(cm$^{-3}$)}             
}
\startdata

050709  & 0 & Y & $2.31 \pm 0.13$ & 0.1 & 0.09 & $2.6^{+0.8}_{-0.5} \times 10^{-3}$ & 0.97 & $1.0^{+0.5}_{-0.4}$ \\
	& 0 & Y & $2.31 \pm 0.13$ & 0.01 & 0.09 & $6.2^{+0.4}_{-0.6} \times 10^{-3}$ & 0.93 & \sbo{1.6}{-0.2}{+0.2} \\
		
050724A & 0 & N & $2.29 \pm 0.10$ & $10^{-4}$ & 0.24 & \sbo{0.18}{-0.05}{+0.08} & 0.58 & \sbo{0.89}{-0.49}{+0.58} \\

051221A & 0 & Y & $2.24 \pm 0.07$ & 0.1 & 1.3 & \sbo{0.16}{-0.01}{+0.01} & 0.89 & \sbo{0.03}{-0.005}{+0.006} \\
	& 0 & Y & $2.24 \pm 0.07$ & 0.01 & 1.3 & \sbo{0.27}{-0.03}{+0.03} & 0.83 & \sbo{0.14}{-0.04}{+0.05}  \\

060121  & 1.6 & Y & $2.24 \pm 0.20$ & 0.1 & 4.5 & \sbo{0.20}{-0.04}{+0.05} & 0.96 & \snbth{5.4}{-2.2}{+5.6} \\
	& 1.6 & Y & $2.24 \pm 0.20$ & 0.01 & 4.5 & \sbo{0.23}{-0.04}{+0.05} & 0.95 & \sbo{0.16}{-0.06}{+0.17} \\

060313$^{ab}$ & 0 & Y & $2.03 \pm 0.20$ & 0.1 & 2.9 & \sbo{0.45}{-0.05}{+0.05} & 0.87 & \snbth{3.3}{-0.5}{+1.0} \\

061006 & 0 & N & $2.39 \pm 0.31$ & 0.1 & 1.1 & \sbo{0.64}{-0.37}{+0.85} & 0.63 & \snbfi{2.2}{-1.9}{+16} \\
	& 0 & N & $2.39 \pm 0.31$ & 0.01 & 1.1 & \sbo{1.1}{-0.7}{+2.1} & 0.50 & \snbfo{1.2}{-1.1}{+29} \\

061201 & 0 & N & $2.35 \pm 0.24$ & 0.1 & 0.05 & \sbo{0.05}{-0.03}{+0.10} & 0.47 & \snbfi{5.0}{-4.6}{+66} \\
	& 0 & N & $2.35 \pm 0.24$ & 0.01 & 0.05 & \sbo{0.1}{-0.09}{+0.4} & 0.29 & \snbfo{2.7}{-2.6}{+120} \\

070714B$^{bc}$ & $0.5$ & Y & $2.30 \pm 0.35$ & 0.1 & 1.7 & \sbo{0.1}{-0.02}{+0.02} & 0.94 & \sbo{0.056}{-0.011}{+0.024} \\

070724A & 1.5 & N & $2.24 \pm 0.33$ & 0.1 & 0.03 & \sbo{0.35}{-0.20}{+0.49} & 0.07 & \snbfi{1.9}{-1.6}{+12} \\
	& 2.0 & N & $2.24 \pm 0.33$ & 0.01 & 0.03 & \sbo{1.1}{-0.8}{+3.0} & 0.02 & \snbfi{9.3}{-9.2}{+210} \\

070809  & 0 & N & $2.12 \pm 1.47^{d}$ & 0.1 & 0.09 & \sbo{0.5}{-0.3}{+0.7} & 0.14 & \snbfi{2.2}{-1.9}{+15} \\
	& 0 & N & $2.12 \pm 1.47^{d}$ & 0.01 & 0.09 & \sbo{1.1}{-0.8}{+2.7} & 0.07 & \snbfo{1.2}{-1.1}{+30} \\

071227  & 0 & Y & $1.92 \pm 0.31$ & 0.1 & 0.14 & \snbth{8.4}{-1.9}{+2.5} & 0.94 & \sbo{1.9}{-1.1}{+2.4} \\
	& 0 & Y & $1.92 \pm 0.31$ & 0.01 & 0.14 & \snbth{8.9}{-1.8}{+2.6} & 0.94 & \sbo{60}{-33}{+75} \\

080426  & 0 & Y & $2.29 \pm 0.26$ & 0.1 & 0.35 & \sbo{0.05}{-0.01}{+0.01} & 0.87 & \sbo{0.04}{-0.02}{+0.04} \\
	& 0 & Y & $2.29 \pm 0.26$ & 0.01 & 0.35 & \sbo{0.06}{-0.01}{+0.01} & 0.85 & \sbo{1.2}{-0.5}{+1.2} \\

080905A & 0 & N & $2.06 \pm 0.36$ & 0.1 & 0.02 & \sbo{0.04}{-0.03}{+0.12} & 0.34 & \snbfo{1.3}{-1.2}{+33} \\
	& 0 & N & $2.06 \pm 0.36$ & 0.01 & 0.02 & \sbo{0.08}{-0.07}{+0.44}& 0.21 & \snbfo{7.1}{-7.1}{+610} \\

080919  & $\gtrsim 6$ & Y & $2.97 \pm 0.68$ & 0.1 & 0.07 & \sbo{0.019}{-0.004}{+0.005} & 0.78 & \sbo{0.20}{-0.07}{+0.19} \\
	& $\gtrsim 6$ & Y & $2.97 \pm 0.68$ & 0.01 & 0.07 & \sbo{0.029}{-0.006}{+0.007} & 0.70 & \sbo{5.1}{-1.7}{+4.5} \\

081024A & 0 & N & $2.40 \pm 0.76$ & 0.1 & 0.11 & \sbo{0.11}{-0.07}{+0.22} & 0.51 & \snbfi{8.1}{-7.7}{+150} \\
	& 0 & N & $2.40 \pm 0.76$ & 0.01 & 0.11 & \sbo{0.25}{-0.20}{+0.92} & 0.31 & \snbfo{4.3}{-4.2}{260} \\

081226A & 1.0 & N & $2.27 \pm 0.39$ & 0.1 & 0.09 & \sbo{0.20}{-0.13}{+0.36} & 0.32 & \snbfi{3.2}{-2.9}{+29} \\
	& 1.0 & N & $2.27 \pm 0.39$ & 0.01 & 0.09 & \sbo{0.43}{-0.33}{+1.5} & 0.18 & \snbfo{1.7}{-1.6}{+54} \\

090426A & 0 & Y & $2.13 \pm 0.14$ & 0.1 & 2.0 & \sbo{1.4}{-0.3}{+0.4} & 0.59 & \sbo{0.04}{-0.02}{+0.04} \\
	& 0 & Y & $2.13 \pm 0.14$ & 0.01 & 2.0 & \sbo{1.5}{-0.3}{+0.4} & 0.57 & \sbo{1.2}{-0.6}{+1.4}\\

090510 & 0 & N & $2.65 \pm 0.08$ & 0.1 & 0.77 & \sbo{0.77}{-0.33}{+0.57} & 0.50 & \snbfi{1.2}{-1.0}{+5.5} \\
	& 0 & N & $2.65 \pm 0.08$ & 0.01 & 0.77 & \sbo{1.9}{-1.2}{+3.0} & 0.29 & \snbfi{6.4}{-6.0}{+100} \\

090607  & 0 & Y & $2.40 \pm 0.76$ & 0.1 & 0.10 & \snbth{2.1}{-0.3}{+0.4} & 0.98 & \sbo{0.84}{-0.22}{+0.52} \\
	& 0 & Y & $2.40 \pm 0.76$ & 0.01 & 0.10 & \snbth{2.6}{-0.4}{+0.5} & 0.98 & \sbo{24}{-6.3}{+15} \\

090621B$^{c}$ & $2.5$ & Y & $2.64 \pm 0.72$ & 0.1 & 0.07 & \sbo{0.023}{-0.006}{+0.007} & 0.75 & \sbo{0.05}{-0.02}{+0.06} \\
	& $2.5$ & Y & $2.64 \pm 0.72$ & 0.01 & 0.07 & \sbo{0.031}{-0.007}{+0.009} & 0.68 & \sbo{1.0}{-0.27}{+0.52} \\

091109B & 0 & N & $2.40 \pm 0.32$ & 0.1 & 0.18 & \sbo{0.25}{-0.16}{+0.18} & 0.42 & \snbfi{2.8}{-2.5}{+24} \\*
	& 0 & N & $2.40 \pm 0.32$ & 0.01 & 0.18 & \sbo{0.8}{-0.6}{+2.2} & 0.18 & \snbfi{9.4}{-9.0}{200} \\

100117A & 0 & Y & $2.36 \pm 0.30$ & 0.1 & 0.22 & \sbo{0.019}{-0.003}{+0.003} & 0.92 & \sbo{0.04}{-0.01}{+0.03} \\
	& 0 & Y & $2.36 \pm 0.30$ & 0.01 & 0.22 & \sbo{0.023}{-0.004}{+0.004} & 0.90 & \sbo{1.2}{-0.3}{+0.9} \\

101219A & $\gtrsim 2.5$ & N & $2.73 \pm 0.13$ &  0.1 & 0.74 & \sbo{0.3}{-0.2}{+0.5} & 0.68 & \snbfi{4.6}{-4.3}{+59} \\
	& $\gtrsim 2.5$ & N & $2.73 \pm 0.13$ & 0.01 & 0.74 & \sbo{0.87}{-0.64}{+2.3} & 0.46 & \snbfo{2.4}{-2.3}{+97} \\

110112A$^{b}$ & 0 & Y & $2.49 \pm 0.07$ &  0.1 & 0.03 & \sbo{0.064}{-0.005}{+0.008} & 0.31 & \snbt{2.4}{-0.4}{+0.4} \\

111020A$^{abc}$ & $0.5$ & Y & $2.08 \pm 0.32$ & 0.1 & 0.17 & \sbo{0.48}{-0.08}{+0.09} & 0.26 & \snbth{4.5}{-3.8}{+6.0} \\

111117A & 0 & Y & $2.27 \pm 0.07$ & 0.1 & 0.55 & \sbo{0.06}{-0.01}{+0.01} & 0.90 & \snbth{8.3}{-2.3}{+5.9} \\
	& 0 & Y & $2.27 \pm 0.07$ & 0.01 & 0.55 & \sbo{0.07}{-0.01}{+0.02} & 0.89 & \sbo{0.25}{-0.07}{+0.16} \\

111121A & N/A & N & $2.87 \pm 0.21$ & 0.1 & 2.1 & \sbo{3.1}{-1.2}{+1.9} & 0.40 & \snbsi{8.2}{-6.3}{+27} \\
	& N/A & N & $2.87 \pm 0.21$ & 0.01 & 2.1 & \sbo{8.3}{-4.8}{+11.1} & 0.20 & \snbfi{4.2}{-3.8}{+48} \\

120804A$^{c}$ & $2.5$ & Y & $2.08 \pm 0.11$ & 0.1 & 3.4 & \sbo{1.1}{-0.2}{+0.3} & 0.76 & \snbth{3.2}{-1.5}{+3.1} \\
	& $2.5$ & Y & $2.08 \pm 0.11$ & 0.01 & 3.4 & \sbo{2.3}{-0.2}{+0.2} & 0.60 & \sbo{0.014}{-0.001}{+0.002} \\

121226A$^{ab}$ & 1 & Y & $2.50 \pm 0.37$ & 0.1 & 0.37 & \sbo{0.6}{-0.09}{+0.08} & 0.37 & \snbth{4.0}{-0.6}{+1.0} \\

130603B & 1.2 & Y & $2.70 \pm 0.06$ & 0.1 & 0.37 & \sbo{0.11}{-0.01}{+0.02} & 0.77 & \sbo{0.09}{-0.03}{+0.04} \\
	& 0.3 & Y & $2.70 \pm 0.06$ & 0.01 & 0.37 & \sbo{0.15}{-0.02}{+0.02} & 0.72 & \sbo{0.31}{-0.04}{+0.08} \\

130912A$^{c}$ & $1.3$ & N & $2.49 \pm 0.17$ & 0.1 & 0.16 & \sbo{0.50}{-0.20}{+0.37} & 0.25 & \snbfi{1.7}{-1.5}{+9.9} \\ 
	    & $1.3$ & N & $2.49 \pm 0.17$ & 0.01 & 0.16 & \sbo{1.4}{-0.8}{+0.2} & 0.11 & \snbfi{5.2}{-4.9}{+7.6} \\

131004A & 0 & N & $2.57 \pm 0.48$ & 0.1 & 0.45 & \sbo{1.2}{-0.36}{+0.56} & 0.28 & \snbfi{1.2}{-1.0}{+5.7} \\
	& 0 & N & $2.57 \pm 0.48$ & 0.01 & 0.45 & \sbo{2.8}{-1.5}{+3.4} & 0.45 & \snbfo{6.5}{-6.2}{+11}  \\
	
140129B & 0 & N & $3.00 \pm 0.19$ & 0.1 & 0.07 & \sbo{0.98}{-0.53}{+1.14} & 0.06 & \snbfi{3.1}{-2.8}{+29.4} \\
	& 0 & N & $3.00 \pm 0.19$ & 0.01 & 0.07 & \sbo{3.8}{-2.9}{+12.0} & 0.02 & \snbfo{1.6}{-1.5}{+47} \\

140516A & 0 & N & $2.40 \pm 0.88$ & 0.1 & 0.02 & \sbo{0.02}{-0.01}{+0.04} & 0.54 & \snbfo{1.0}{-0.99}{+24} \\
	& 0 & N & $2.40 \pm 0.88$ & 0.01 & 0.02 & \sbo{0.04}{-0.03}{+0.18} & 0.34 & \snbfo{5.5}{-5.4}{+41} \\

140622A & 0 & N & $2.10 \pm 0.60$ & 0.1 & 0.07 & \sbo{0.12}{-0.08}{+0.26} & 0.36 & \snbfi{5.8}{-5.4}{+8.8} \\
	& 0 & N & $2.10 \pm 0.60$ & 0.01 & 0.07 & \sbo{0.25}{-0.20}{+0.97} & 0.21 & \snbfo{3.2}{-3.2}{+16.2} \\
	
140903A & 0 & N & $2.27 \pm 0.16$ & $10^{-3}$ & 0.08 & \sbo{2.9}{-0.74}{+0.92} & 0.03 & \snbth{3.4}{-1.6}{+2.9} \\

140930B & 0 & N & $2.67 \pm 0.19$ & 0.1 & 0.40 & \sbo{0.28}{-0.17}{+0.42} & 0.58 & \snbfi{4.6}{-4.3}{+58} \\
	& 0 & N & $2.67 \pm 0.19$ & 0.01 & 0.40 & \sbo{1.8}{-0.9}{+1.7} & 0.18 & \snbfi{1.8}{-1.5}{11} \\
	
150101B$^{c}$ & 0.5 & N & $2.40 \pm 0.17$ & 0.1 & \snth{4.0} & \sbo{0.61}{-0.25}{+0.45} & \snth{6.6} & \snbsi{8.0}{-6.1}{+24} \\
	& 0.5 & N & $2.40 \pm 0.17$ & 0.1 & \snth{4.0} & \sbo{3.0}{-1.1}{+1.7} & \snth{1.3} & \snbsi{5.3}{-3.6}{+11}
\enddata
\tablecomments{Quoted uncertainties are $1\sigma$. All solutions presented here are for a fixed $\epsilon_e=0.1$ and assume a lower density bound of $n_{\rm min}=10^{-6}$~cm$^{-3}$.  \\
$^{a}$ We assume a redshift of $z=1$ for this burst. \\
$^{b}$ No valid solution is found for $\epsilon_B=0.01$. \\
$^{c}$ Value of $A_V^{\rm host}$ is determined from a comparison of the optical and X-ray bands, and not directly from the optical/NIR SED. \\
$^{d}$ Determined from $\alpha_X$ alone. \\
}
\end{deluxetable*}
\linespread{1}

\begin{figure}
\centering
\includegraphics*[width=\columnwidth,clip=]{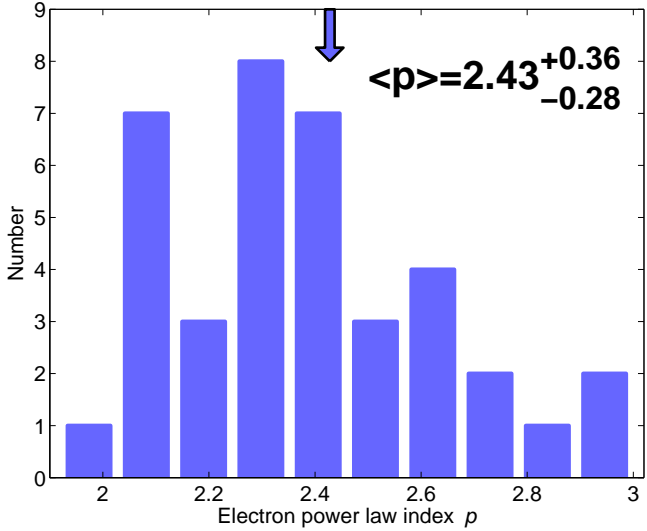} 
\caption{Histogram of electron power law index, $p$, for 38 short GRBs as inferred from the X-ray temporal and spectral indices. The weighted mean and $1\sigma$ uncertainties for the population (blue arrow) is $\langle p \rangle=2.43^{+0.36}_{-0.28}$. These values correspond to those listed in Table~\ref{tab:prop}.
\label{fig:phist}}
\end{figure}

To determine the electron power law index $p$, we use $\alpha_X$ and $\beta_X$ to constrain the location of the cooling frequency, $\nu_c$, with respect to the X-ray band. We use the relations given by \citet{gs02} which relate $\alpha_X$ and $\beta_X$ to the value of $p$ for the scenarios $\nu_m<\nu_X<\nu_c$ and $\nu_c<\nu_X$ by

\begin{equation}
\label{eqn:gtnux}
 p=
\begin{cases}
\left.\begin{aligned}
 1-2\beta_X \\
 1-\frac{4}{3}\alpha_X
\end{aligned}
\right\}
\qquad \nu_m<\nu_X<\nu_c
\end{cases}
\end{equation}

\begin{equation}
\label{eqn:ltnux}
 p=
\begin{cases}
\left.\begin{aligned}
 -2\beta_X \\
 \frac{2-4\alpha_X}{3}
\end{aligned}
\right\}
\qquad \nu_c<\nu_X.
\end{cases}
\end{equation}

\noindent For a given burst, we calculate the values of $p$ and $1\sigma$ uncertainties using Equations~\ref{eqn:gtnux} and \ref{eqn:ltnux} and standard propagation of errors. We select the valid scenario under the condition that the values of $p$ independently determined from $\alpha_X$ and $\beta_X$ for a given scenario agree within the $1\sigma$ uncertainties. Following this condition, we can constrain the location of $\nu_c$ with respect to the X-ray band for 38 bursts (Table~\ref{tab:prop}). Using Equations~\ref{eqn:gtnux} and \ref{eqn:ltnux}, we calculate the weighted mean for the value of $p$ in the valid scenario; the resulting values and uncertainties are listed in Table~\ref{tab:prop} and displayed in Figure~\ref{fig:phist}. We note that in the case of GRB\,071227, the condition is satisfied for $\nu_c<\nu_X$, but has a median of $p=1.92 \pm 0.31$, which yields a divergent total integrated energy. Thus for this burst, we employ $p=2.05$ in our subsequent analysis (Table~\ref{tab:prop}).

Under the reasonable assumption that the optical band lies between the peak frequency, $\nu_m$, and the cooling frequency, $\nu_c$ (i.e., $\nu_m<\nu_{\rm opt}<\nu_c$), we use the available values for $\alpha_{\rm opt}$ and $\beta_{\rm opt}$ and Equation~\ref{eqn:gtnux} to independently determine the value of $p$. We then include this in our weighted average of $p$ for each burst (Figure~\ref{fig:phist}). The weighted mean for the sample of $38$ bursts is $\left\langle p \right\rangle = 2.43^{+0.36}_{-0.28}$ ($1\sigma$; Figure~\ref{fig:phist}). 

For the remaining bursts in the afterglow catalog, there is not enough information to determine the location of the cooling frequency with respect to the X-ray band, and therefore the correct value of $p$. We thus concentrate on the subset of $38$ bursts with determined values of $p$ for our subsequent analysis. We find $\nu_c<\nu_X$ for 18 cases, while $\nu_c>\nu_X$ for 20 cases.

\subsection{The Isotropic-Equivalent Kinetic Energies and Densities for Individual Bursts}
\label{sec:ipd}

In the standard synchrotron model from \citet{gs02}, the dependencies on the isotropic-equivalent kinetic energy, circumburst density, and the microphysical parameters are as follows for a given flux density, $F_{\nu,i}$ and observing band, $\nu_i$:

\begin{equation}
\label{eqn:fnu}
F_{\nu,i} \propto
\begin{cases}
 n_0^{1/2} E_{\rm K,iso,52}^{5/6} \epsilon_{e,-1}^{-2/3} \epsilon_{B,-1}^{1/3} & \nu_a<\nu_i<\nu_m \\
\smallskip
 n_0^{1/2} E_{\rm K,iso,52}^{\frac{3+p}{4}} \epsilon_{e,-1}^{p-1} \epsilon_{B,-1}^{\frac{1+p}{4}} & \nu_m<\nu_i<\nu_c \\
\smallskip
 E_{\rm K,iso,52}^{\frac{2+p}{4}} \epsilon_{e,-1}^{p-1} \epsilon_{B,-1}^{\frac{p-2}{4}} & \nu_i>\nu_c \\
\end{cases}
\end{equation}

\noindent where $n_0$ is in units of cm$^{-3}$, $E_{\rm K,iso,52}$ is in units of $10^{52}$~erg, and $\epsilon_{e,-1}$ and $\epsilon_{B,-1}$ are in units of 0.1. In addition to these four parameters, $F_{\nu,i}$ is also dependent on the redshift, luminosity distance, $\nu_i$, time after the burst, $\delta t$, and the value of $p$; the exact dependencies are provided in \citet{gs02}. We note that for $\nu_i>\nu_c$, the flux density is independent of circumburst density. For bursts with no spectroscopic redshift, we assume $z=0.5$. In all cases, we cannot independently constrain $\epsilon_e$ and $\epsilon_B$ since this requires knowledge of the locations of the three break frequencies ($\nu_a$, $\nu_m$, and $\nu_c$), which generally necessitates well-sampled light curves and spectra in multiple bands. Thus, in order to determine ranges for $E_{\rm K,iso}$ and $n$, we fix the values of the microphysical parameters. We first consider the fiducial case that $\epsilon_e=0.1$ and $\epsilon_B=0.1$.

For each burst, we determine the constraints on $E_{\rm K,iso}$ and $n$ by computing individual probability distributions for each observation. We then assign the probabilities to a grid of values, and use a joint probability analysis to calculate the distributions in each parameter. For the grid, the ranges of the density and isotropic-equivalent kinetic energy are $n_i=10^{-6}-10^{3}$~cm$^{-3}$ and $E_i=10^{46}-10^{54}$~erg, with $1000$ logarithmically, uniformly-spaced steps in each parameter. We choose the lower bound of the density range, $n_{\rm min}=10^{-6}$~cm$^{-3}$ to match the typical density of the intergalactic medium (IGM).

\begin{figure*}
\begin{minipage}[c]{\textwidth}
\tabcolsep0.0in
\includegraphics*[width=0.5\textwidth,clip=]{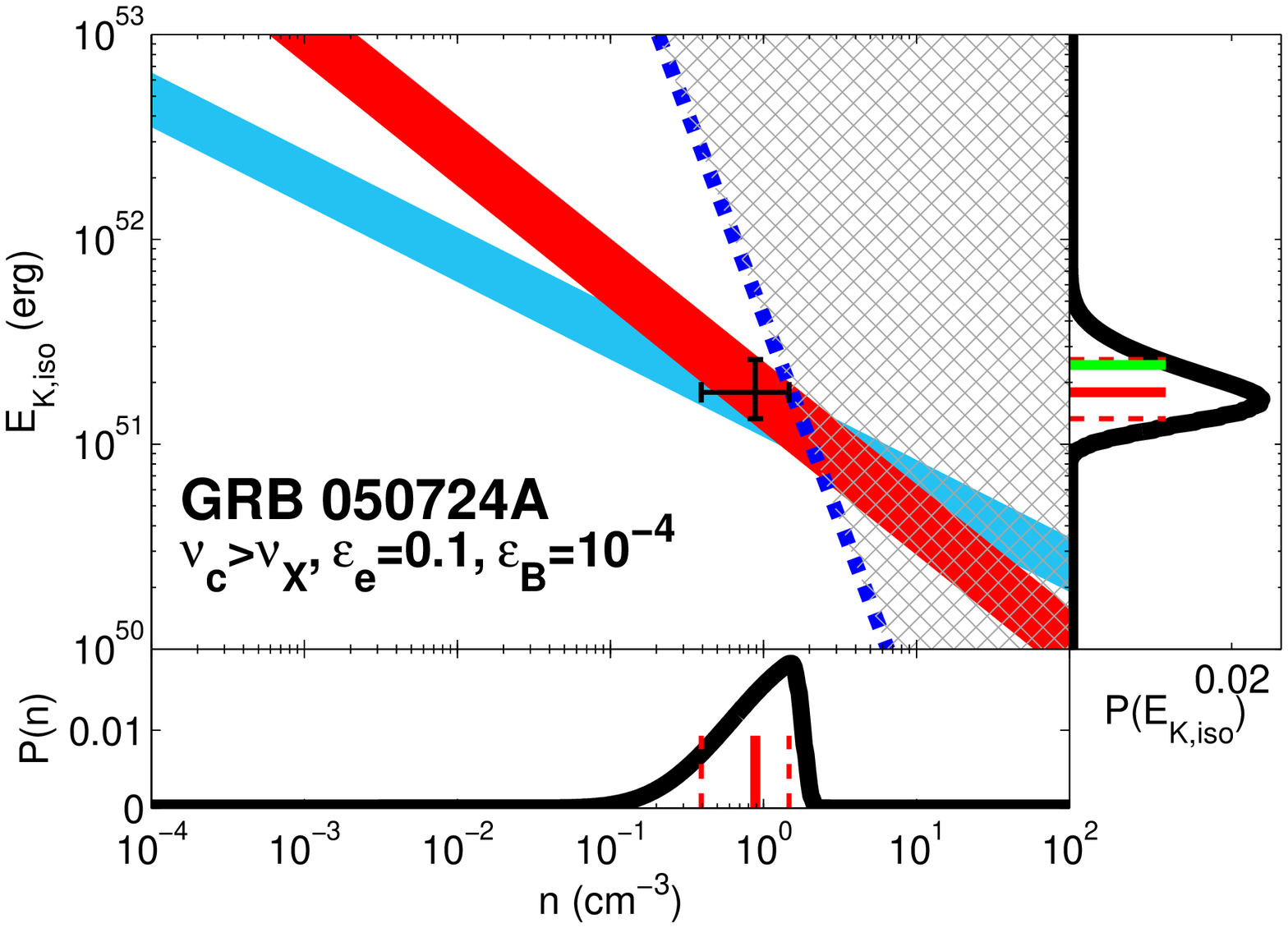}
\includegraphics*[width=0.5\textwidth,clip=]{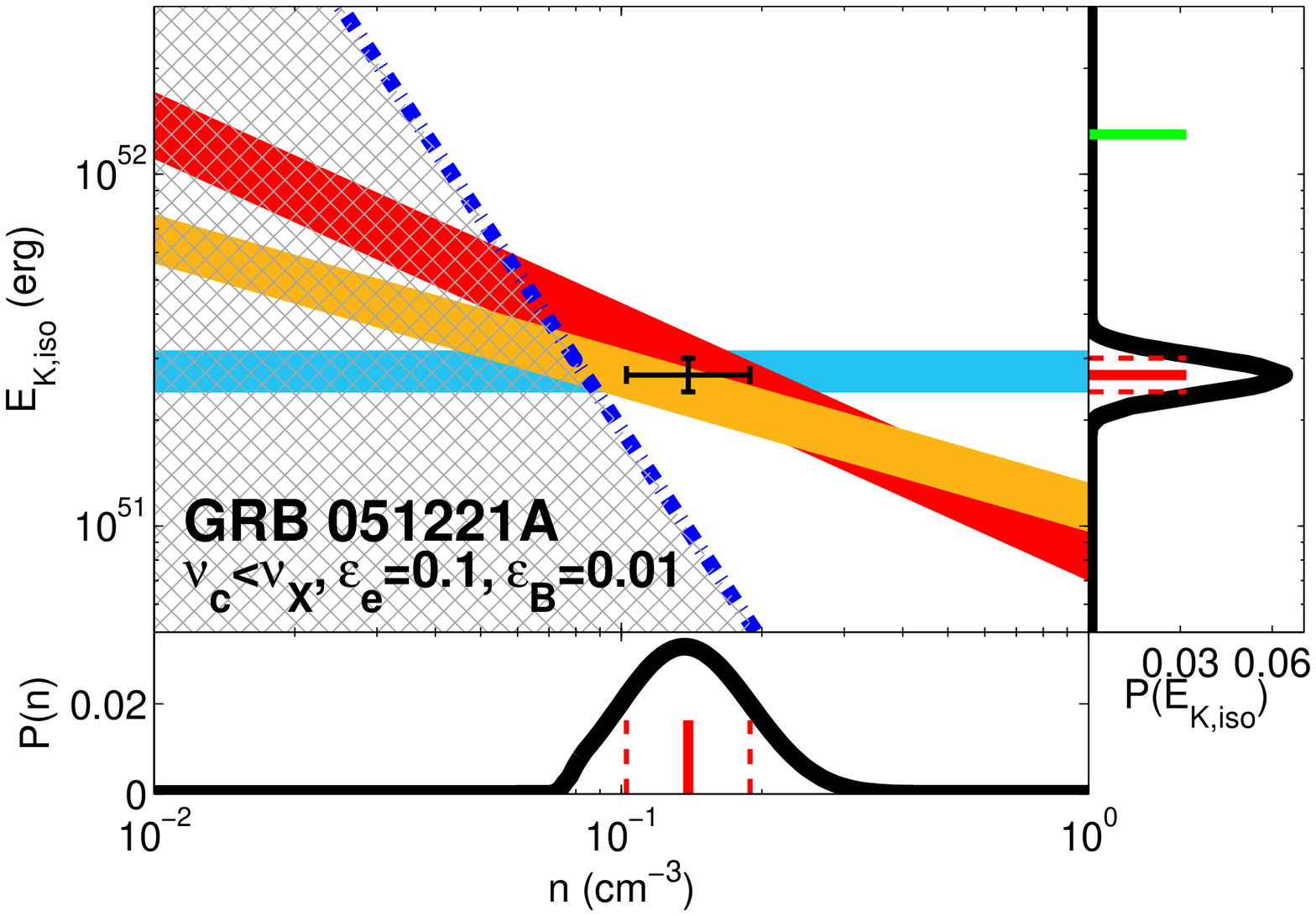} \\
\includegraphics*[width=0.5\textwidth,clip=]{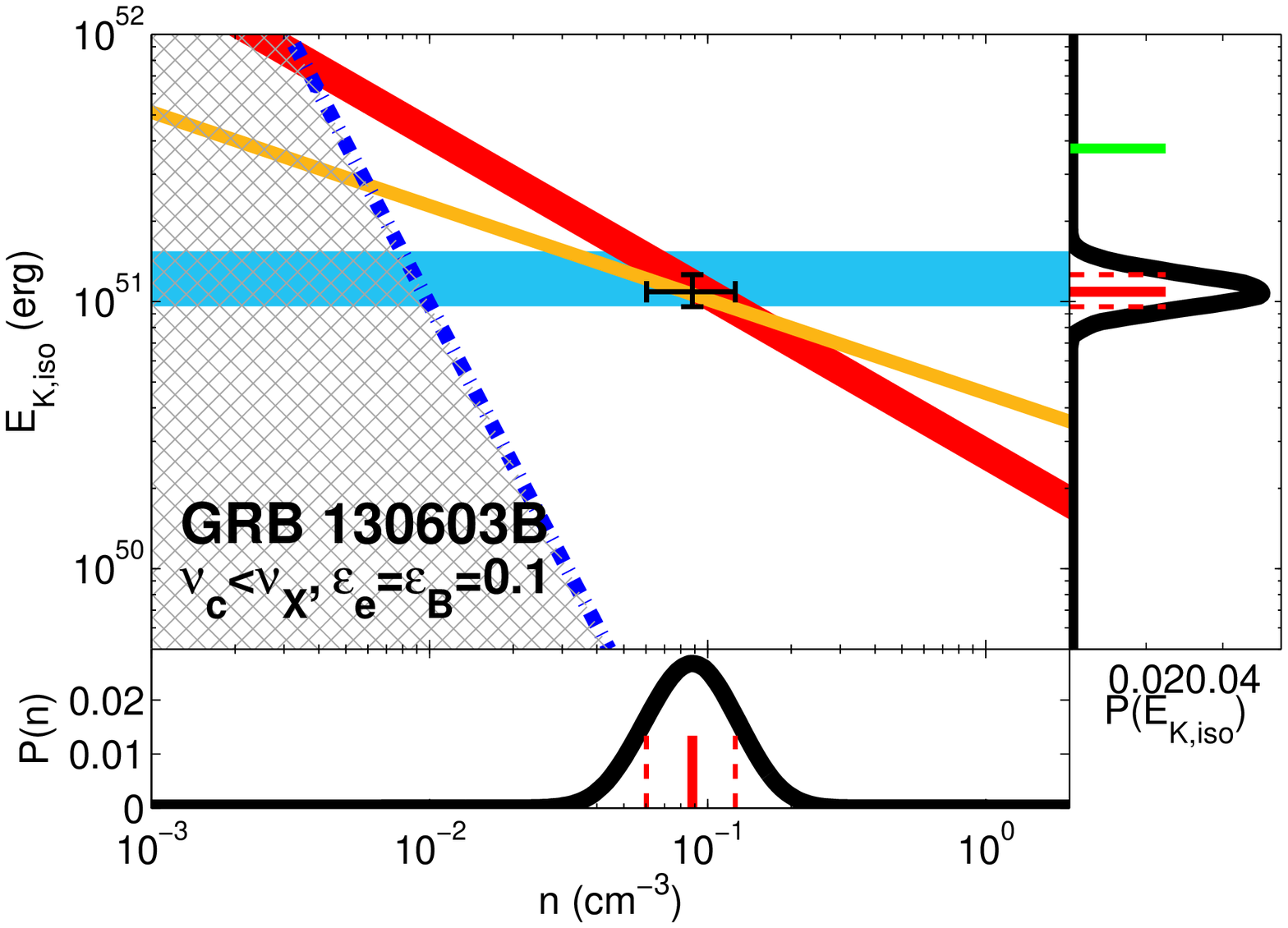}
\includegraphics*[width=0.5\textwidth,clip=]{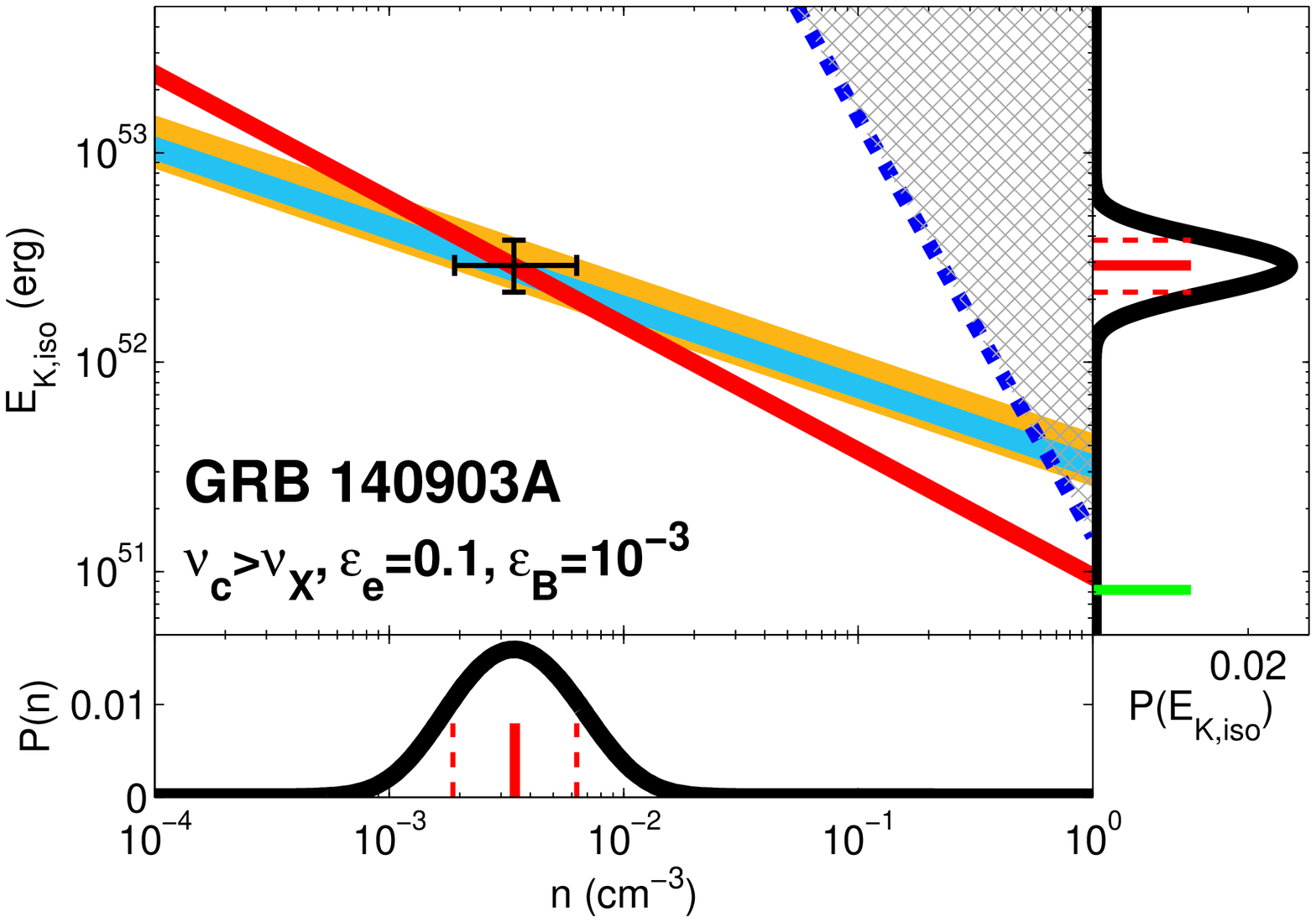}
\end{minipage}
\caption{Isotropic-equivalent kinetic energy versus circumburst density for four short GRBs with radio afterglow detections. In each panel, the X-rays (light blue), optical (orange) and radio (red) provide independent constraints on the parameter space. Measurements are shown as solid regions, where the width of the region corresponds to the $1\sigma$ uncertainty. Upper (lower) limits set by the cooling frequency are denoted by blue dotted (dash-dotted) lines. The regions of parameter space ruled out by the observations are denoted (grey hatched regions). The median solution and $1\sigma$ uncertainty is indicated by the black cross in each panel, corresponding to the values listed in Table~\ref{tab:prop}. For each burst, the joint probability distributions in $n$ (bottom panel) and $E_{\rm K,iso}$ (right panel) are shown. Red lines correspond to the median, and dotted lines are the $1\sigma$ uncertainty about the median. The green line corresponds to $E_{\gamma,{\rm iso}}$. The only optical observations available for GRB\,050724A are during a flare; thus we do not include the optical data in our analysis. The addition of a detection in the radio band is crucial in constraining the best-fit solution, and in three cases constrains $\epsilon_B$.
\label{fig:rad}}
\end{figure*}

\begin{figure*}
\begin{minipage}[c]{\textwidth}
\tabcolsep0.0in
\includegraphics*[width=0.247\textwidth,clip=]{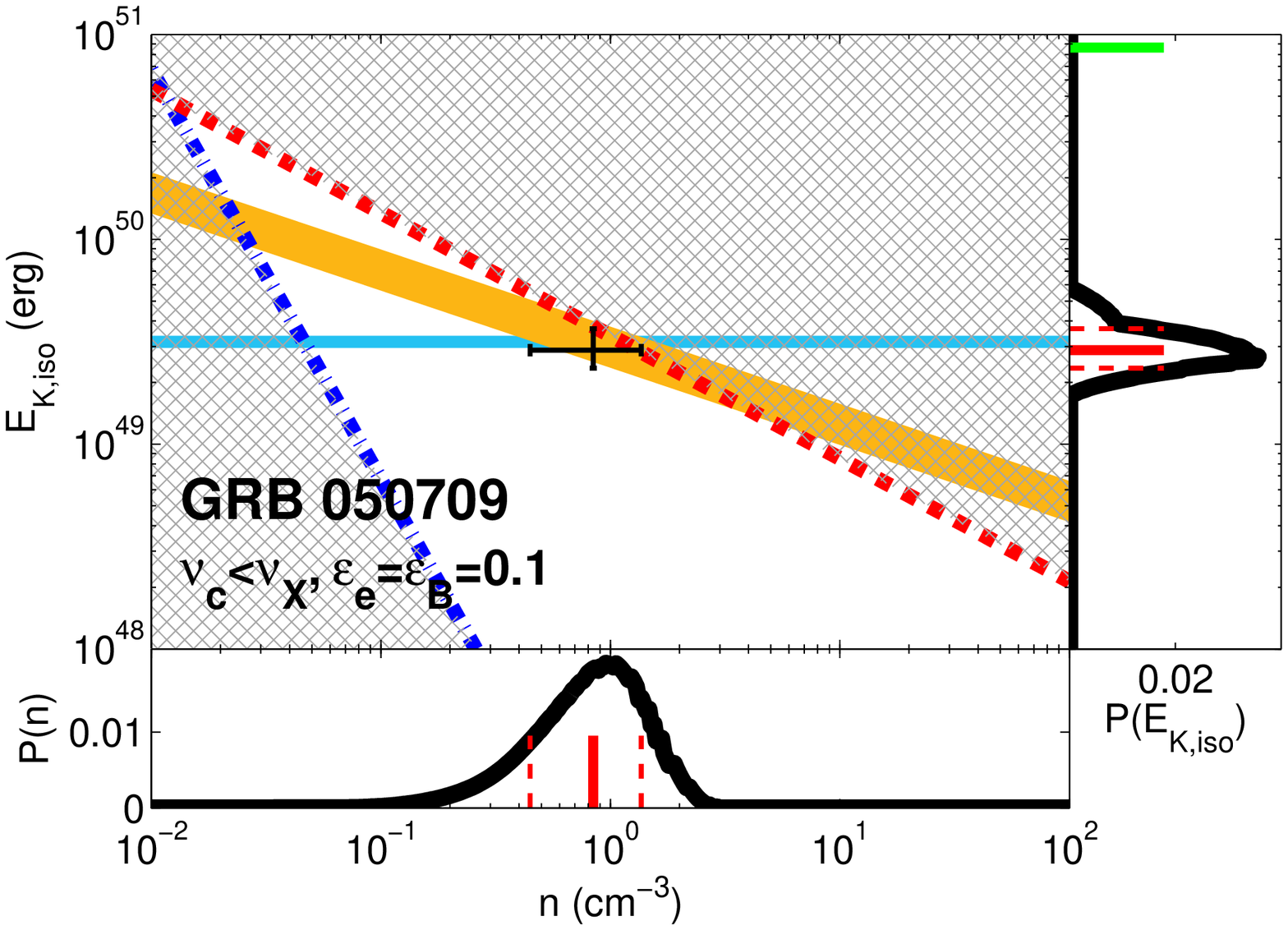} 
\includegraphics*[width=0.247\textwidth,clip=]{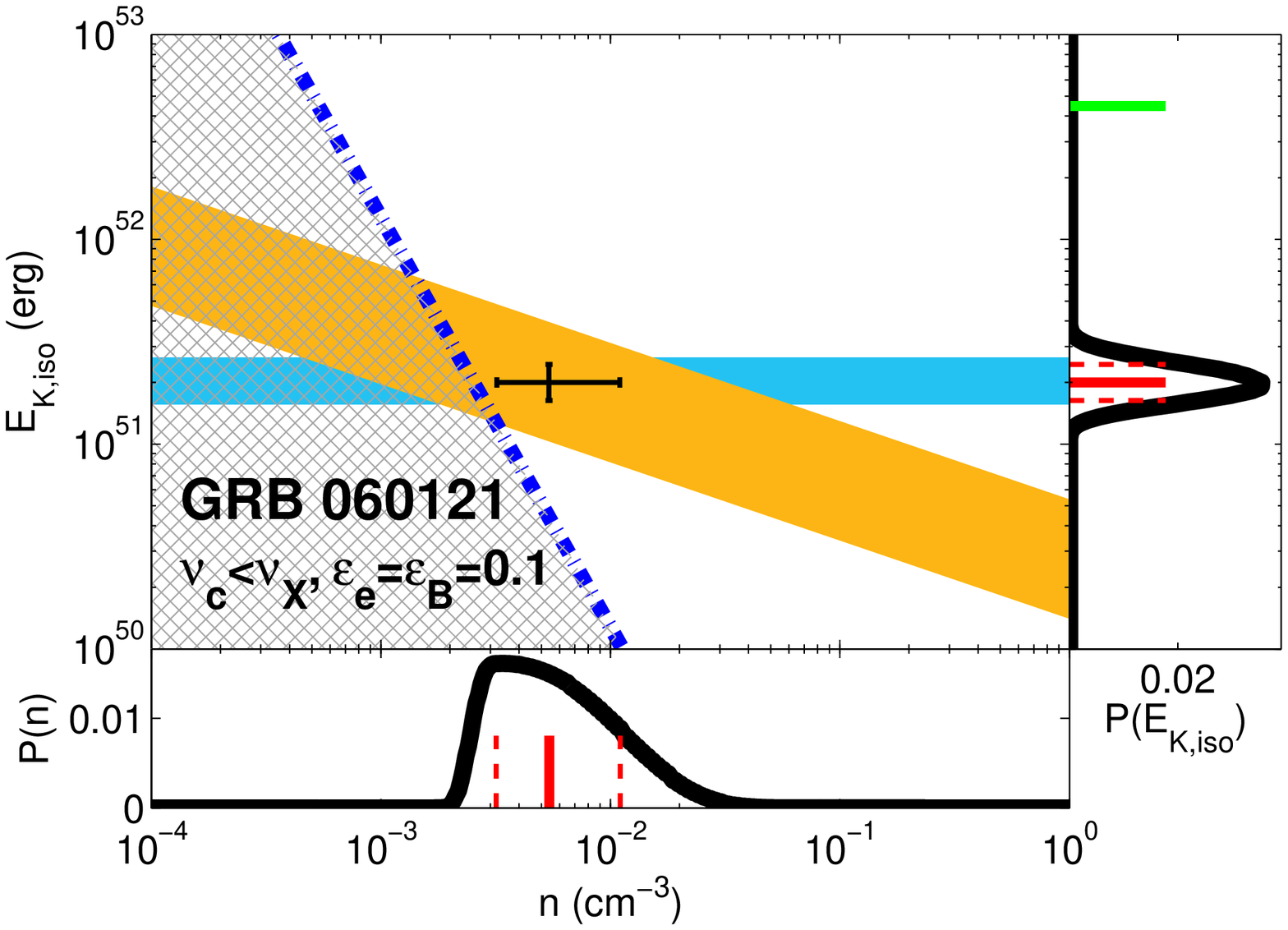} 
\includegraphics*[width=0.247\textwidth,clip=]{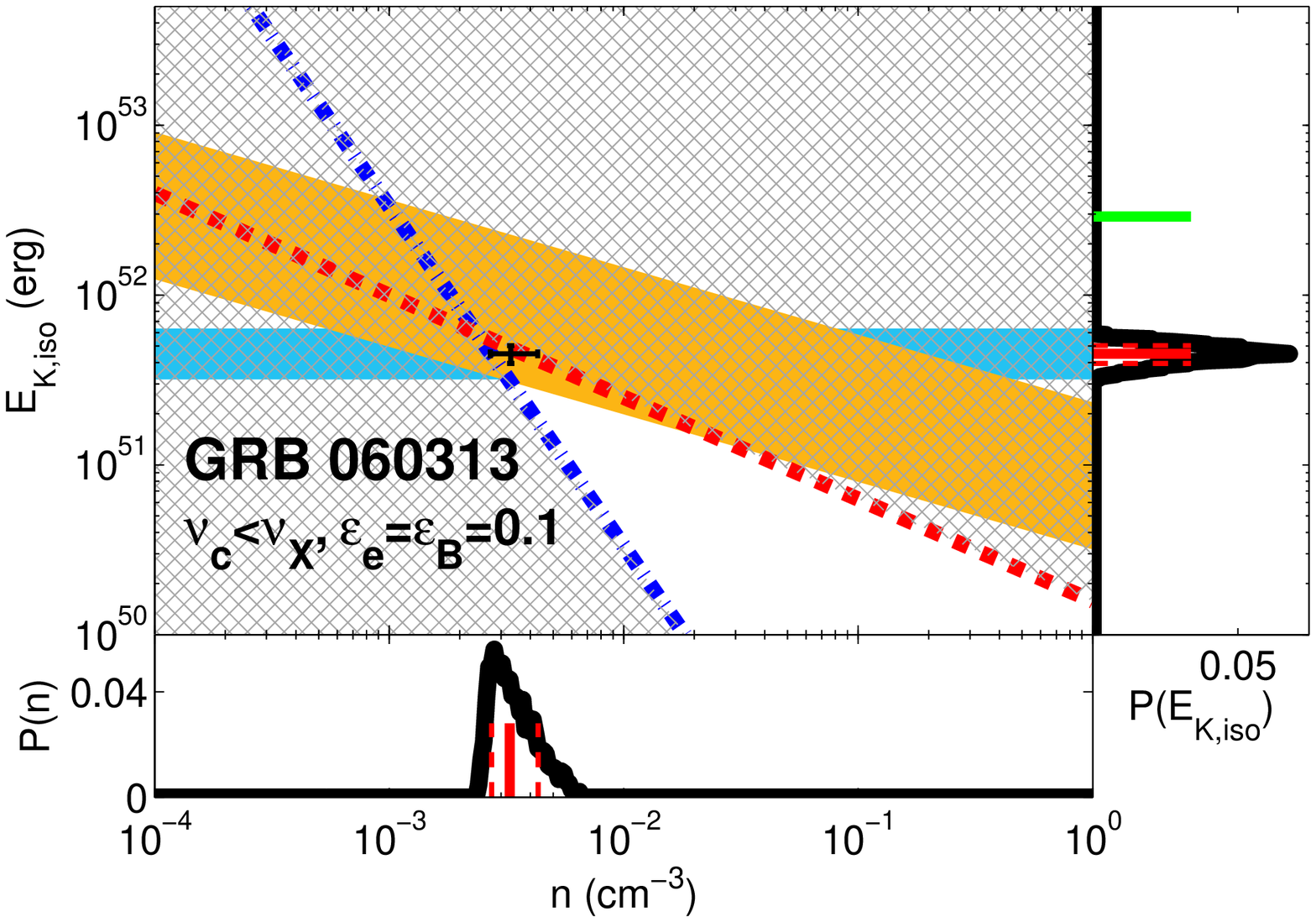} 
\includegraphics*[width=0.247\textwidth,clip=]{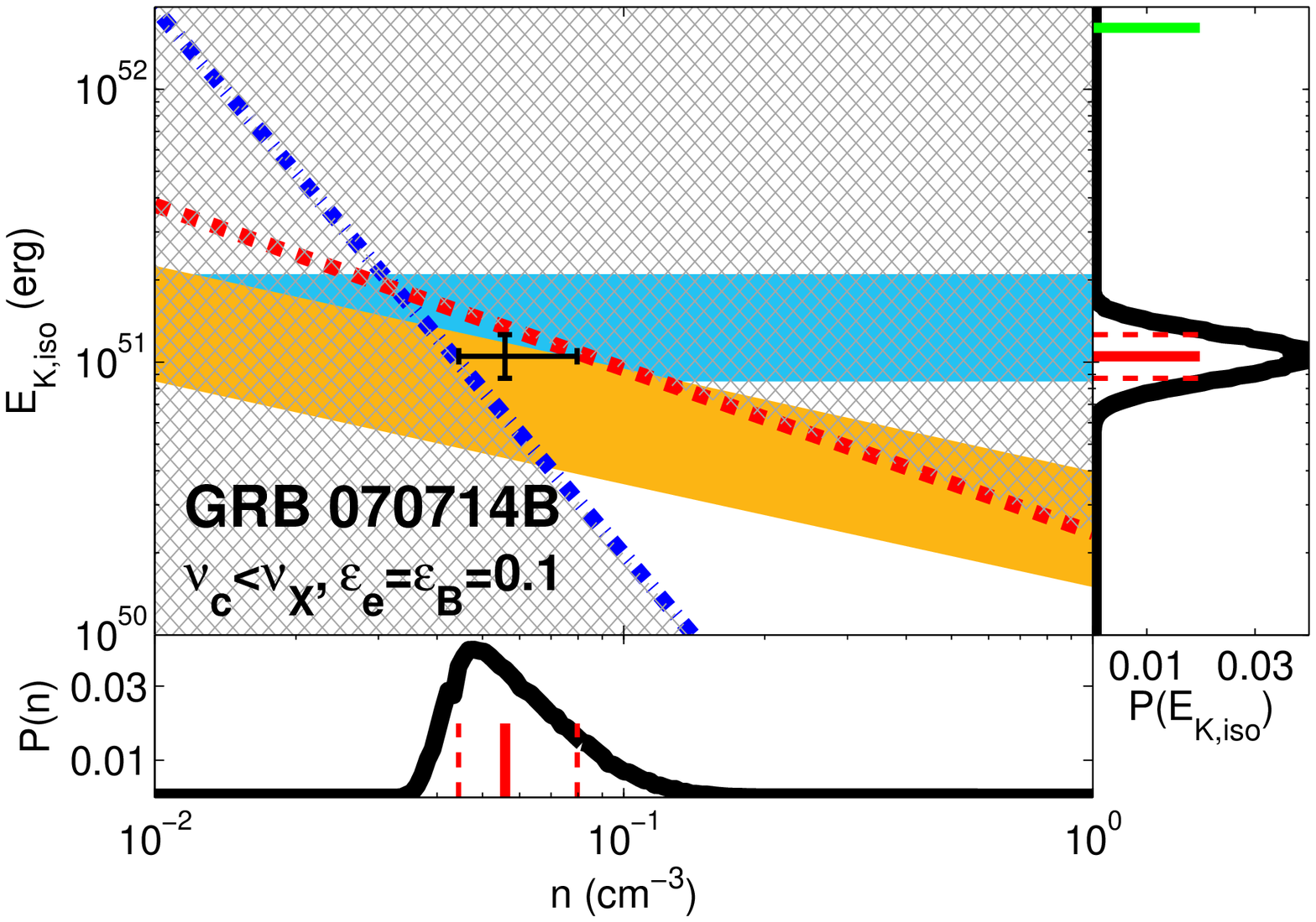} \\  
\includegraphics*[width=0.247\textwidth,clip=]{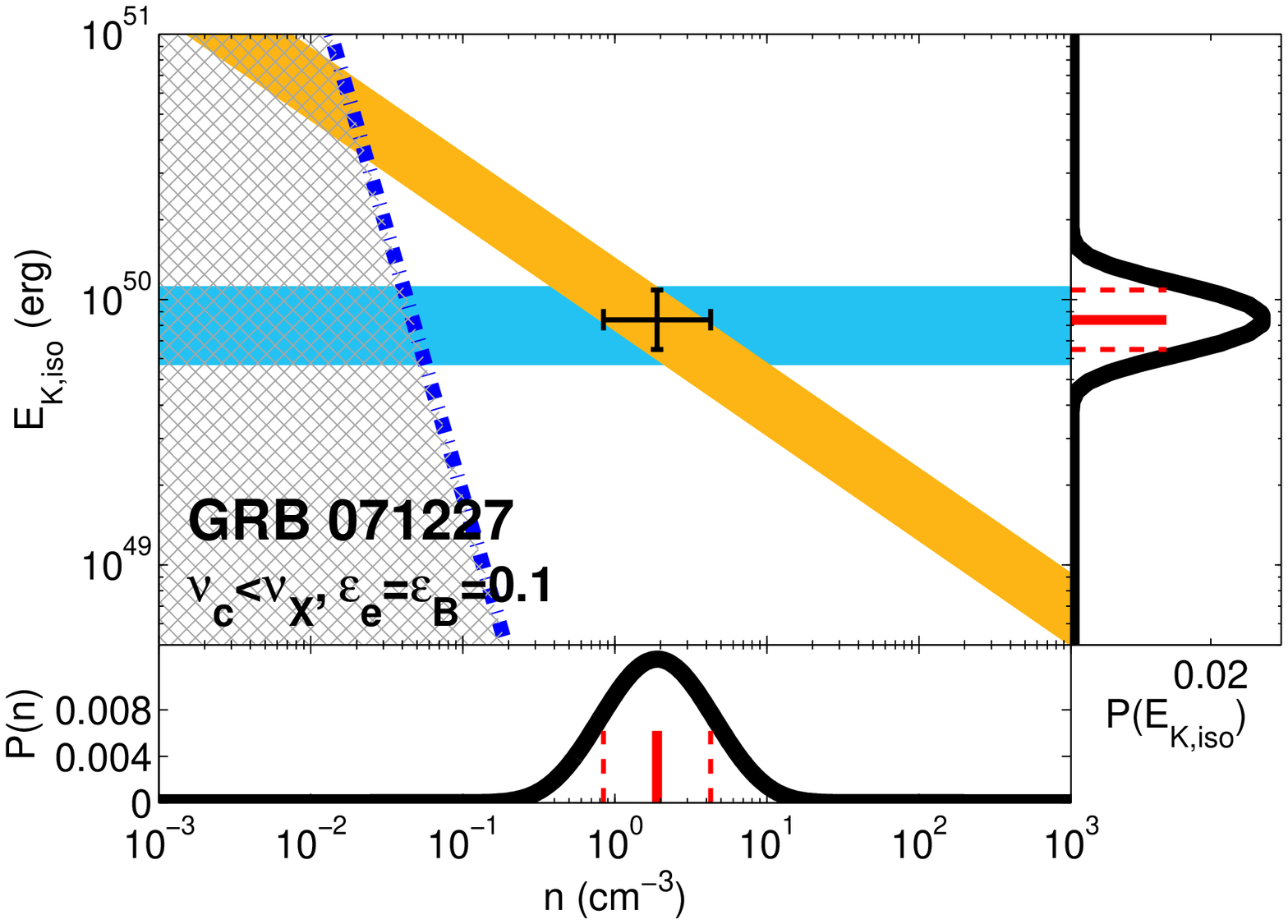} 
\includegraphics*[width=0.247\textwidth,clip=]{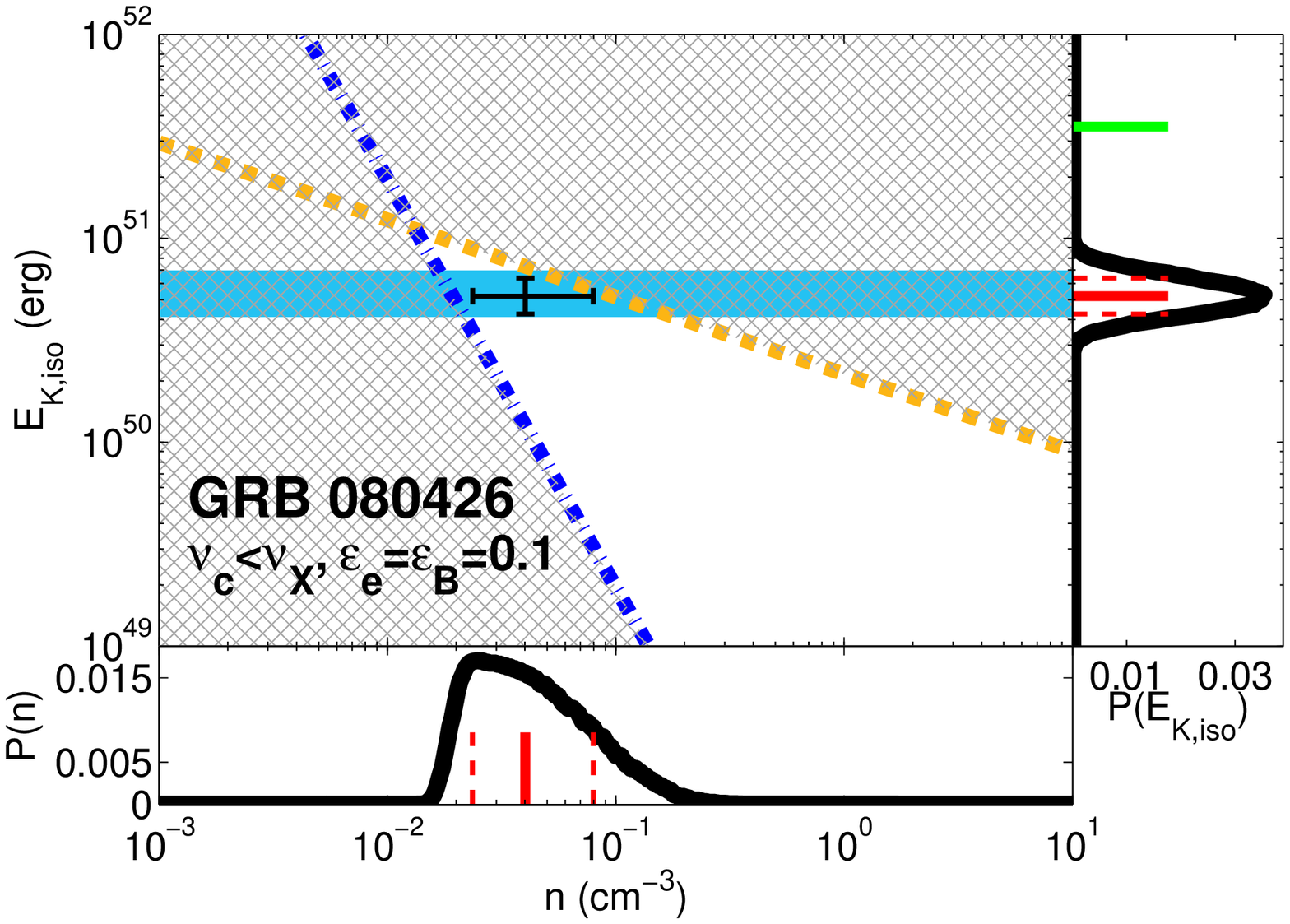} 
\includegraphics*[width=0.247\textwidth,clip=]{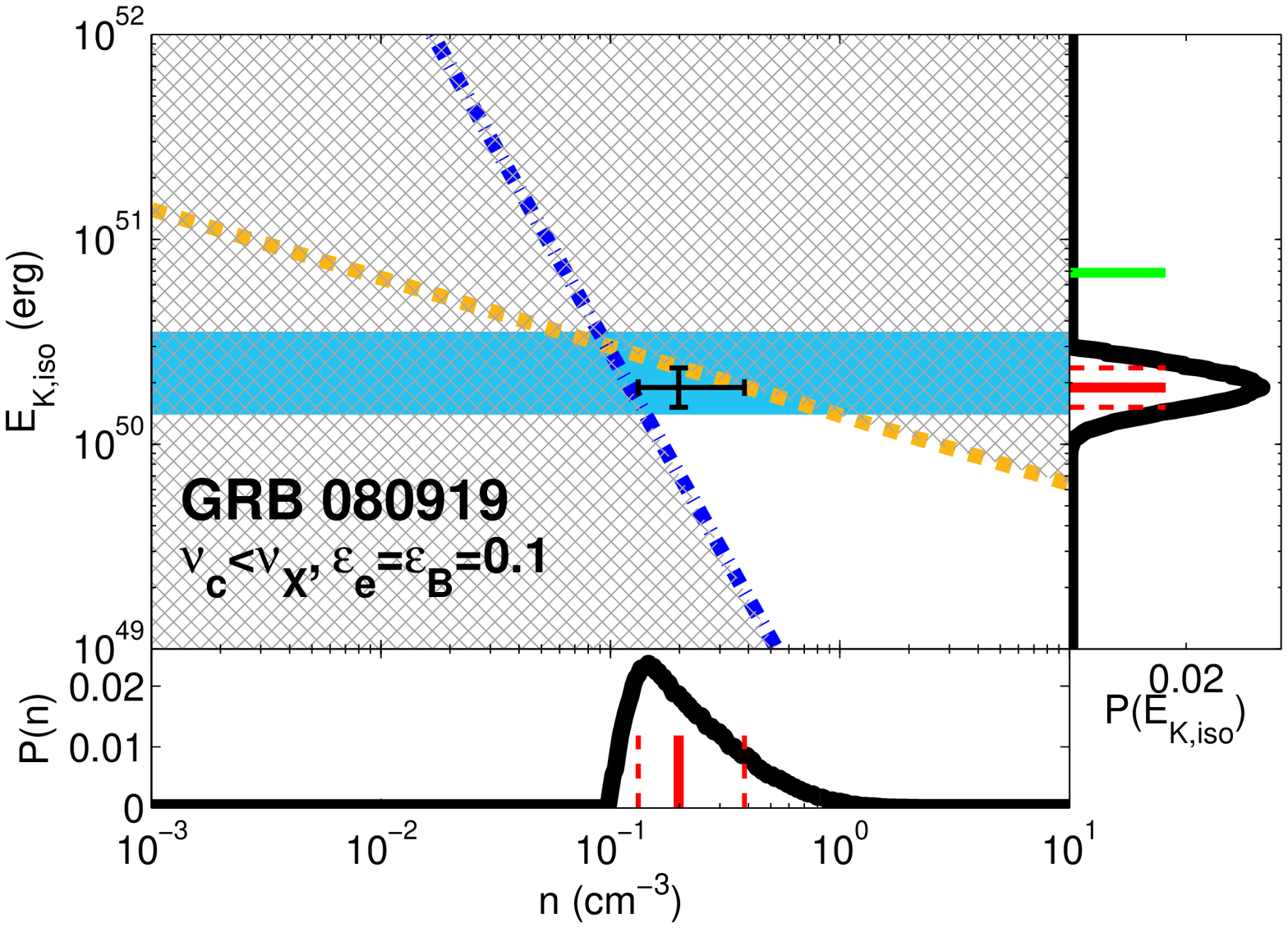} 
\includegraphics*[width=0.247\textwidth,clip=]{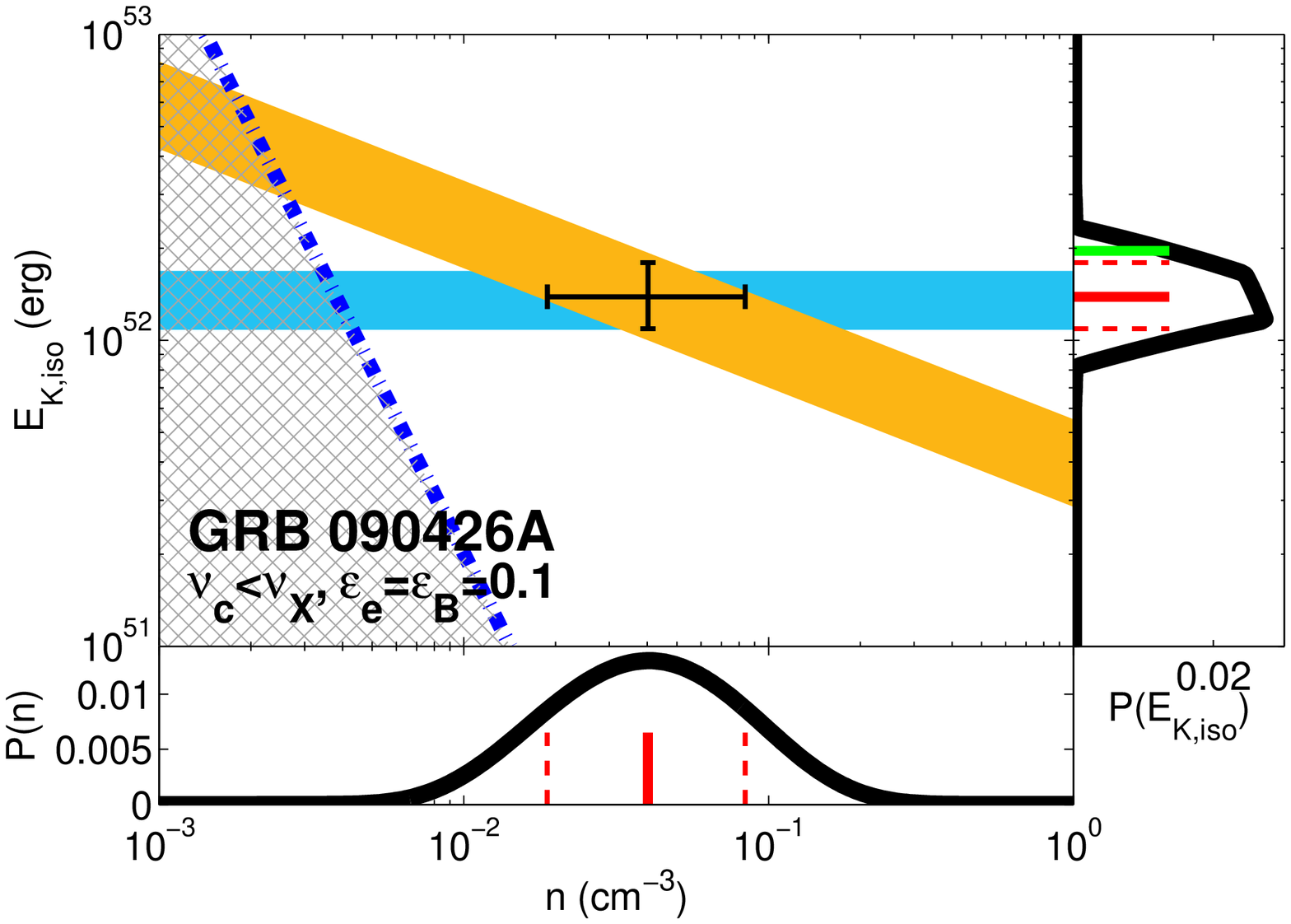} \\
\includegraphics*[width=0.247\textwidth,clip=]{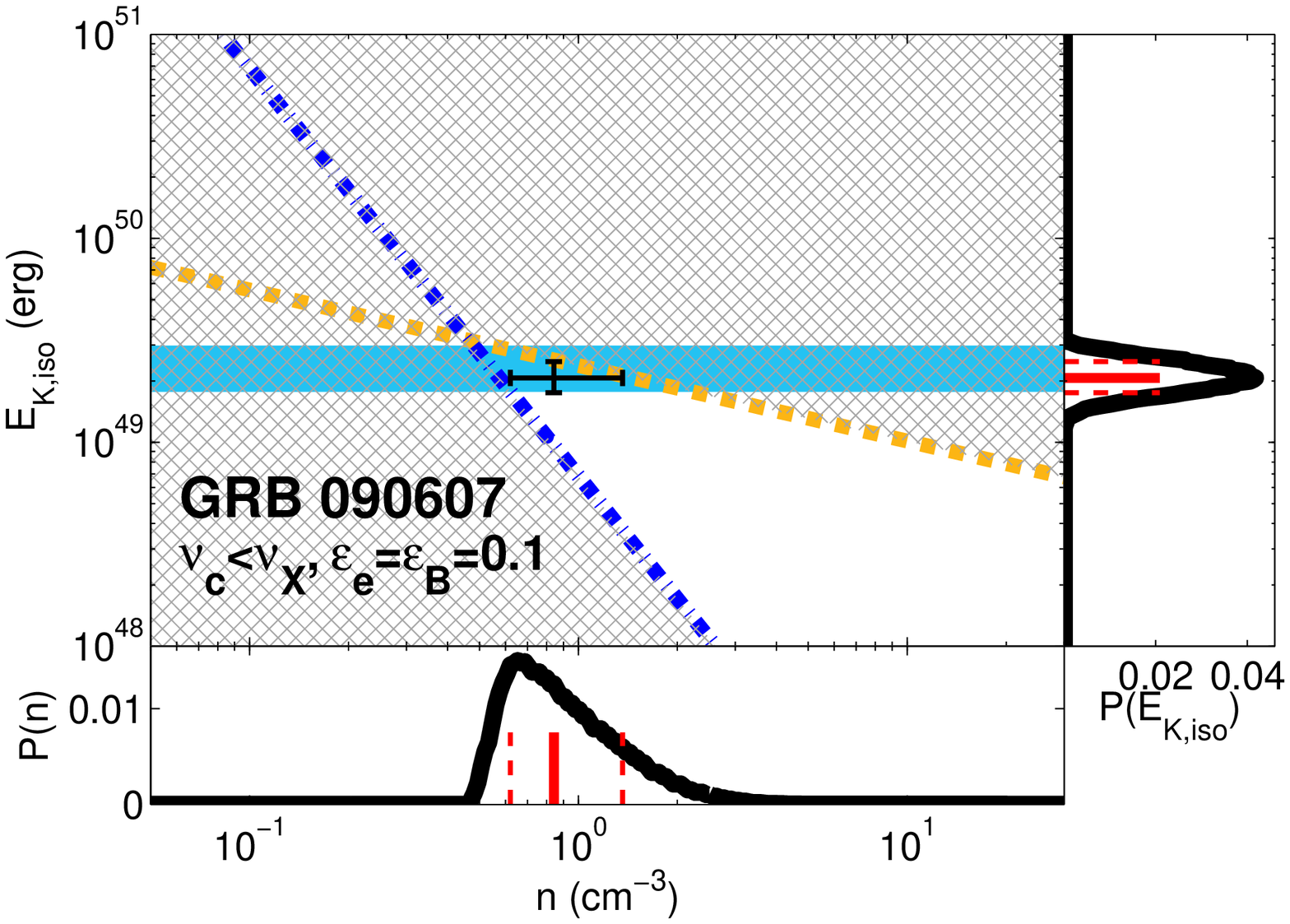} 
\includegraphics*[width=0.247\textwidth,clip=]{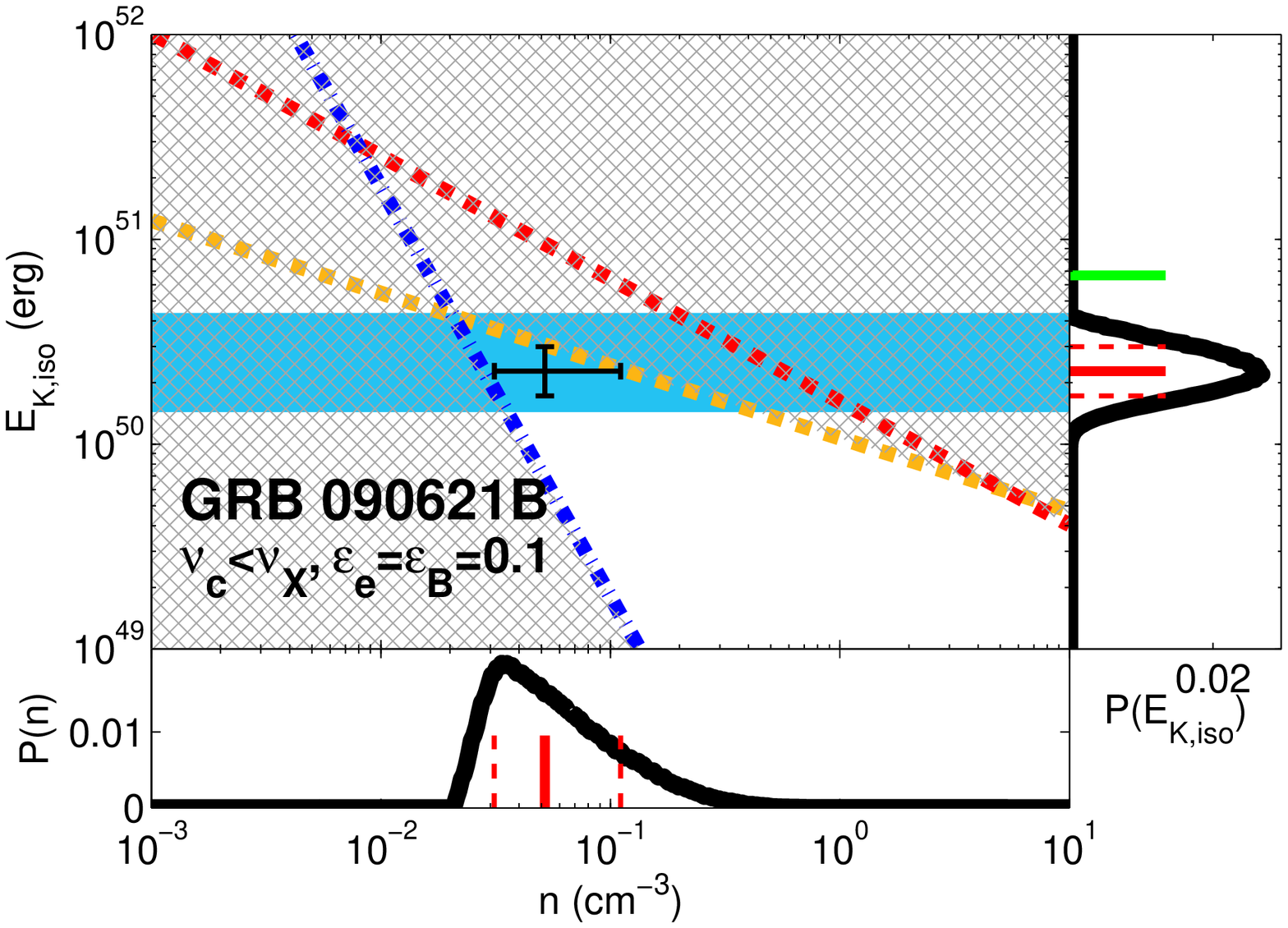} 
\includegraphics*[width=0.247\textwidth,clip=]{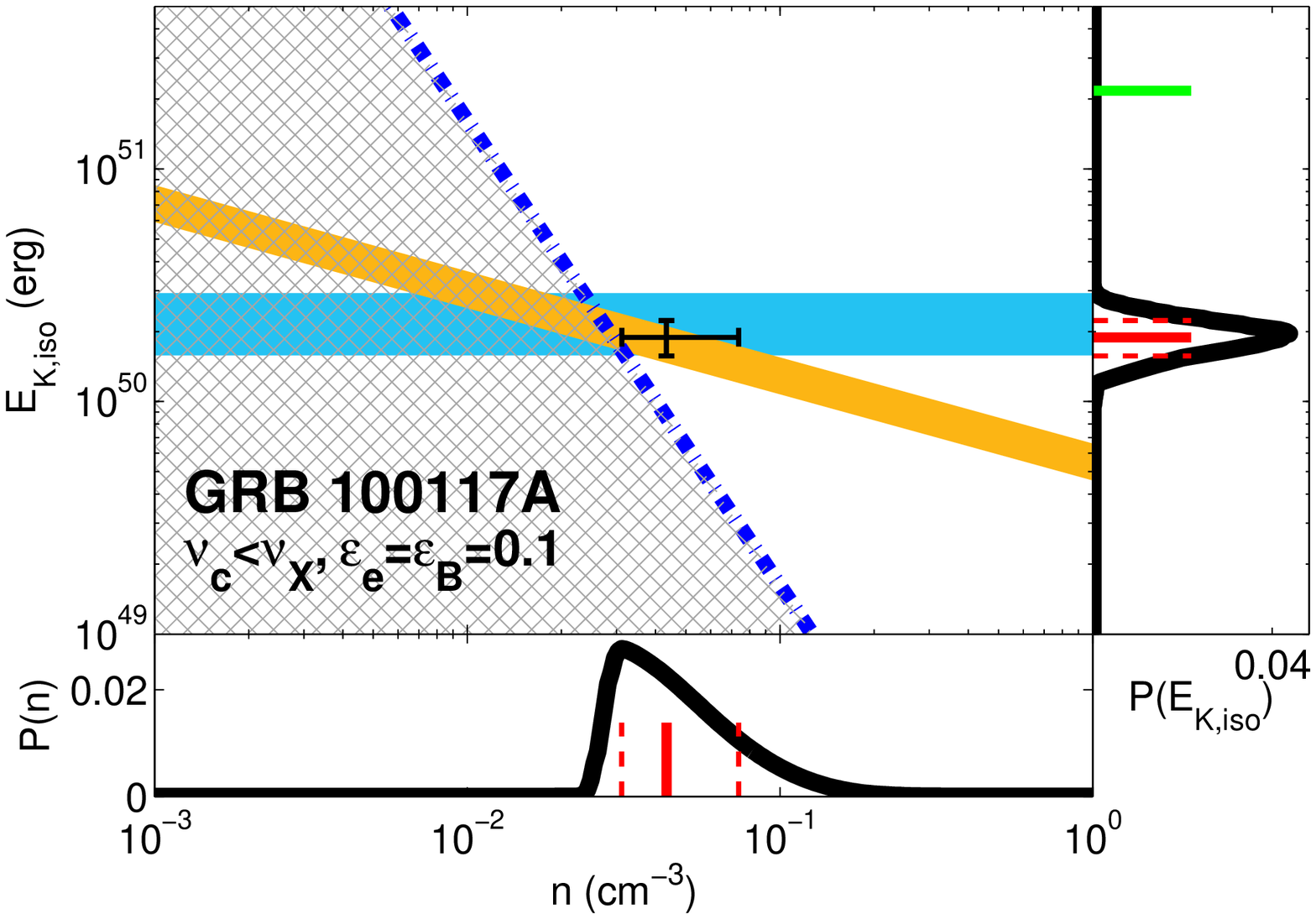} 
\includegraphics*[width=0.247\textwidth,clip=]{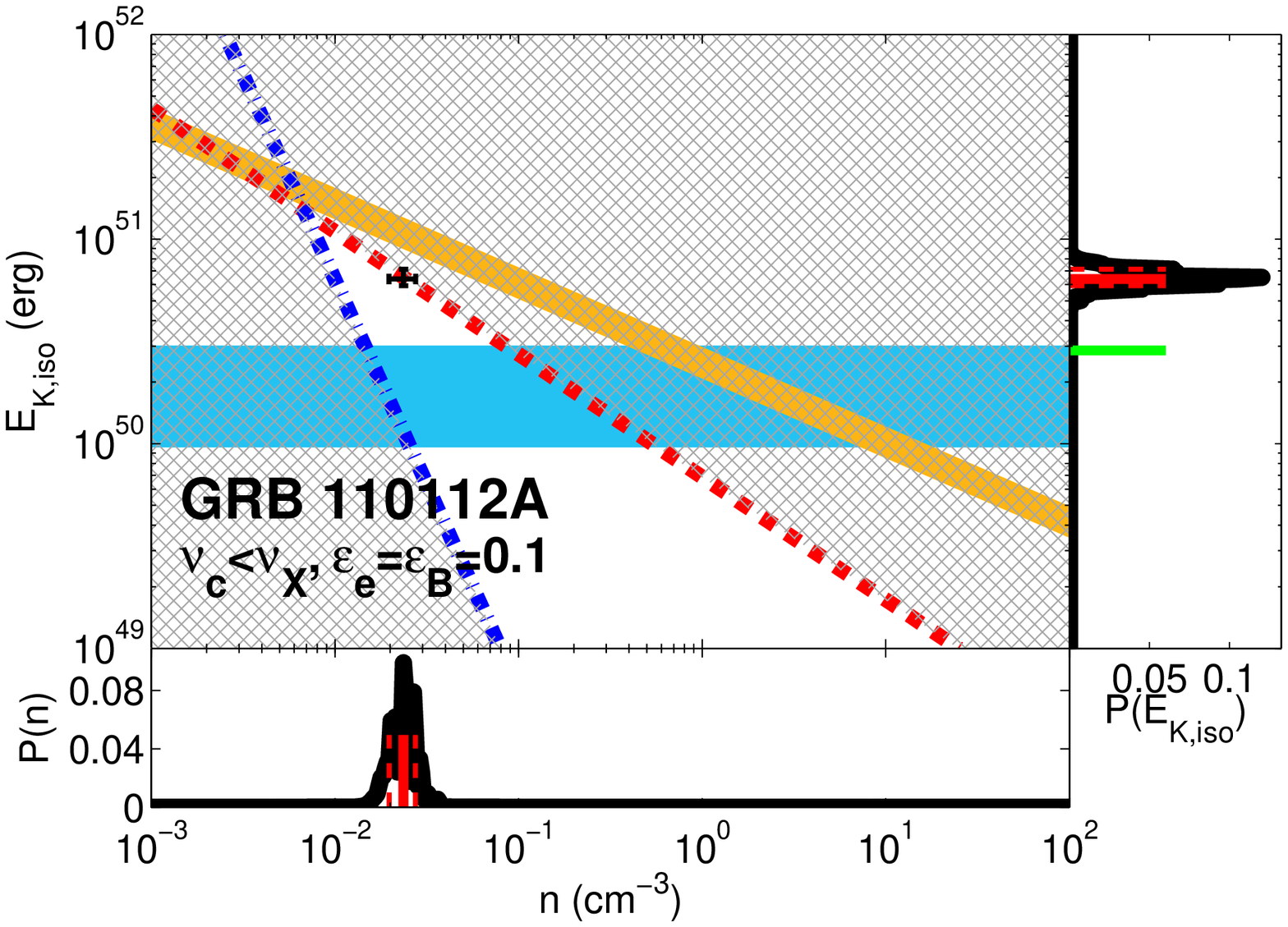} \\   
\includegraphics*[width=0.247\textwidth,clip=]{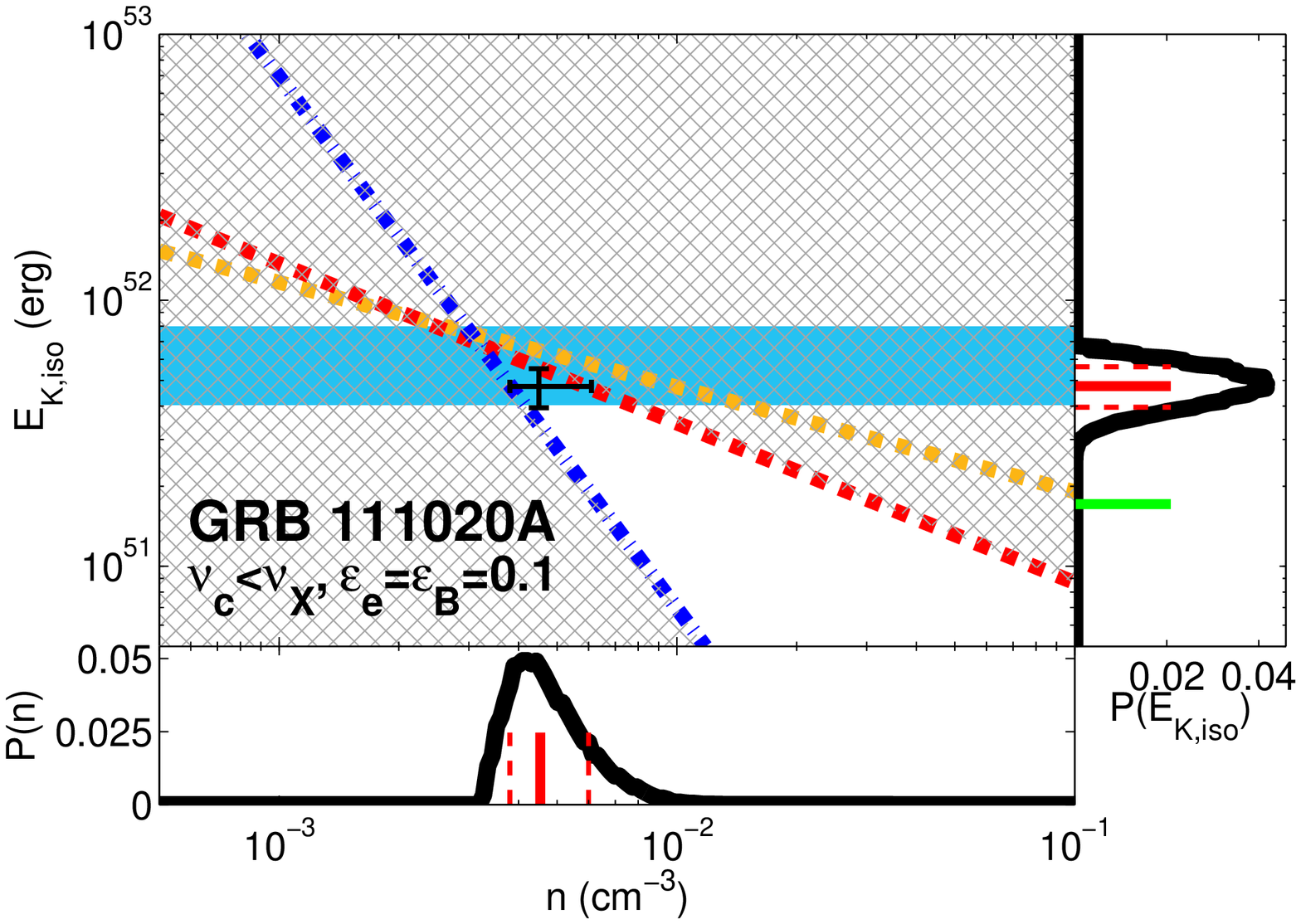} 
\includegraphics*[width=0.247\textwidth,clip=]{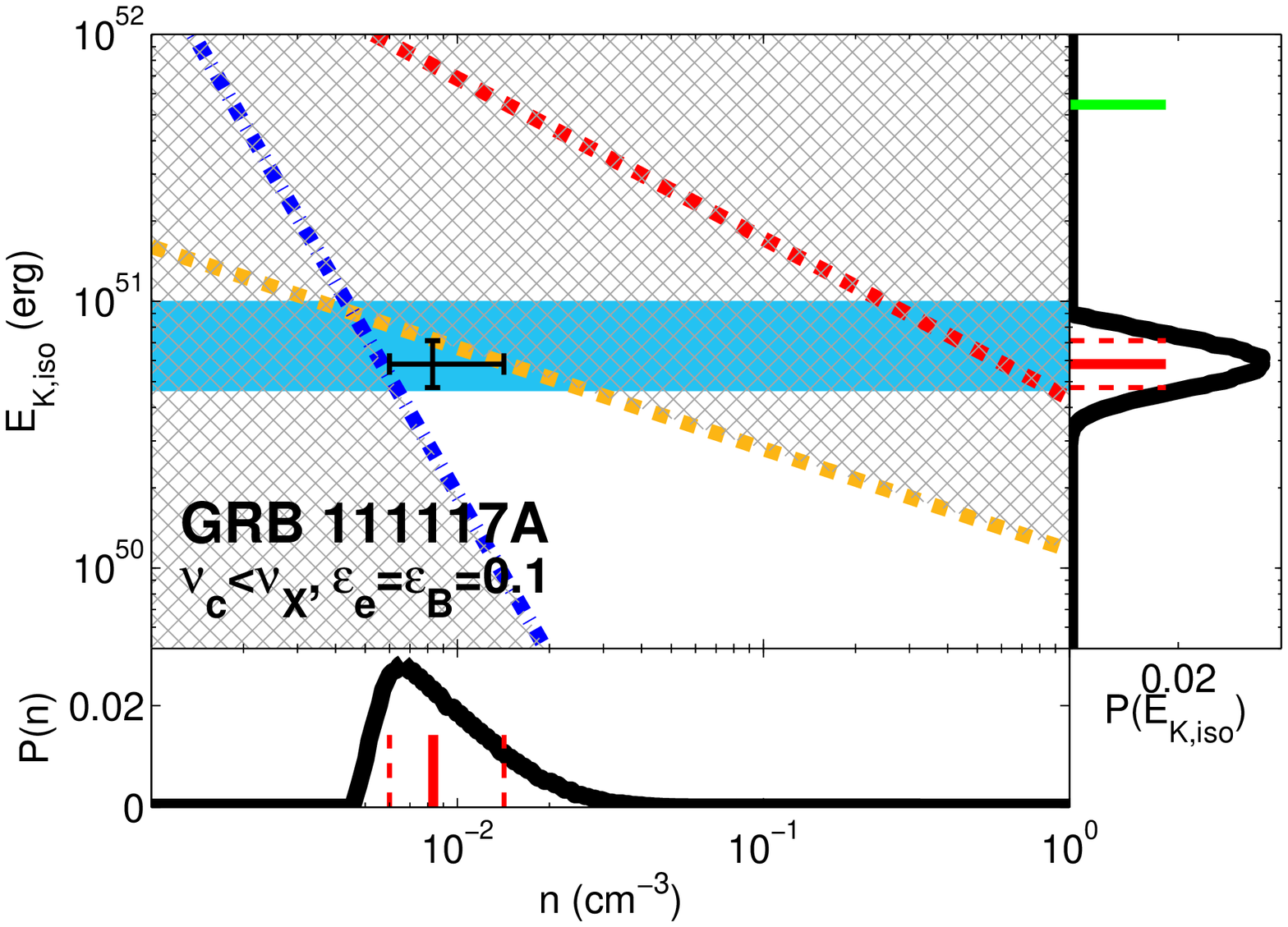} 
\includegraphics*[width=0.247\textwidth,clip=]{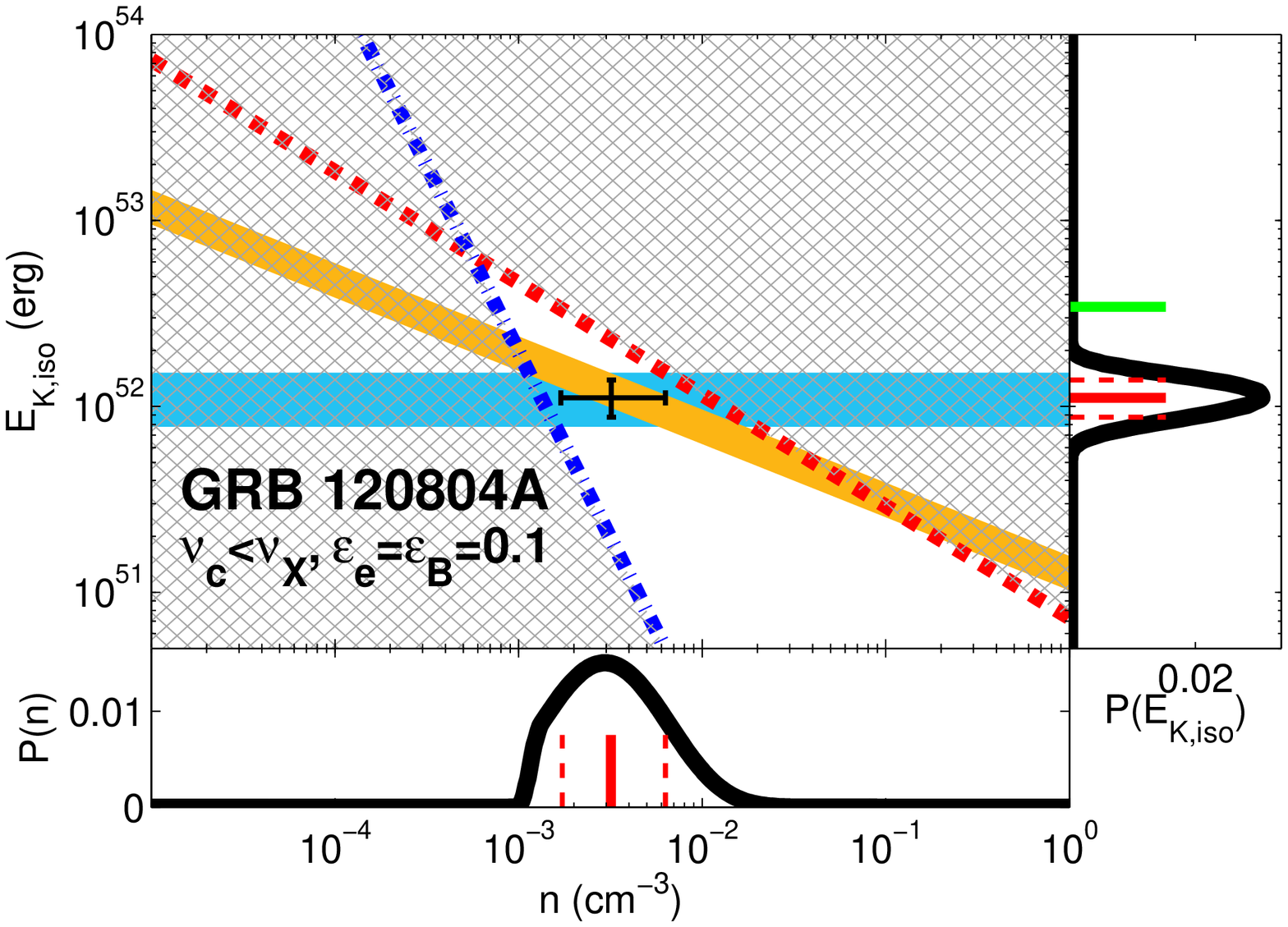}
\includegraphics*[width=0.247\textwidth,clip=]{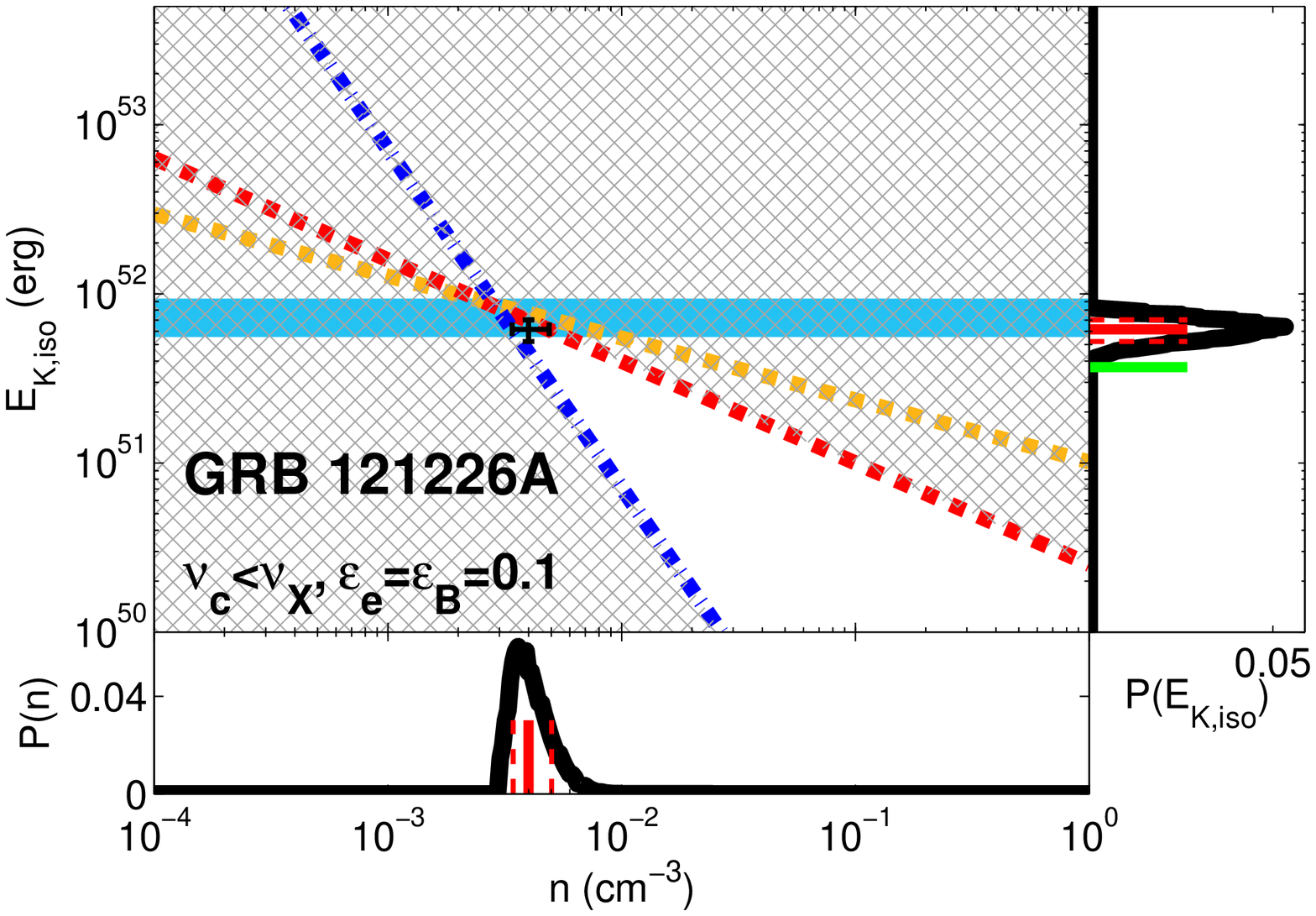}
\end{minipage}
\caption{Isotropic-equivalent kinetic energy versus circumburst density for 16 short GRBs with solutions for $\nu_c<\nu_X$ assuming fiducial values for the microphysical parameters of $\epsilon_e=\epsilon_B=0.1$. In each panel, the X-rays (light blue), optical (orange) and radio (red) provide independent constraints on the parameter space. In particular, the X-ray band provides an estimate of $E_{\rm K,iso}$. Measurements are shown as solid regions, where the width of the region corresponds to the $1\sigma$ uncertainty, while upper limits are denoted as dashed lines. Setting the cooling frequency to a maximum value of $\nu_{c,{\rm max}}=7.2 \times 10^{16}$~Hz (0.3~keV) provides an additional constraint (dark blue dot-dashed line). The regions of parameter space ruled out by the observations are denoted (grey hatched regions). The median solution and $1\sigma$ uncertainty is indicated by the black cross in each panel, corresponding to the values listed in Table~\ref{tab:prop}. For each burst, the joint probability distributions in $n$ (bottom panel) and $E_{\rm K,iso}$ (right panel) are shown. Red lines correspond to the median, and dotted lines are the $1\sigma$ uncertainty about the median. The green line corresponds to $E_{\gamma,{\rm iso}}$. 
\label{fig:wm}}
\end{figure*}

\begin{figure*}
\begin{minipage}[c]{\textwidth}
\tabcolsep0.0in
\includegraphics*[width=0.247\textwidth,clip=]{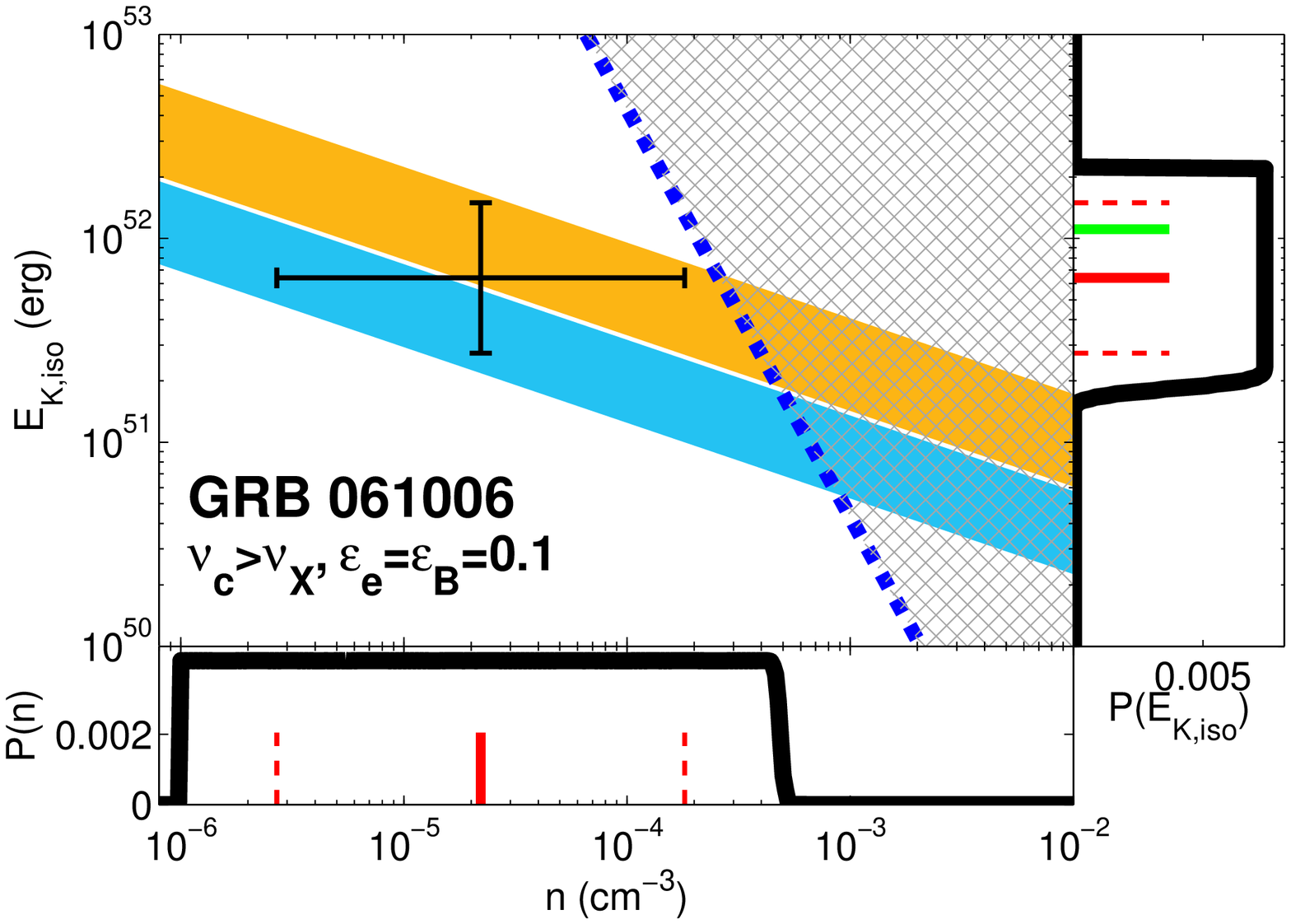} 
\includegraphics*[width=0.247\textwidth,clip=]{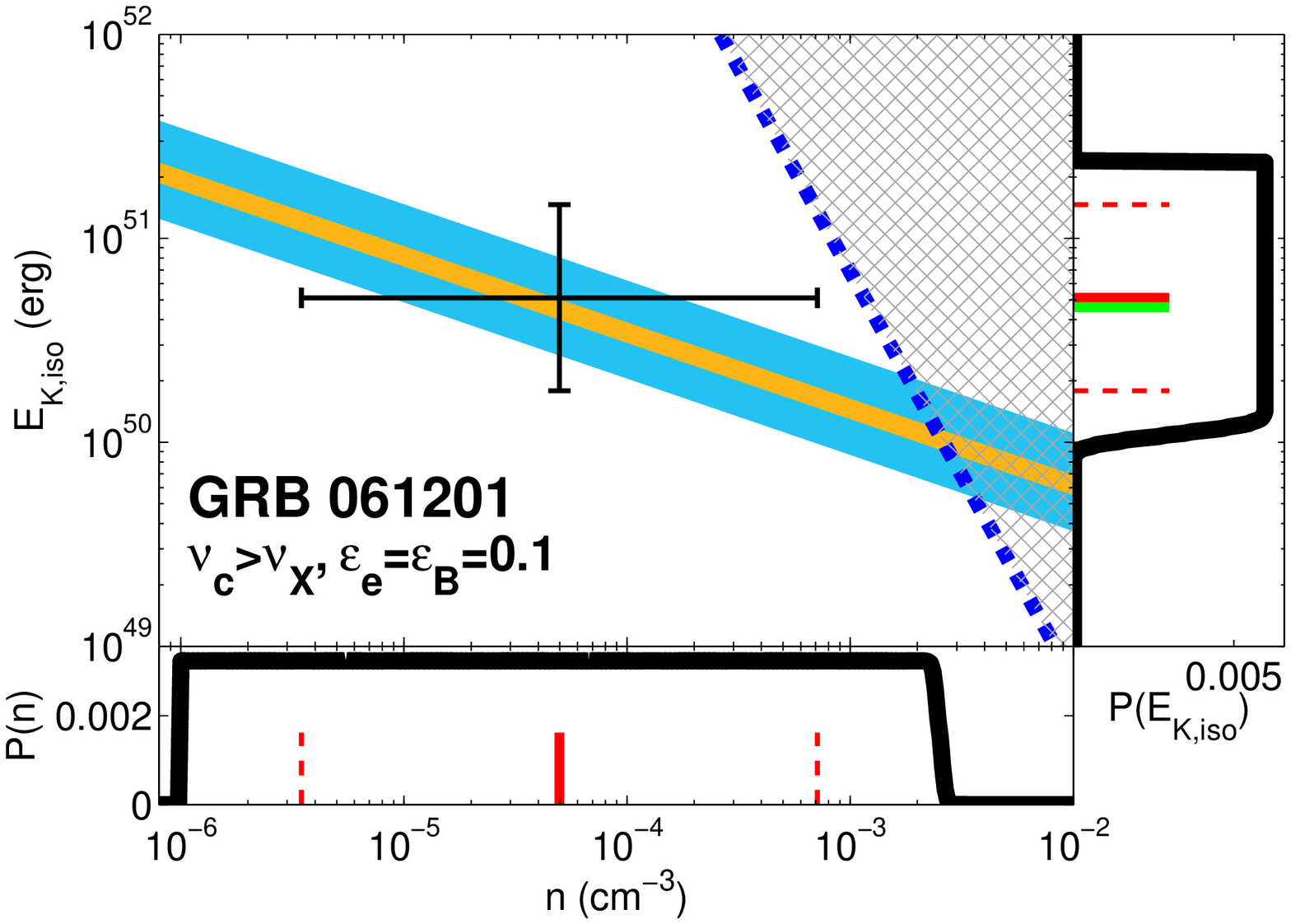} 
\includegraphics*[width=0.247\textwidth,clip=]{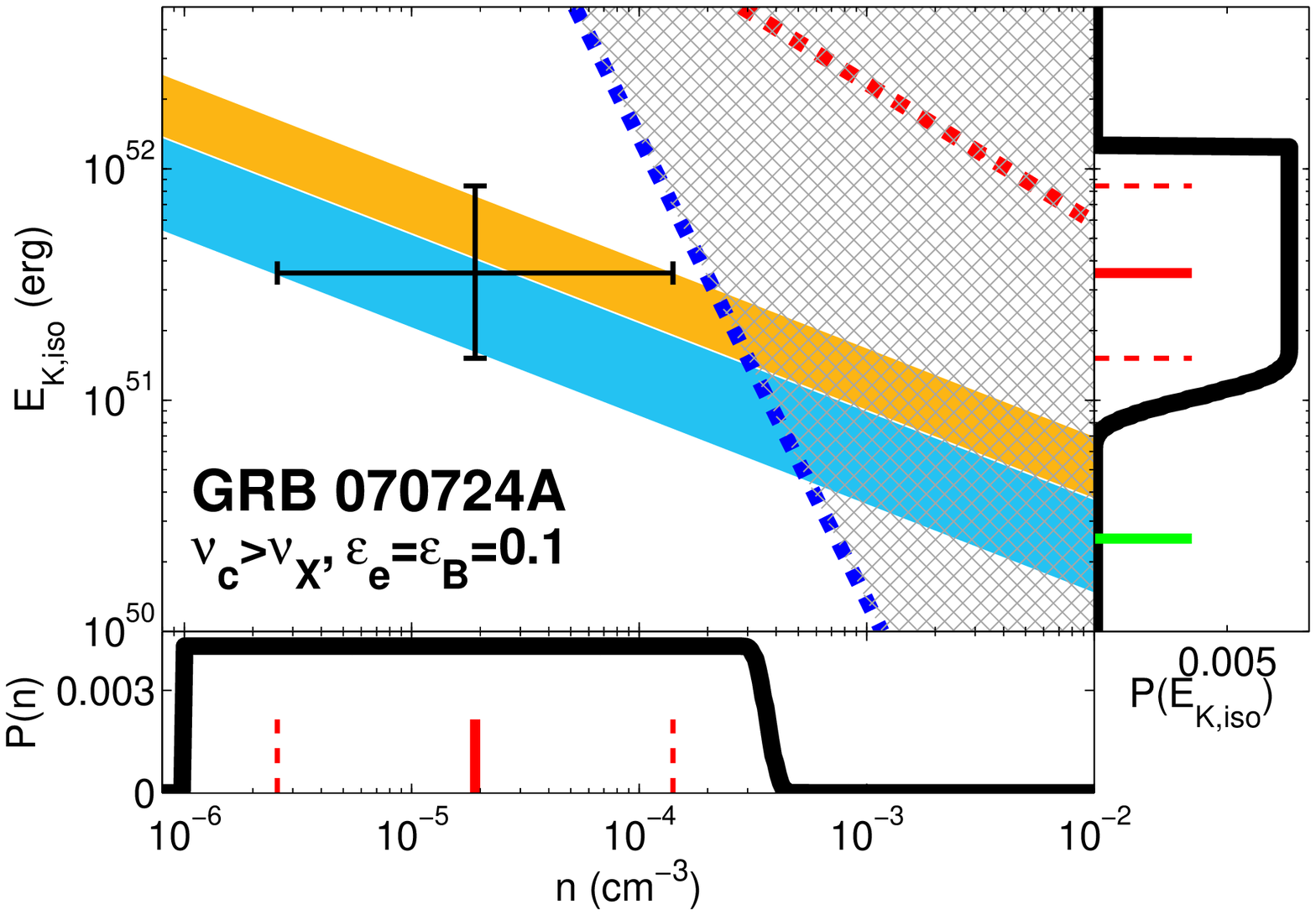} 
\includegraphics*[width=0.247\textwidth,clip=]{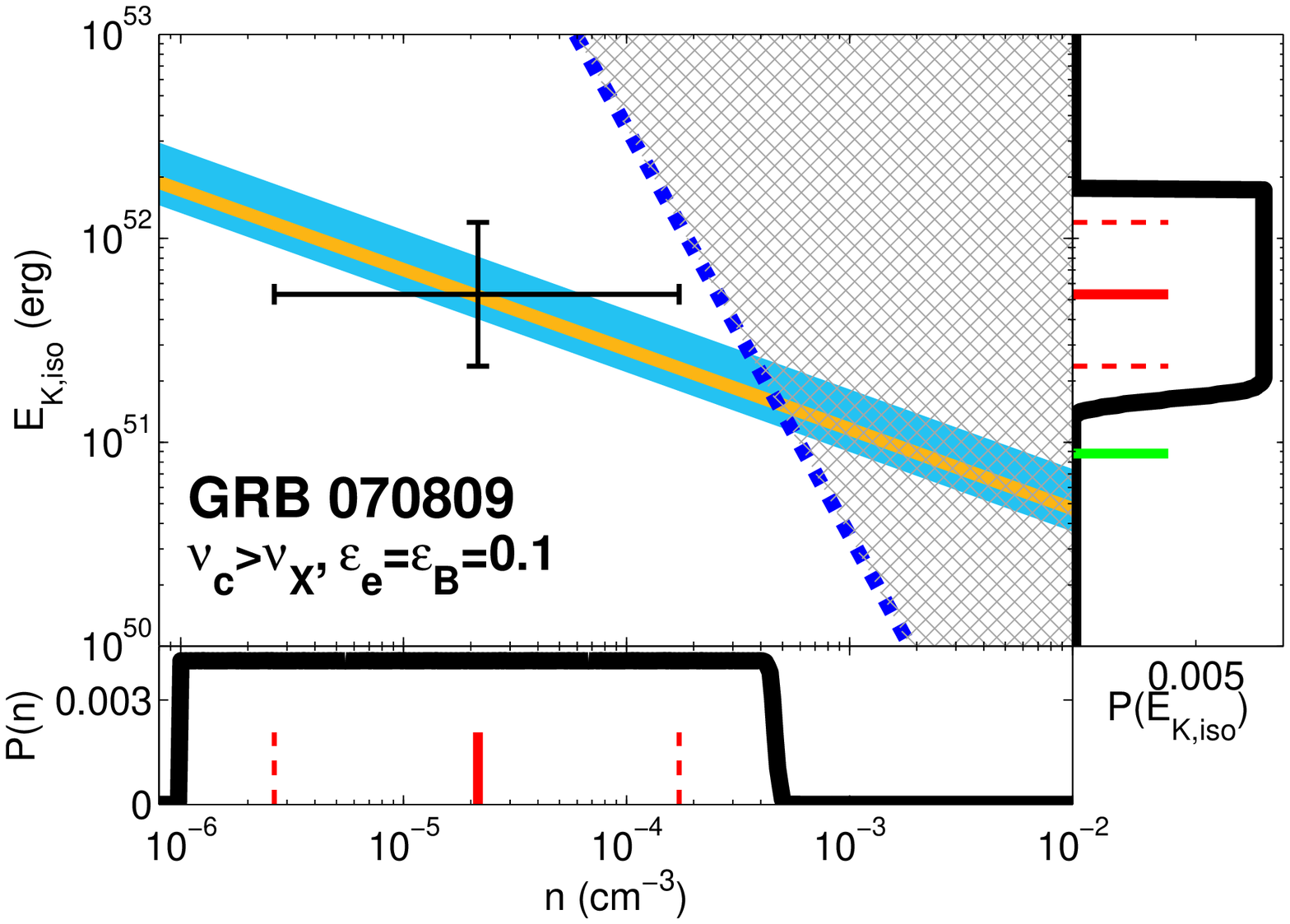} \\
\includegraphics*[width=0.247\textwidth,clip=]{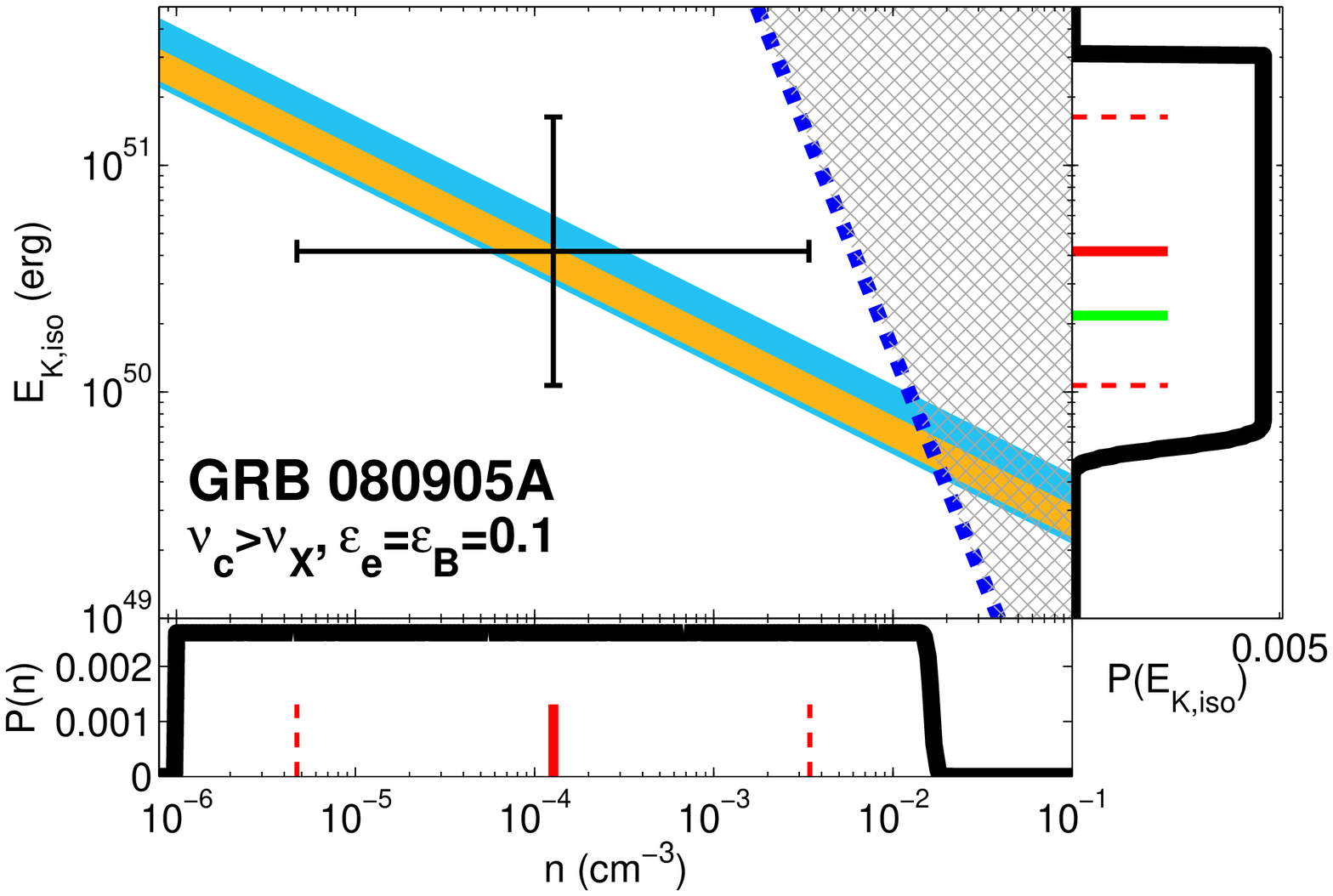}
\includegraphics*[width=0.247\textwidth,clip=]{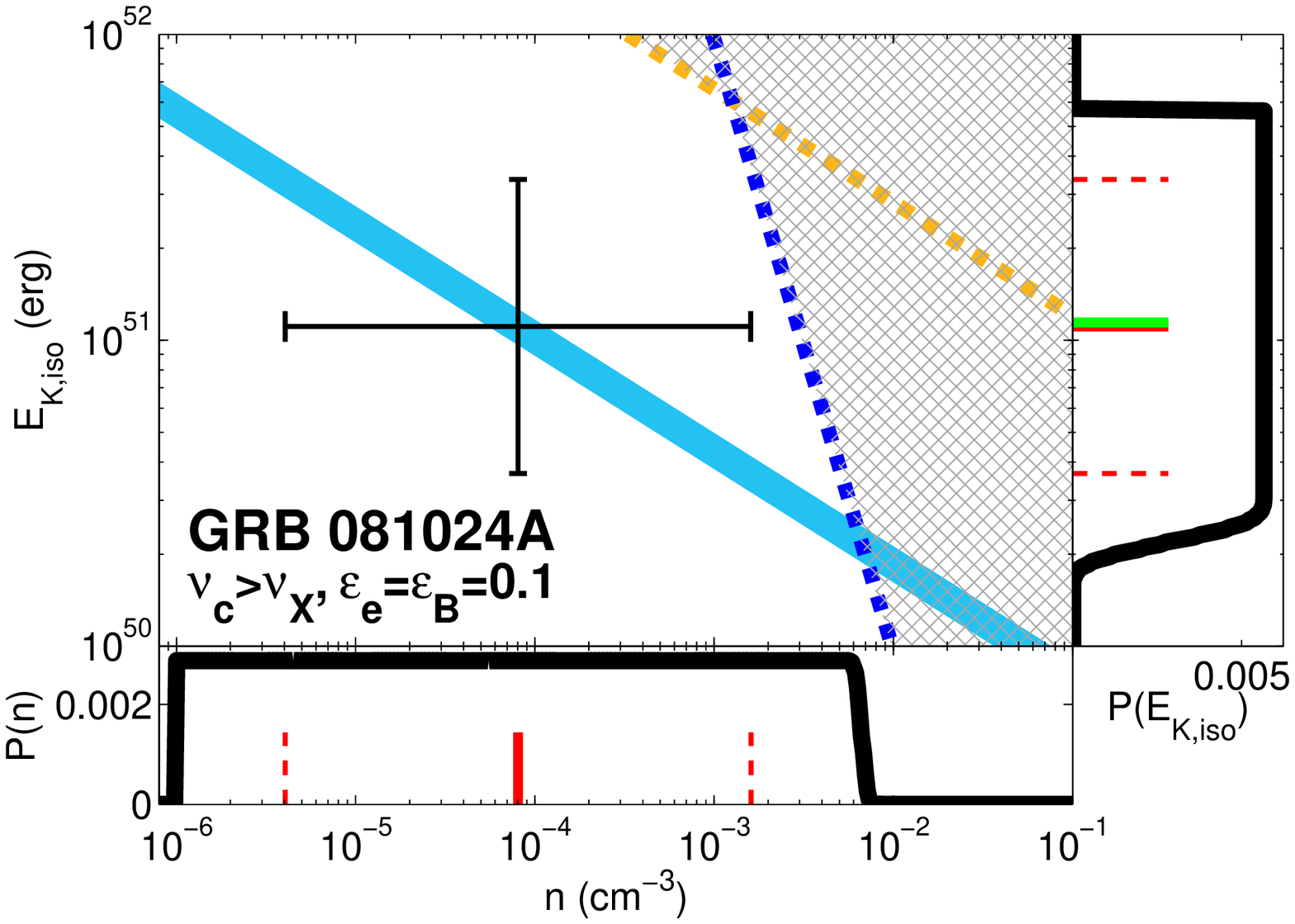} 
\includegraphics*[width=0.247\textwidth,clip=]{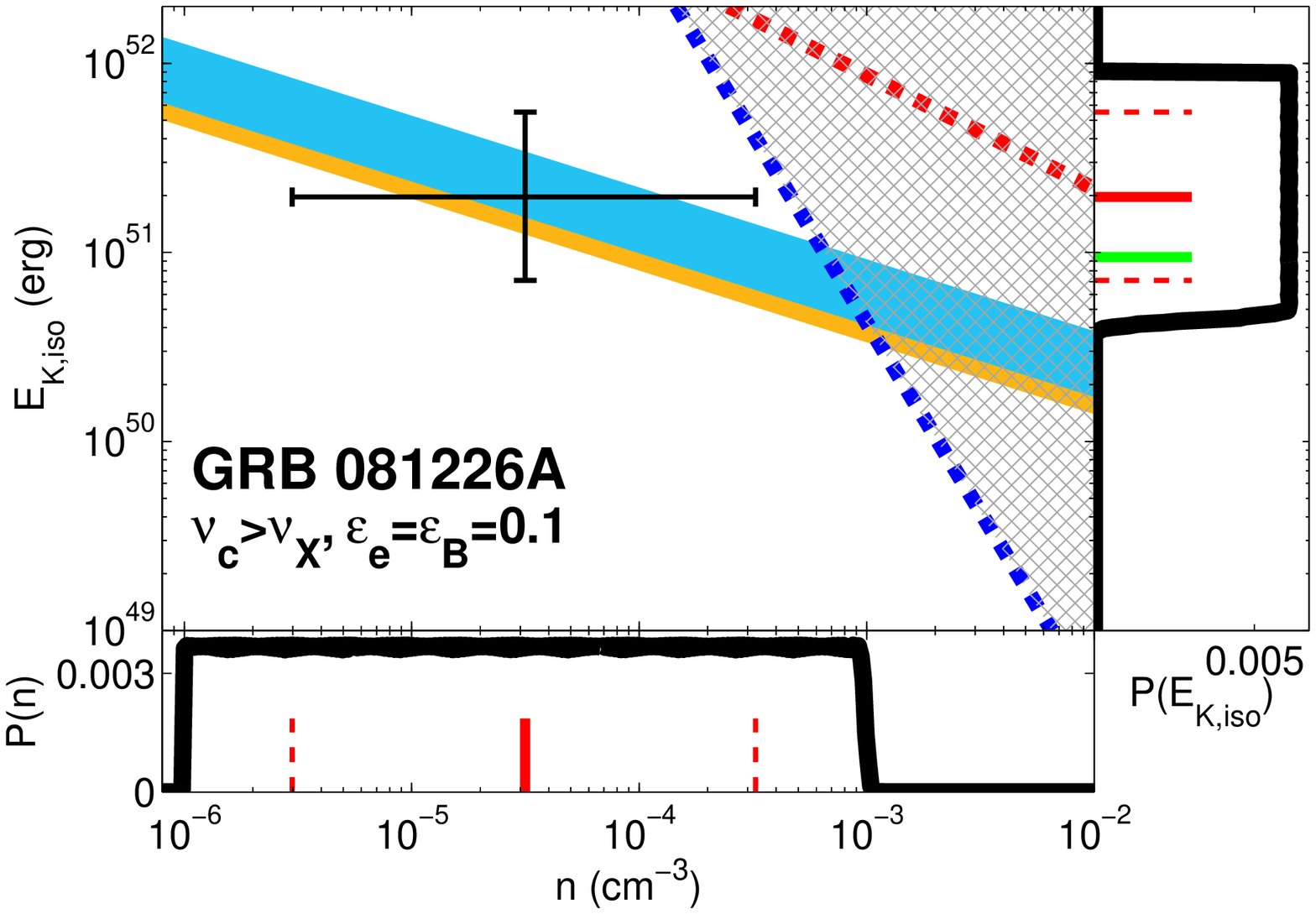}
\includegraphics*[width=0.247\textwidth,clip=]{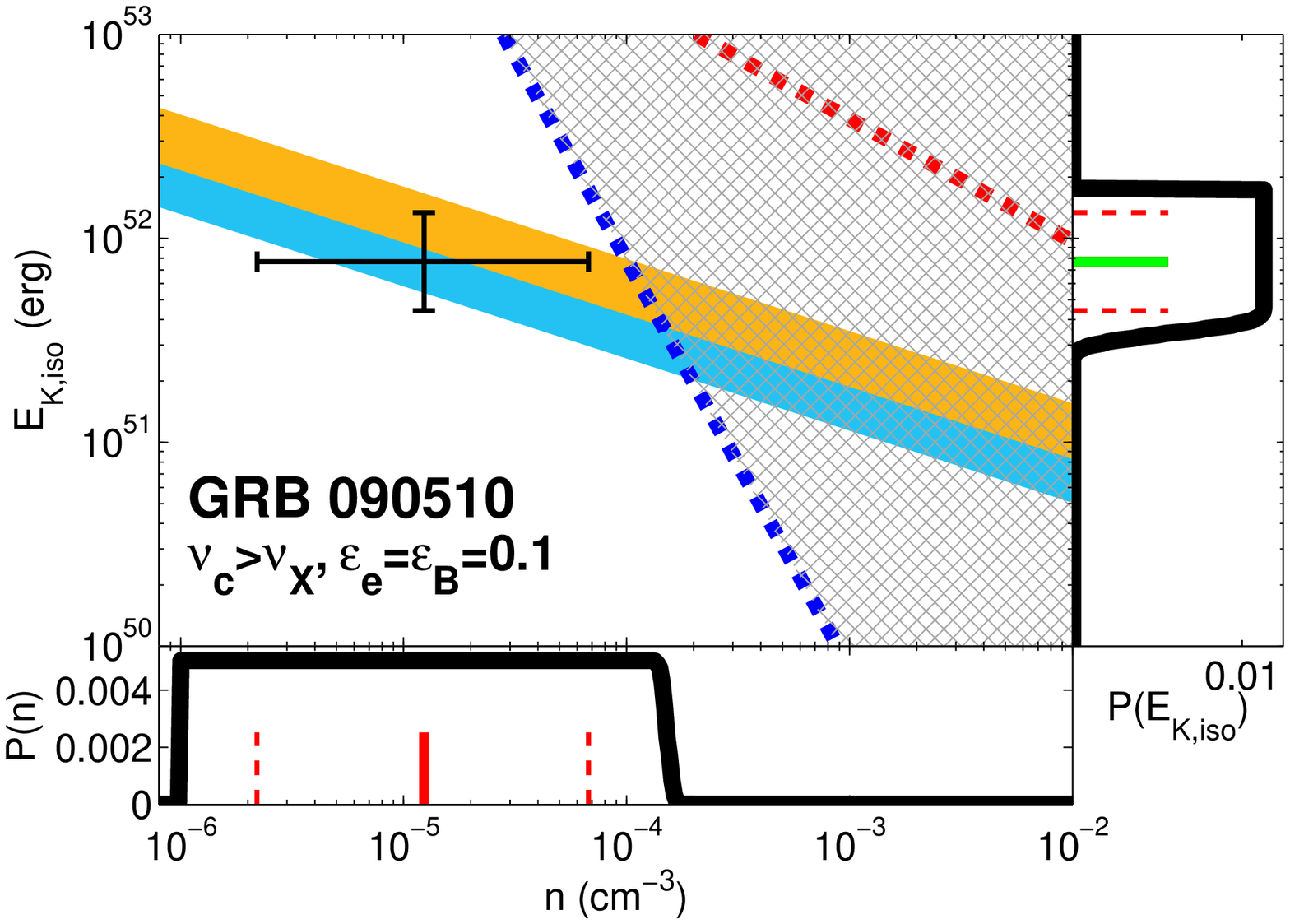} \\
\includegraphics*[width=0.247\textwidth,clip=]{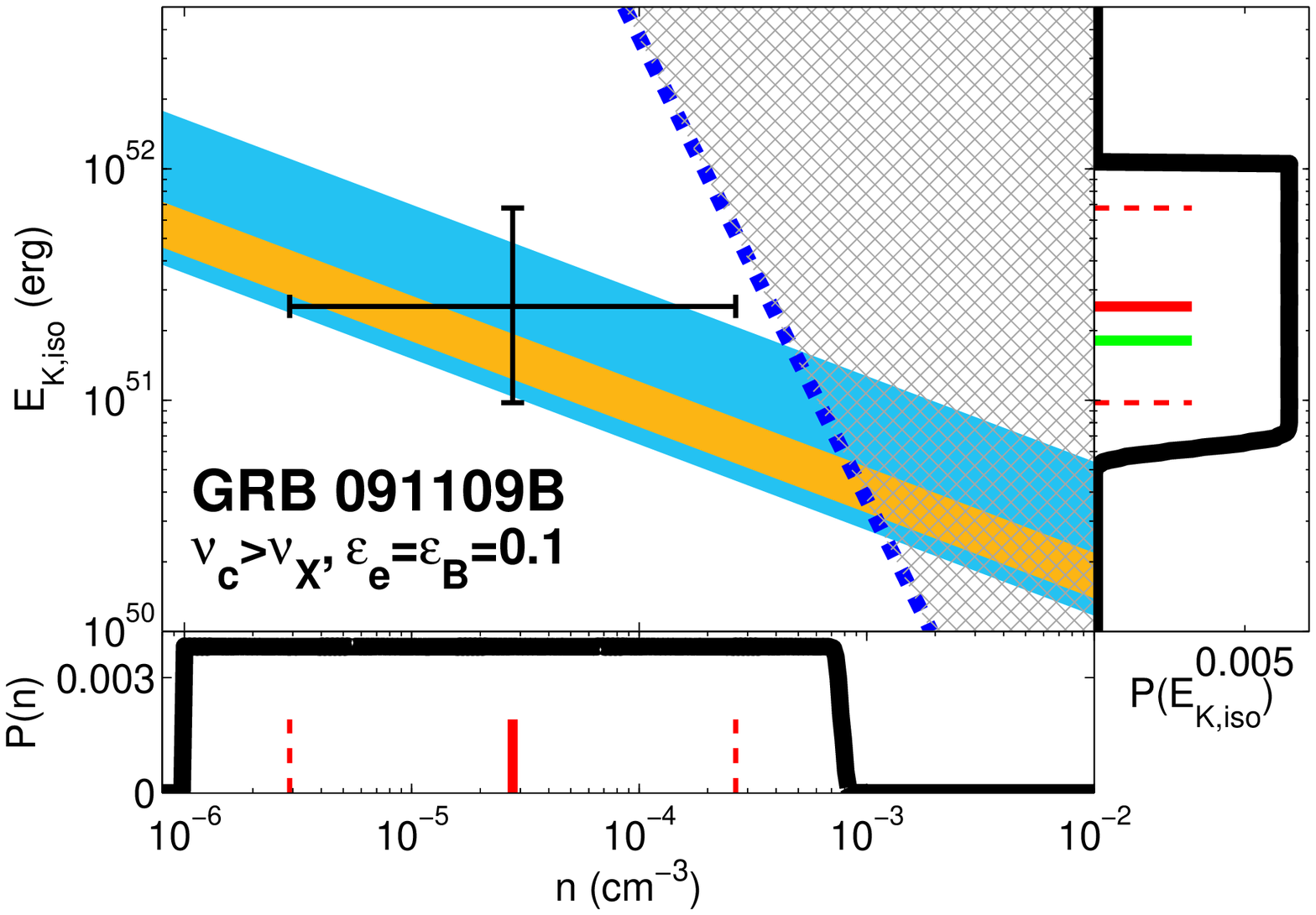} 
\includegraphics*[width=0.247\textwidth,clip=]{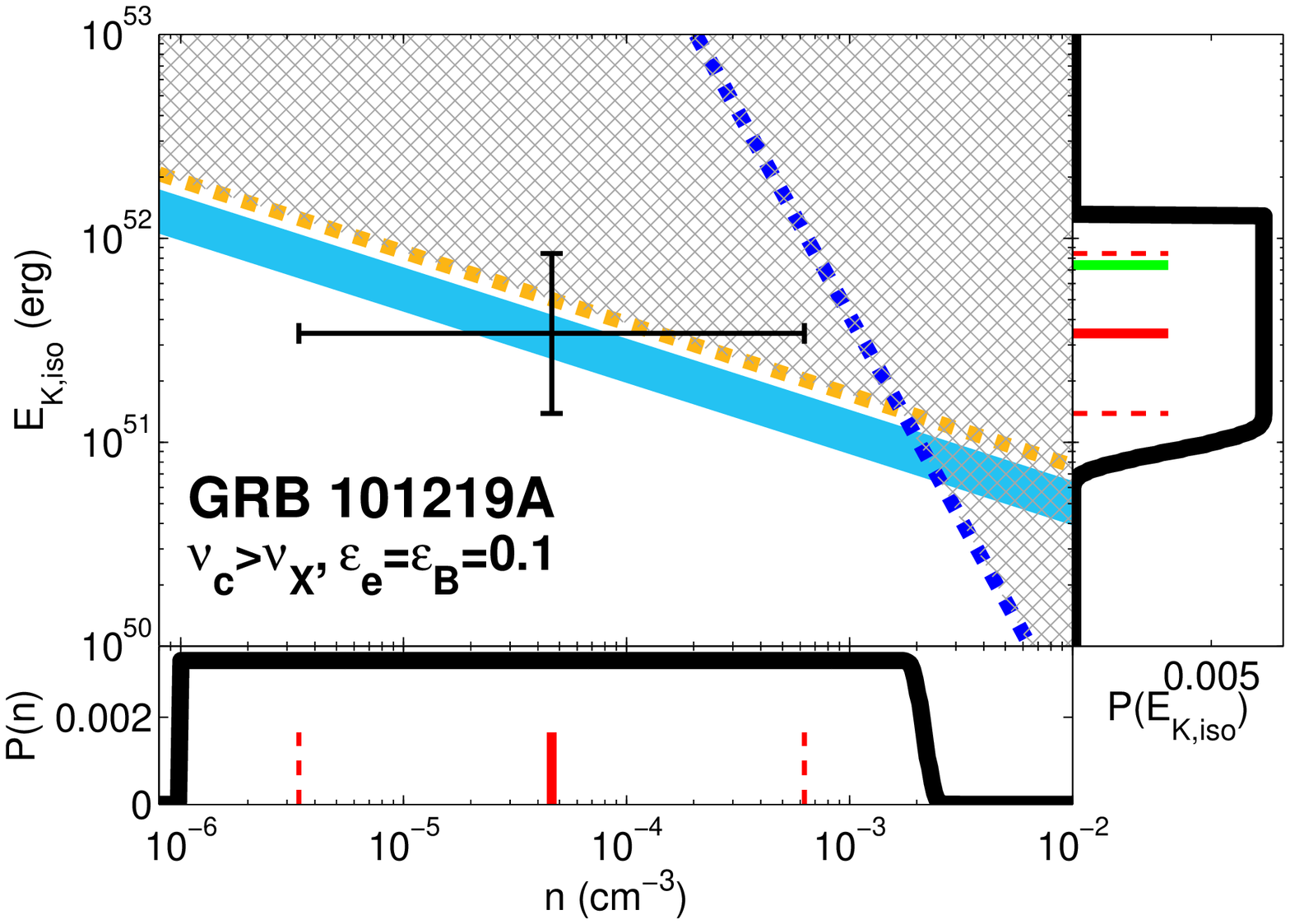}
\includegraphics*[width=0.247\textwidth,clip=]{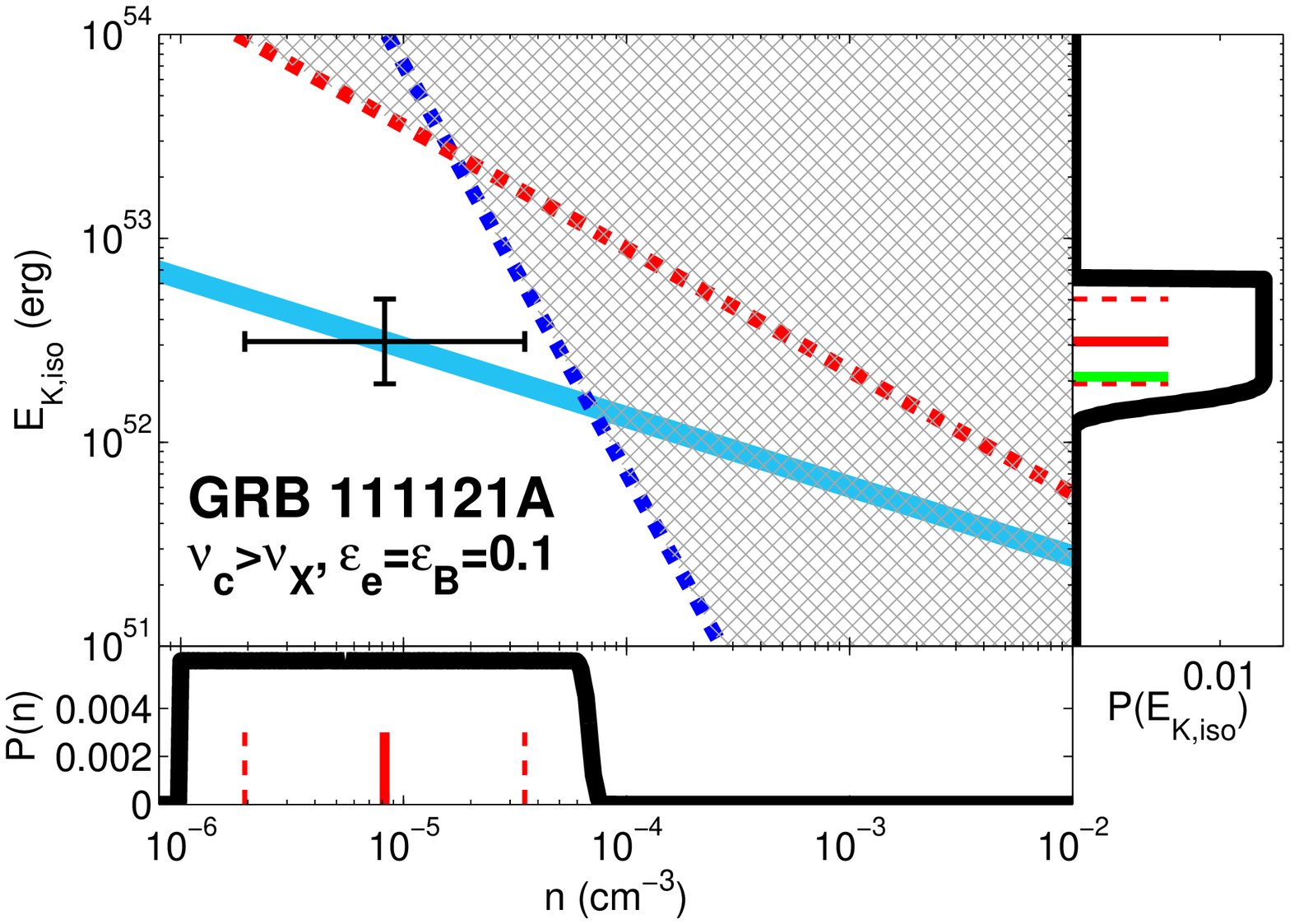} 
\includegraphics*[width=0.247\textwidth,clip=]{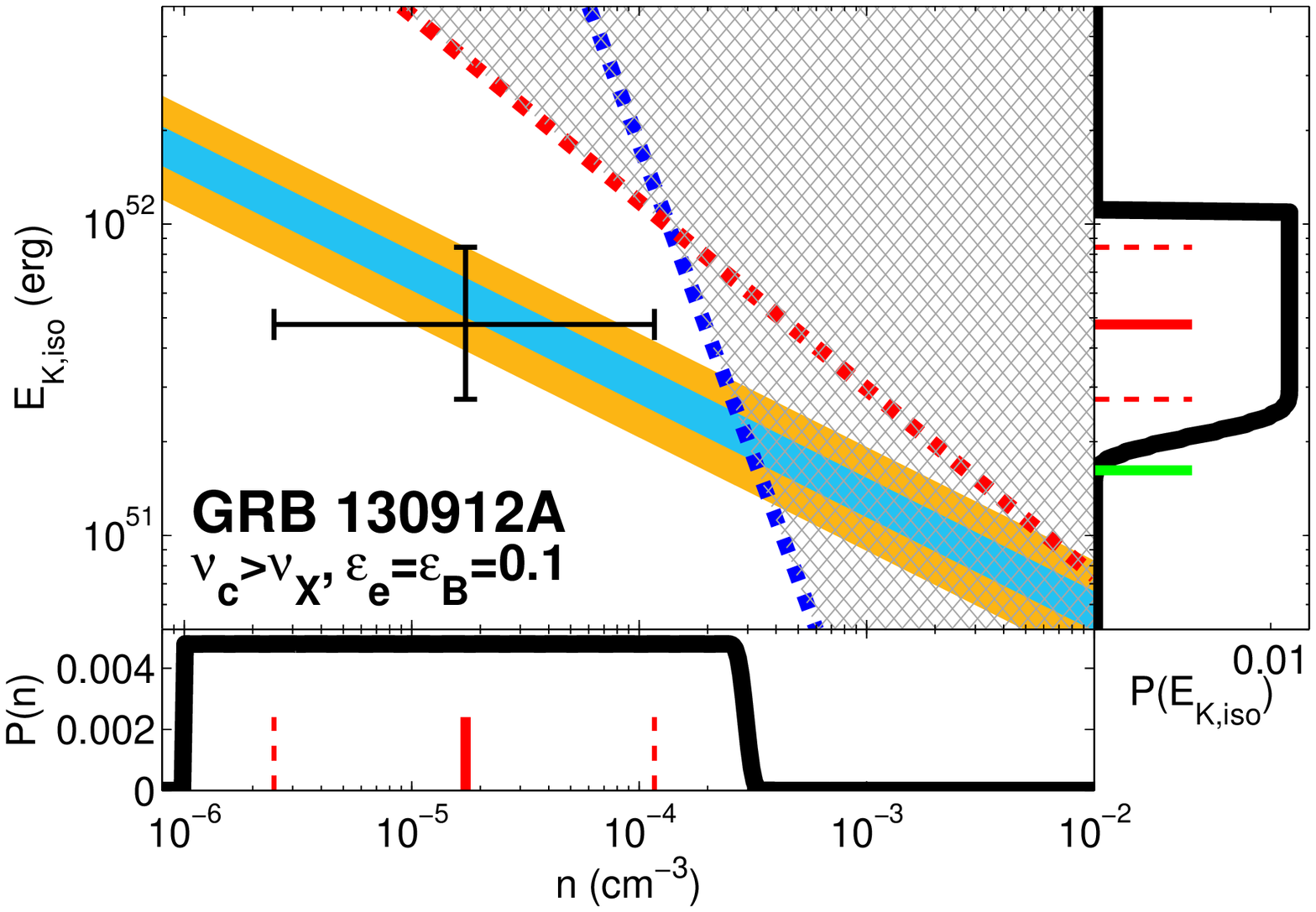} \\
\includegraphics*[width=0.247\textwidth,clip=]{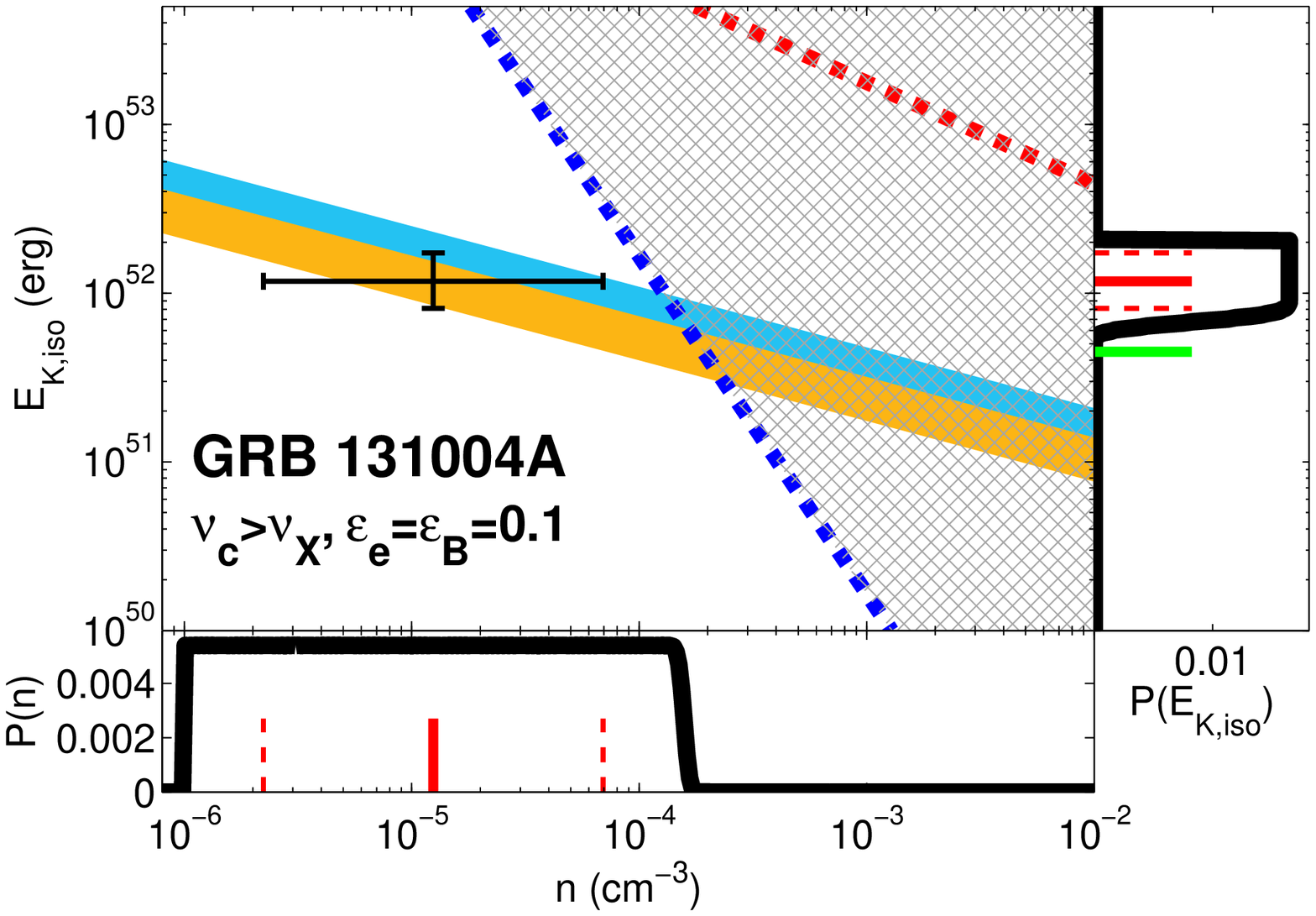} 
\includegraphics*[width=0.247\textwidth,clip=]{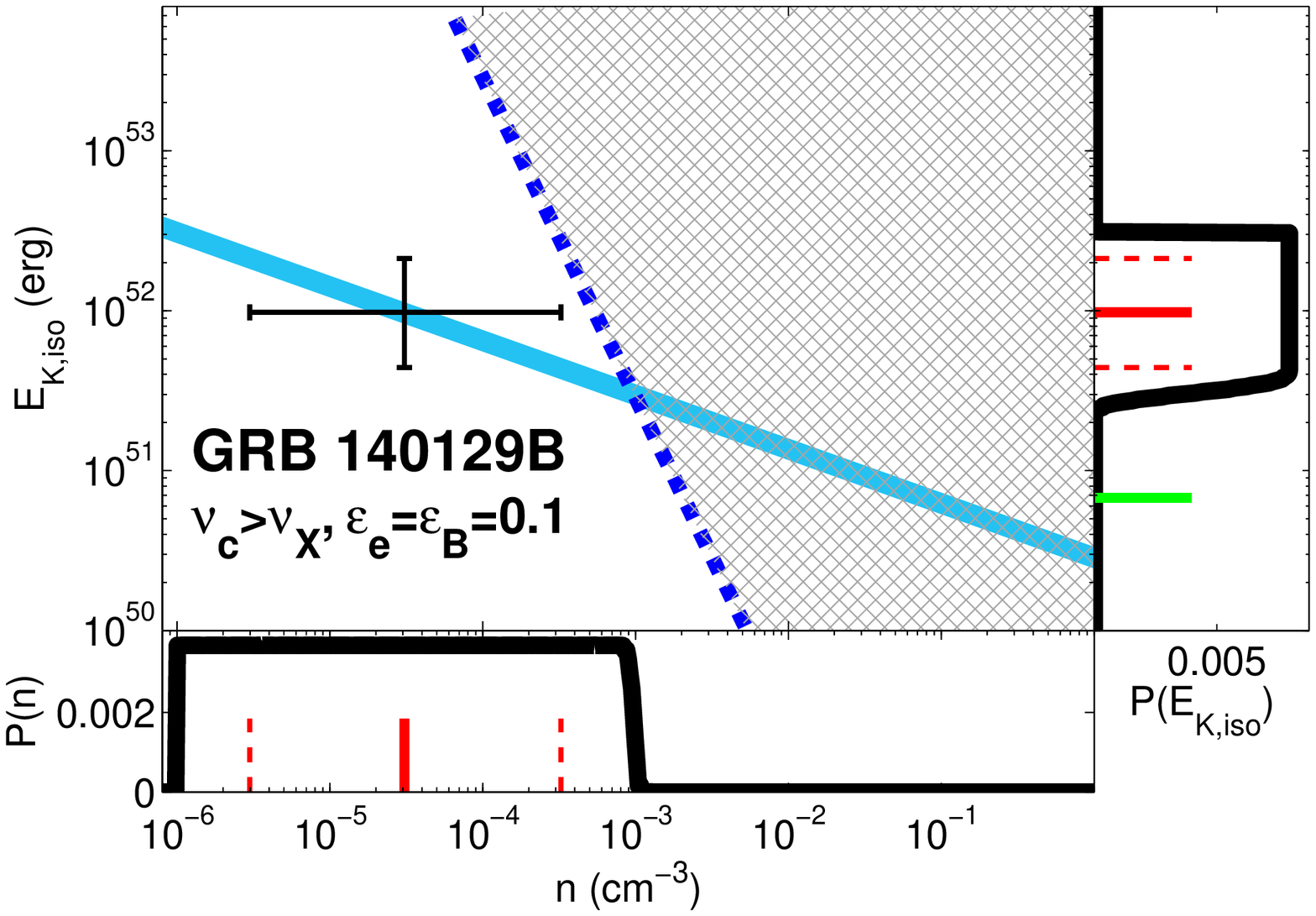}
\includegraphics*[width=0.247\textwidth,clip=]{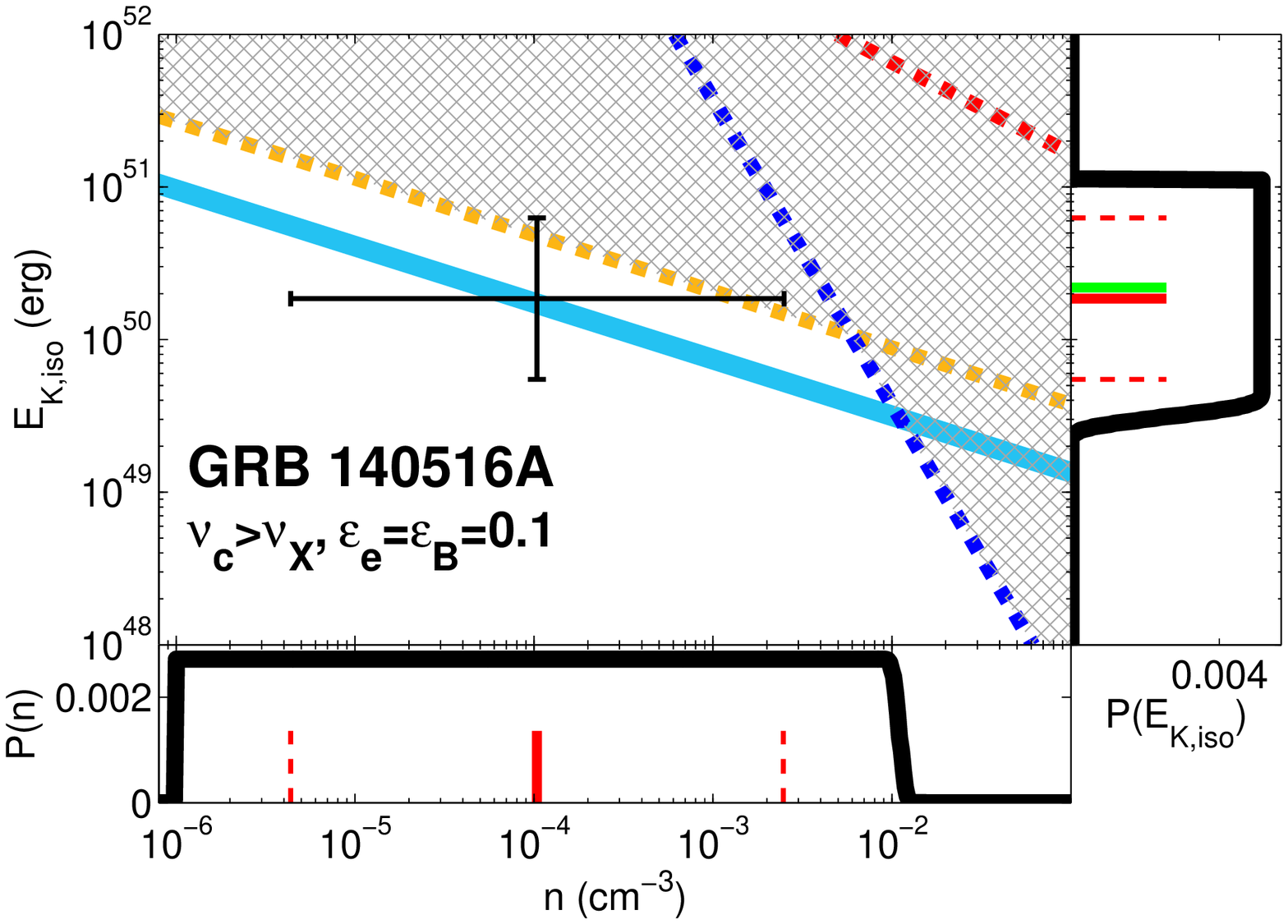} 
\includegraphics*[width=0.247\textwidth,clip=]{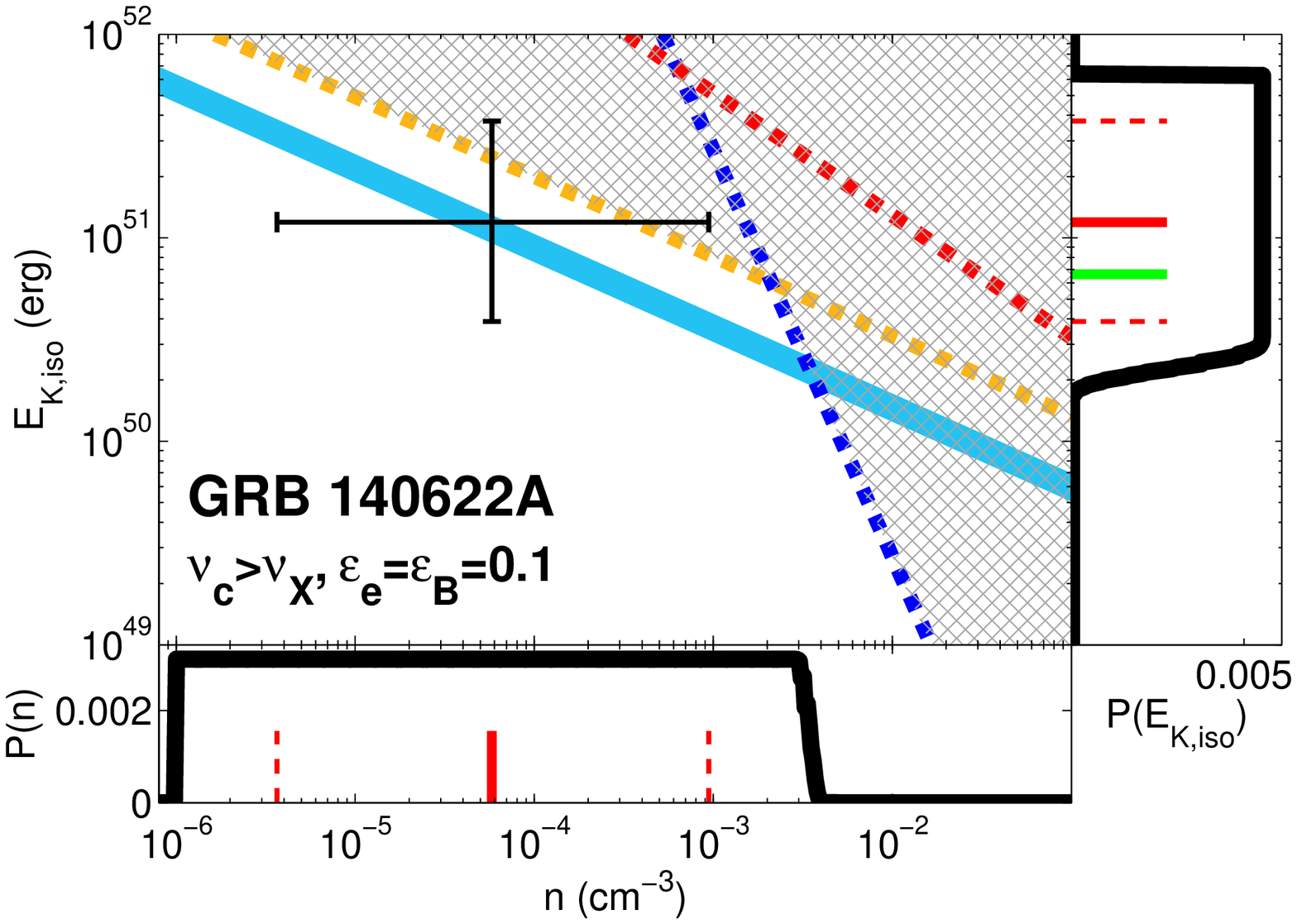} \\
\includegraphics*[width=0.247\textwidth,clip=]{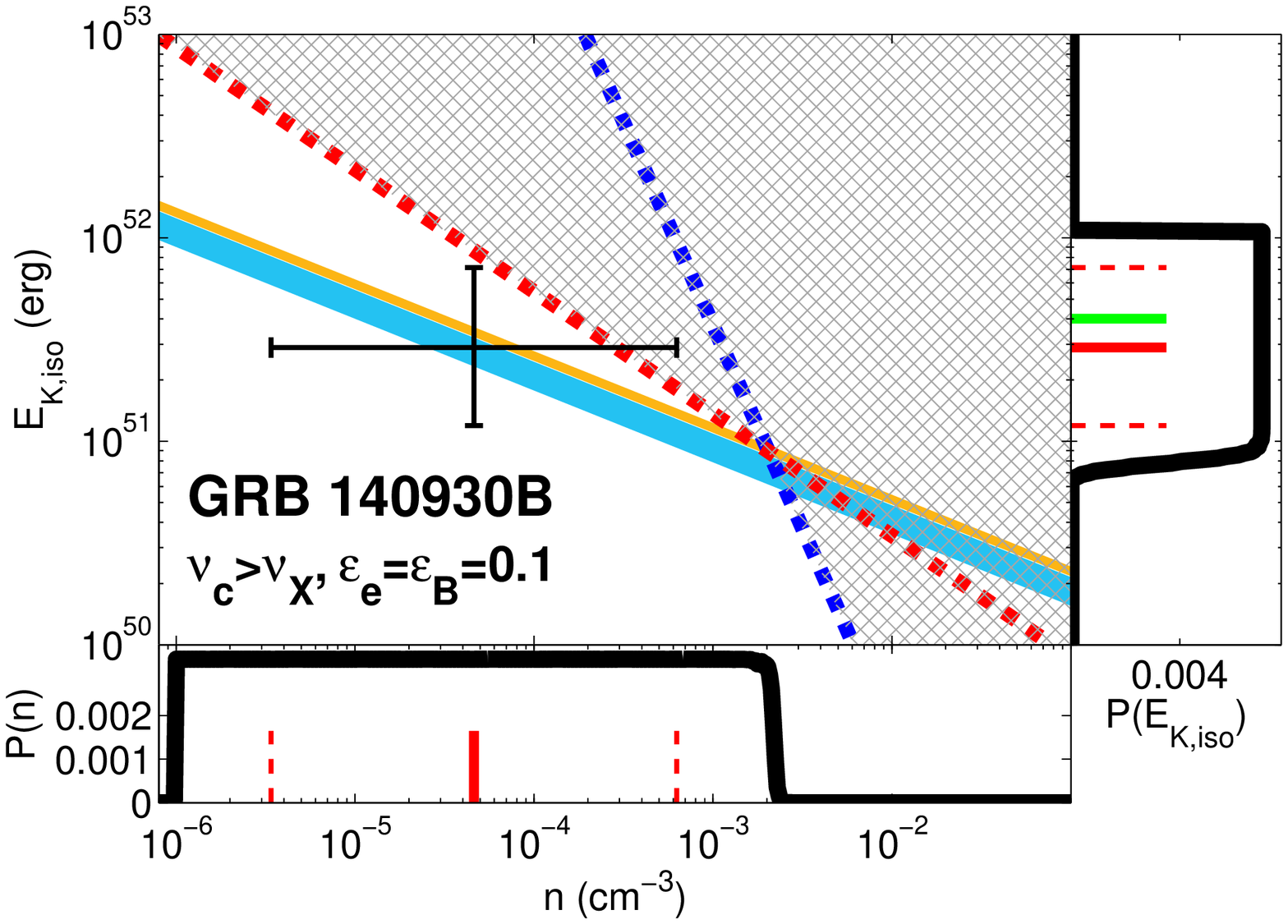} 
\includegraphics*[width=0.247\textwidth,clip=]{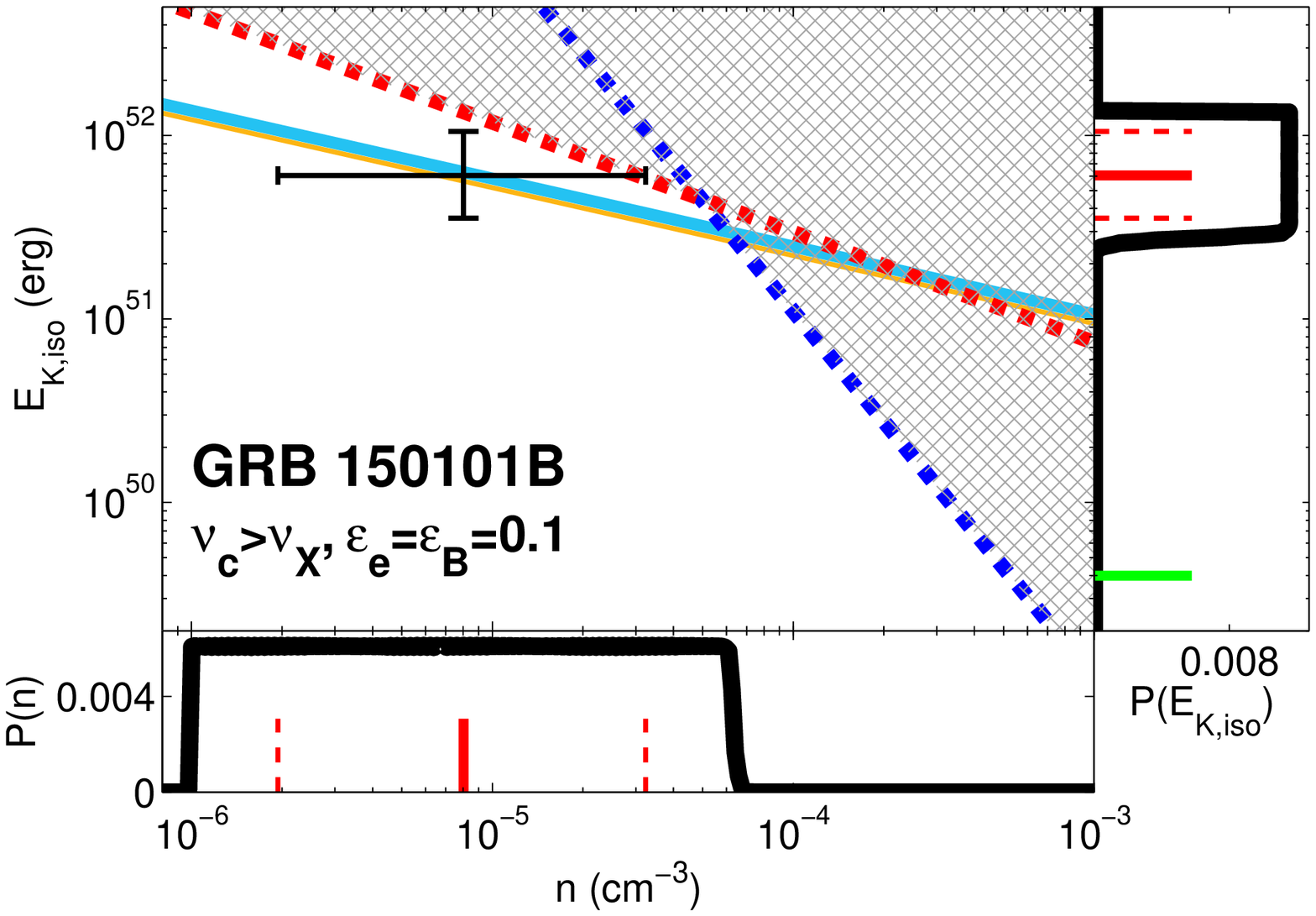} 
\end{minipage}
\caption{Isotropic-equivalent kinetic energy versus circumburst density for 18 short GRBs with solutions for $\nu_c>\nu_X$ assuming fiducial values for the microphysical parameters of $\epsilon_e=\epsilon_B=0.1$. In each panel, the X-rays (light blue), optical (orange) and radio (red) provide independent constraints on the parameter space. Measurements are shown as solid regions, where the width of the region corresponds to the $1\sigma$ uncertainty, while upper limits are denoted as dashed lines. Setting the cooling frequency to a {\it minimum} value of $\nu_{c,{\rm min}}=2.4 \times 10^{18}$~Hz (10~keV) provides an additional constraint (dark blue dashed line). The regions of parameter space ruled out by the observations are denoted (grey hatched regions). The median solution and $1\sigma$ uncertainty is indicated by the black cross in each panel, corresponding to the values listed in Table~\ref{tab:prop}. For each burst, the joint probability distributions in $n$, with an imposed lower bound of $n_{\rm min}=10^{-6}$~cm$^{-3}$ (bottom panel), and $E_{\rm K,iso}$ (right panel) are shown. Red lines correspond to the median, and dotted lines are the $1\sigma$ uncertainty about the median. The green line corresponds to $E_{\gamma,{\rm iso}}$.
\label{fig:th}}
\end{figure*}

\subsubsection{Detections}

To calculate the individual probability distributions for afterglow detections, we apply Equation~\ref{eqn:fnu} to the observations using the relevant regime for each observing band, $\nu_i$. For the X-ray band, we use the location of $\nu_c$ as determined in Section~\ref{sec:xfit} to determine which branch of Equation~\ref{eqn:fnu} to use. Since the value of $p$ is primarily determined from the X-ray band, the $E_{\rm K,iso}-n$ relation remains unchanged when using different X-ray observations that follow the same temporal decline; thus, we only use one X-ray observation per burst. For the optical band, we make the reasonable assumption that $\nu_m<\nu_{\rm opt}<\nu_c$ and utilize the second branch of Equation~\ref{eqn:fnu}. In some cases, individual optical/NIR observations lead to $E_{\rm K, iso}-n$ relations that do not overlap within their $1\sigma$ uncertainties. In these cases, we use the weighted mean and standard deviation of these relations (i.e., systematic uncertainty) as the optical/NIR solution. For the radio band, we assume that $\nu_a<\nu_{\rm radio}<\nu_m$, (first branch of Equation~\ref{eqn:fnu}) to calculate the $E_{\rm K,iso}-n$ relations. Since the radio band lies on a different segment of the synchrotron spectrum than the other bands, it contributes a $E_{\rm K,iso}-n$ relation with a different slope, and thus enables tighter constraints on the physical parameters (Figure~\ref{fig:rad}).

After calculating the unique probability distribution from each of the radio, optical/NIR and X-ray bands, we normalize the area under each of the distributions to unity. We assume that the uncertainties in the flux densities are Gaussian, and thus each band contributes a unique, log-normal distribution.  These distributions are shown for the four bursts with radio afterglow detections (Figure~\ref{fig:rad}), and 34 bursts with radio afterglow non-detections: 16 bursts with $\nu_c<\nu_X$ (Figure~\ref{fig:wm}) and 18 bursts with $\nu_c>\nu_X$ (Figure~\ref{fig:th}).

\subsubsection{Upper Limits}

For flux density upper limits, we use Equation~\ref{eqn:fnu} to determine the $E_{\rm K,iso}-n$ relation at the $3\sigma$ upper limit. We then assign zero probability to the $E_{\rm K,iso}-n$ parameter space above the relationship, and assign a constant probability to the allowed parameter space below the relationship, normalized to unity. The upper limits are denoted by dashed lines in Figures~\ref{fig:rad}--\ref{fig:th}, where the colors of the lines correspond to the relevant observing band (X-rays: blue, optical: orange, radio: red). Regions of parameter space that have been ruled out by the upper limits are marked as hatched regions.

\subsubsection{Cooling Frequency Constraint}

We utilize the relative location of the cooling frequency as a final constraint, since it depends on a combination of energy and density, by

\begin{equation}
\label{eqn:nuc}
\nu_c \propto n_0^{-1} E_{\rm K,iso,52}^{-1/2} \epsilon_{B,-1}^{-3/2},
\end{equation}

\noindent with additional dependencies on $\delta t$ and redshift \citep{gs02}. For the cases in which the X-ray band is located above the cooling frequency ($\nu_c<\nu_X$), we employ a maximum value at the lower edge of the X-ray band, $\nu_{c,{\rm max}}=7.3 \times 10^{16}$~Hz (0.3~keV), to obtain a lower limit on the combination of energy and density. The corresponding probability distribution has zero value in the $E_{\rm K,iso}-n$ parameter space below the relation, and a constant value above the relation, where the area in the allowed parameter space is normalized to unity. The lower limits are shown as as blue dot-dashed lines for GRBs\,051221A and 130603B in Figure~\ref{fig:rad} and for all bursts in Figure~\ref{fig:wm}. In cases where the X-ray band is below the cooling frequency ($\nu_m<\nu_X<\nu_c$), we set the cooling break to a minimum value, $\nu_{c,{\rm min}}=2.4 \times 10^{18}$~Hz (10~keV) at the upper edge of the X-ray band, and determine the $E_{\rm K,iso}-n$ relation for each burst using Equation~\ref{eqn:nuc}. This constraint sets an upper limit on the combination of energy and density. We form the probability distribution in the same manner as for afterglow upper limits. The limits for each burst set by the cooling frequency are shown as blue dashed lines for GRBs\,050724A and 140903A in Figure~\ref{fig:rad} and all bursts in Figure~\ref{fig:th}.

\subsection{Joint Probability Distributions}

Since each of the observing bands, as well as the location of the cooling frequency, contribute an independent probability distribution, we calculate the joint probability from a product of these distributions for each burst. To obtain 1-dimensional probability distributions in $E_{\rm K,iso}$ and $n$, we integrate over each of the parameters. Finally, we normalize the area under each 1-dimensional distribution to unity. The resulting distributions, $P(n)$ and $P(E_{\rm K,iso})$, for 34 bursts are shown in Figures~\ref{fig:wm}--\ref{fig:th} for the fiducial microphysical parameters, $\epsilon_e=\epsilon_B=0.1$. The median values and $1\sigma$ uncertainties in isotropic-equivalent kinetic energy and circumburst density are also shown in these figures and listed in Table~\ref{tab:prop}.

Figure~\ref{fig:rad} shows the probability distributions for the four events with radio afterglow detections. In three of the four cases, we can use the available data to place additional constraints on $\epsilon_B$, fixing $\epsilon_e=0.1$. For GRBs\,050724A and 140903A, the afterglow data require that $\epsilon_B \lesssim 10^{-4}$ and $\lesssim 10^{-3}$, respectively. For larger values of $\epsilon_B$, the constraint from the cooling frequency becomes more stringent and conflicts with the solutions obtained from the afterglow observations on the grid of allowed values. For GRB\,051221A, we find a significantly better fit at $\epsilon_B=0.01$. In all cases, the addition of the radio band enables tighter constraints on both the energy and the density. 

\section{Density and Energy Scale for Short GRBs}
\label{sec:results}

\begin{figure*}[ht]
\begin{minipage}[c]{\textwidth}
\includegraphics*[width=0.48\textwidth,clip=]{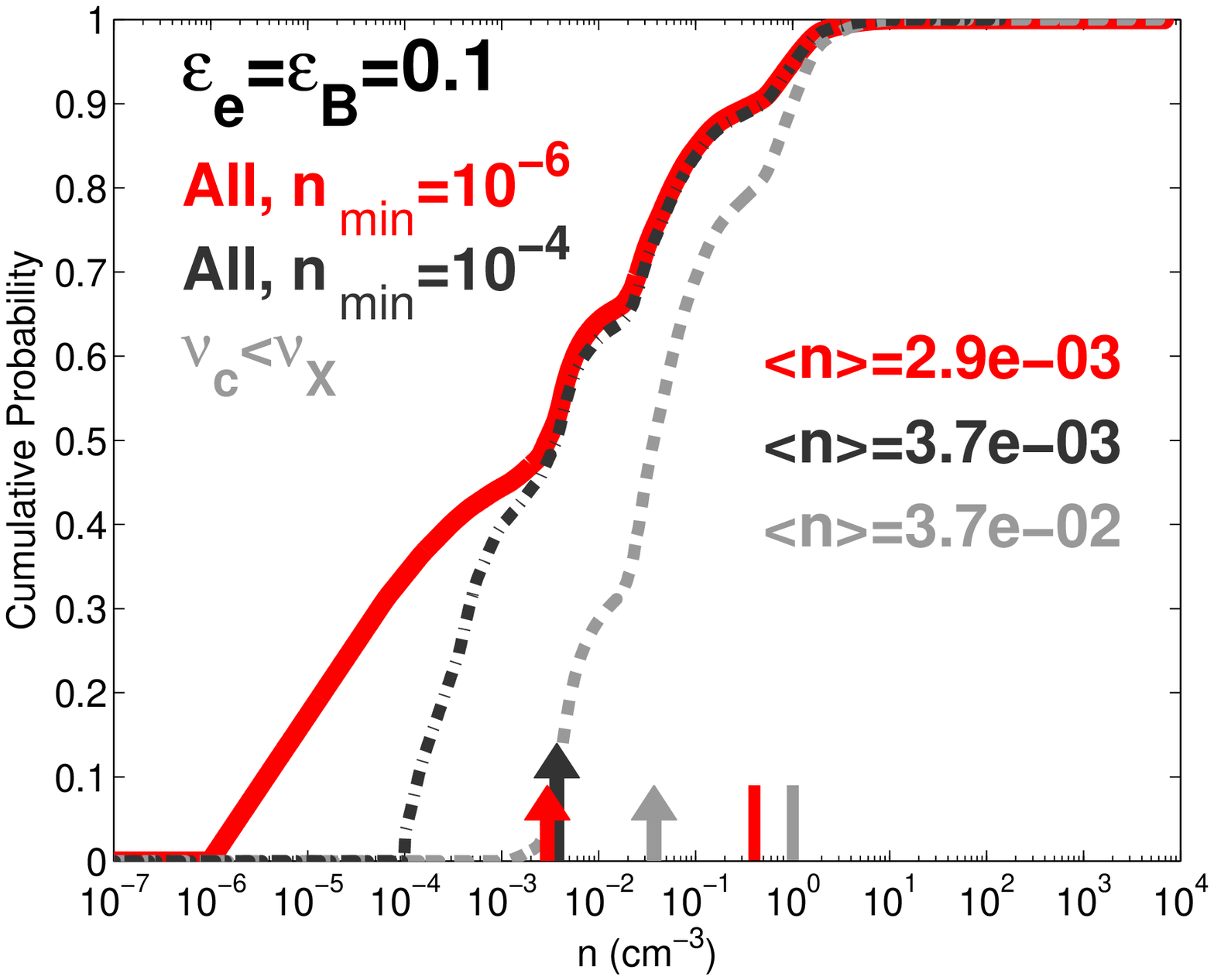}
\includegraphics*[width=0.48\textwidth,clip=]{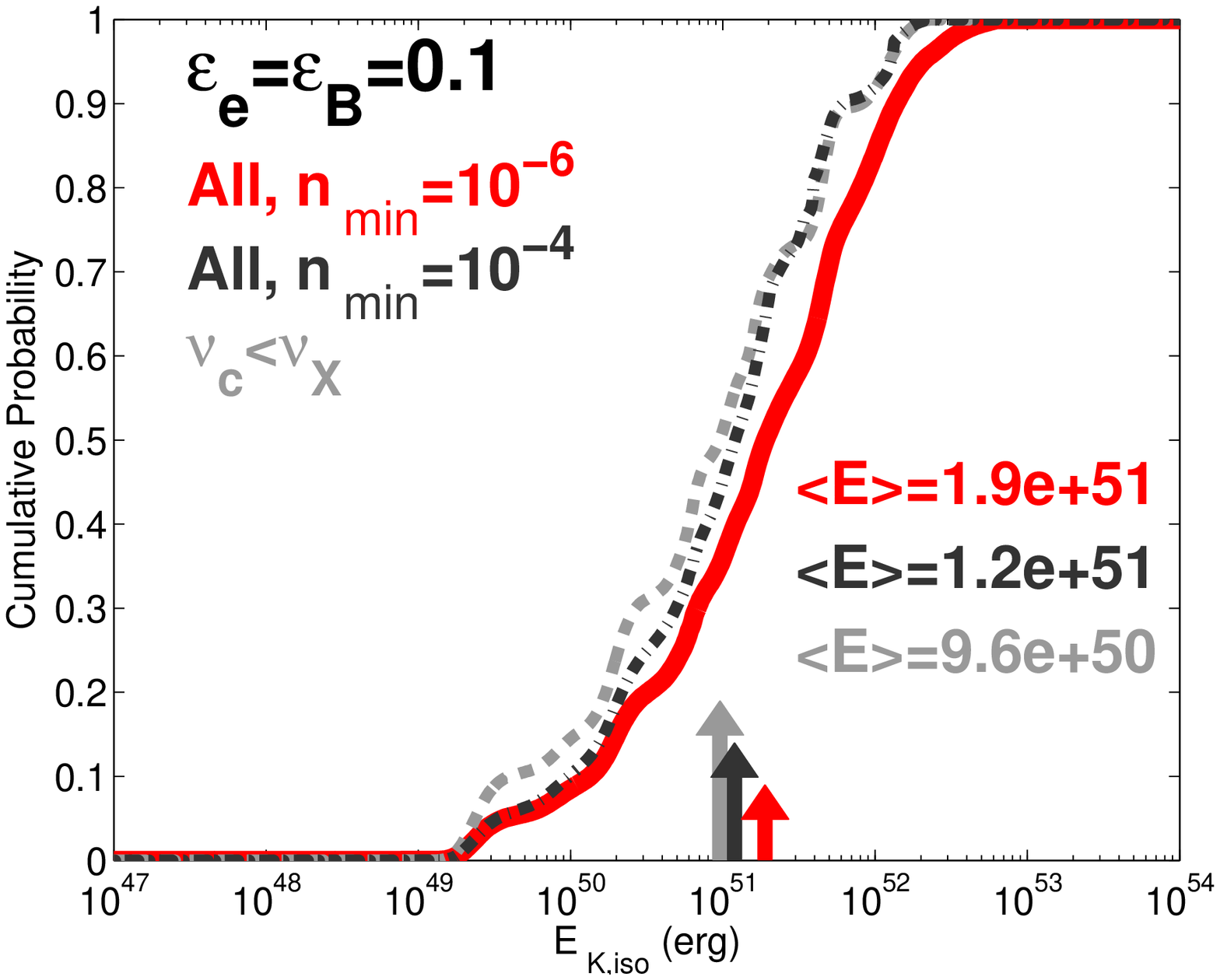}  \\
\includegraphics*[width=0.48\textwidth,clip=]{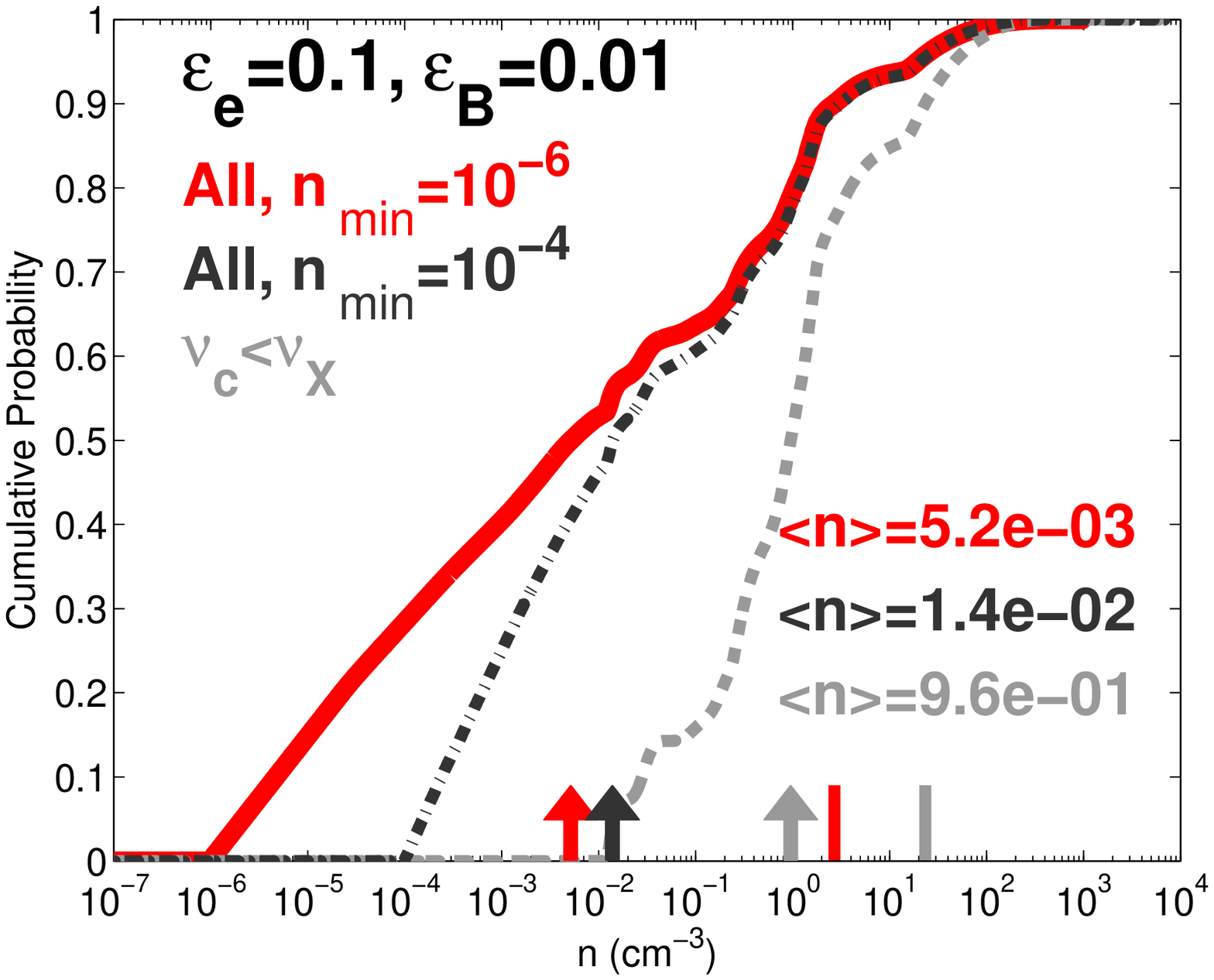}
\includegraphics*[width=0.48\textwidth,clip=]{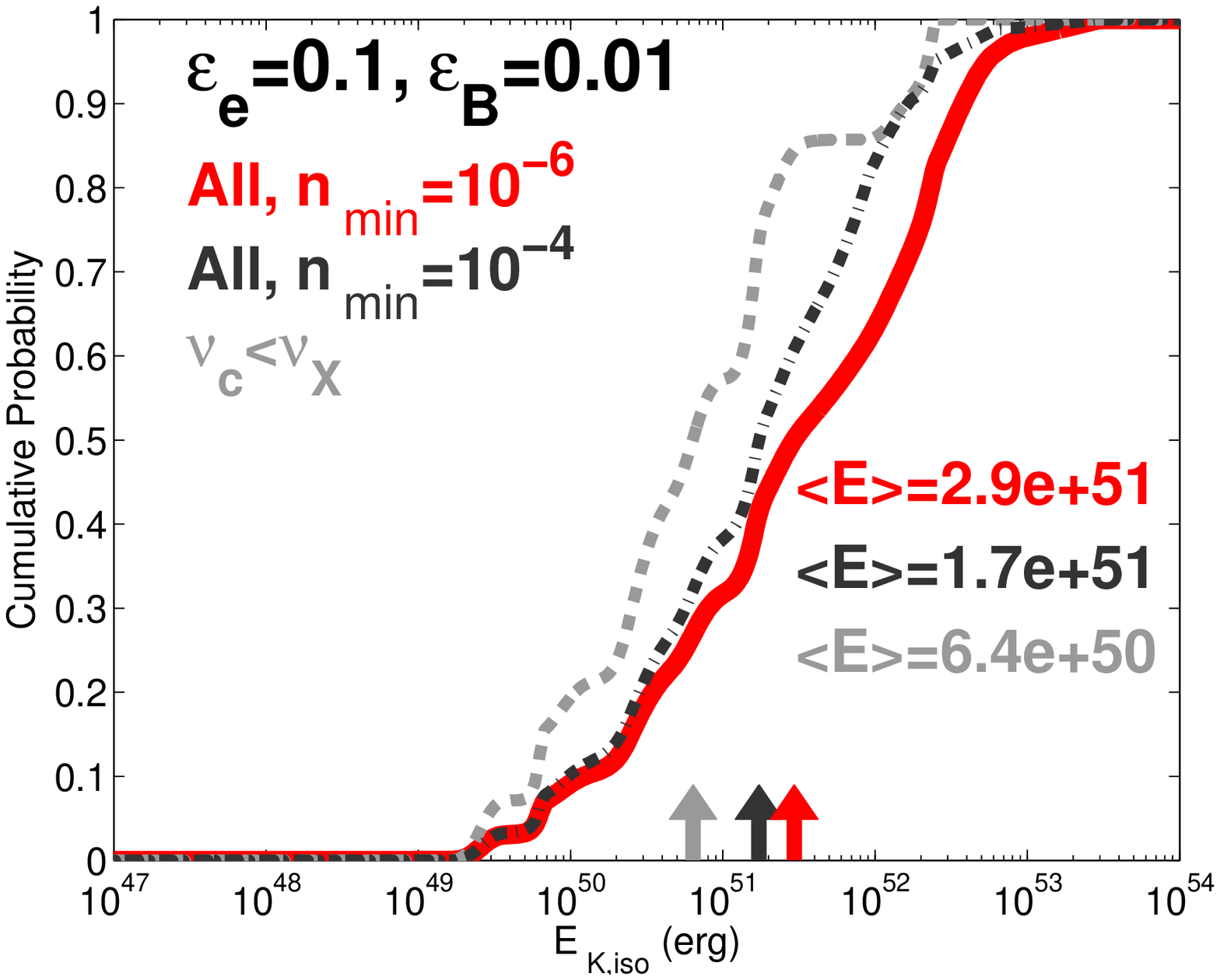} 
\end{minipage}
\caption{Combined and cumulative probability distributions in $n$ and $E_{\rm K,iso}$ assuming $\epsilon_e=\epsilon_B=0.1$ (top) and $\epsilon_e=0.1$, $\epsilon_B=0.01$ (bottom). All scenarios also include GRBs\,050724A and 140903A, with $\epsilon_B=10^{-4}$ and $10^{-3}$, respectively. Each panel shows three populations: all bursts with an imposed lower bound of $n_{\rm min}=10^{-6}$~cm$^{-3}$ (red), all bursts with an imposed lower bound of $n_{\rm min}=10^{-4}$~cm$^{-3}$ (black dot-dashed), and the sub-sample of events with $\nu_c<\nu_X$ (light grey dashed). Color-coded arrows from the bottom denote the median for each distribution (and in some cases, staggered for the sake of clarity), and lines denote $90\%$ upper limits. For $\epsilon_B=0.1$ ($\epsilon_B=0.01$), the cumulative distributions indicate that for $n \gtrsim 3 \times 10^{-3}$~cm$^{-3}$ ($n\gtrsim 0.3$~cm$^{-3}$), the distributions are virtually independent of the choice of $n_{\rm min}$ provided that $n_{\rm min} \gtrsim 10^{-4}$~cm$^{-3}$. This allows us to place robust $90\%$ upper limits of $n\lesssim 0.5-1.0$~cm$^{-3}$ ($n \lesssim 3-23$~cm$^{-3}$) for our sample of bursts. In addition, $\approx 95\%$ ($\approx 80\%$) of the total probability for all events lies below densities of $1$~cm$^{-3}$. The range of median isotropic-equivalent kinetic energies for all scenarios is $\approx (2-20) \times 10^{51}$~erg (c.f., Table~\ref{tab:stat}).
\label{fig:nhist}}
\end{figure*}

\begin{figure*}
\begin{minipage}[c]{\textwidth}
\tabcolsep0.0in
\includegraphics*[width=0.505\textwidth,clip=]{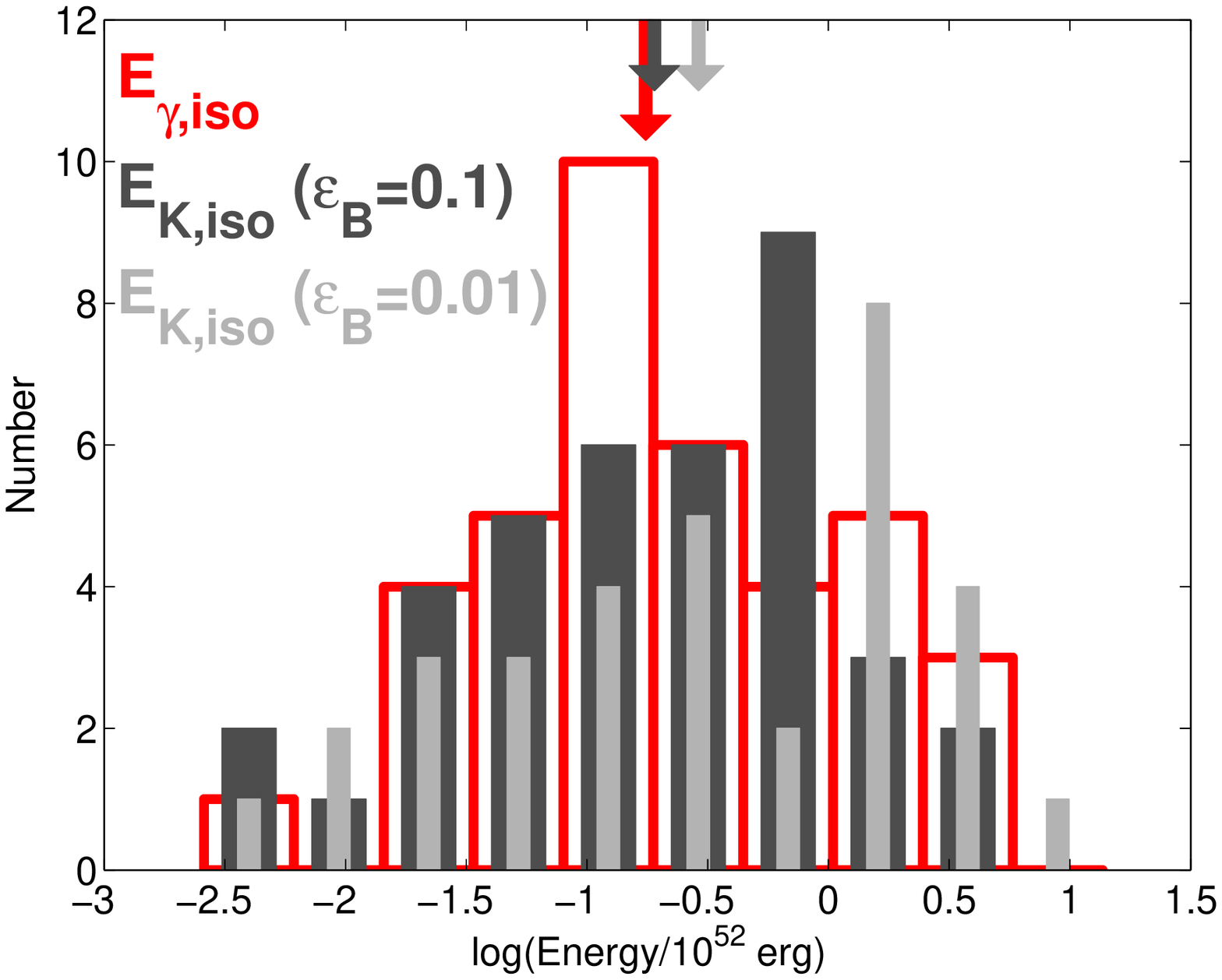}
\includegraphics*[width=0.495\textwidth,clip=]{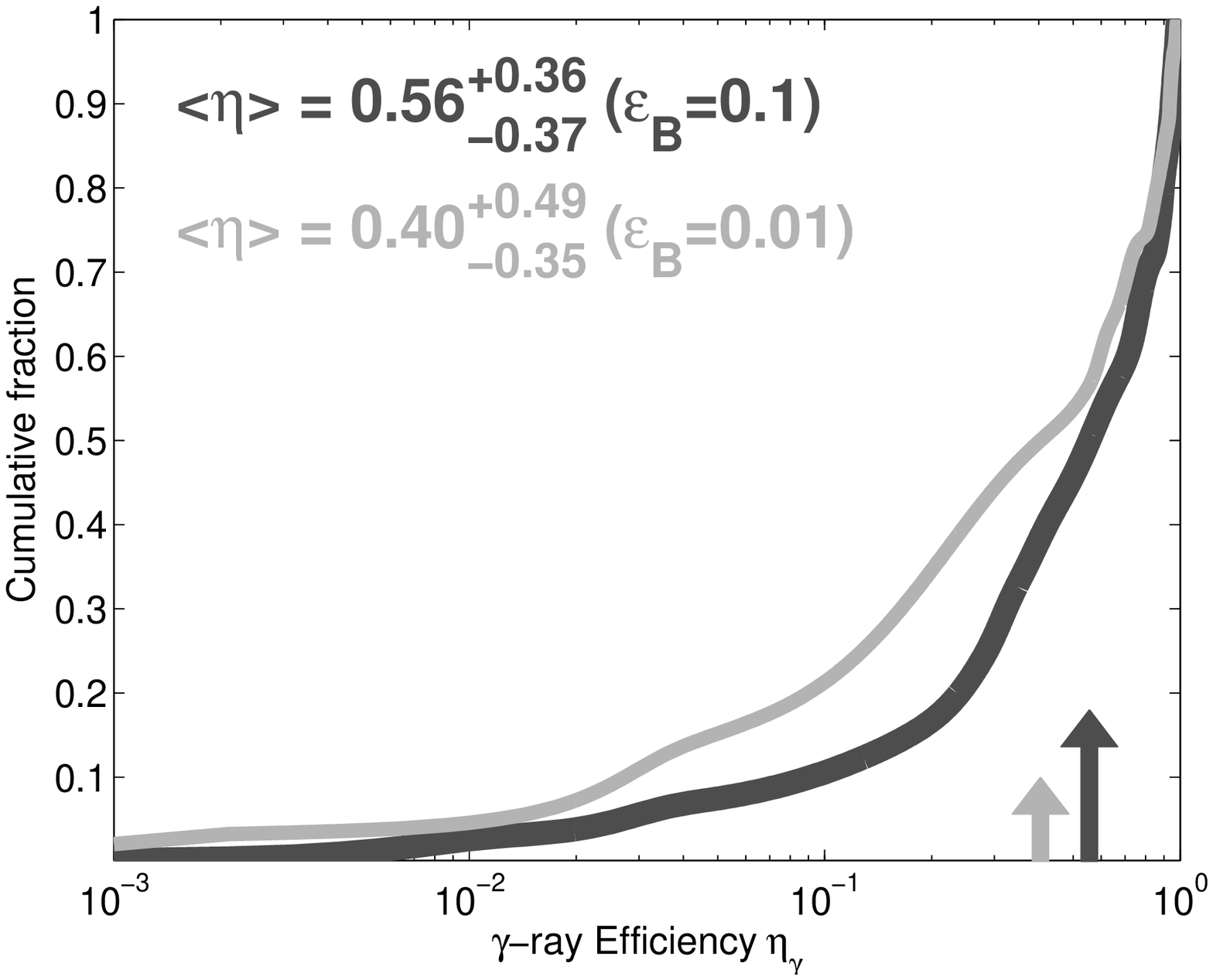}
\end{minipage}
\caption{{\it Left:} Histograms of isotropic-equivalent $\gamma$-ray energy (red), and inferred isotropic-equivalent kinetic energies assuming $\epsilon_B=0.1$ (dark grey, thick bars) and $\epsilon_B=0.01$ (light grey, thin bars) for 38 short GRBs from broad-band afterglow modeling. Median values for each population are denoted by the color-coded arrows from the top. Values for the isotropic-equivalent energies correspond to those listed in Table~\ref{tab:prop}. {\it Right:} Cumulative distributions of $\gamma$-ray efficiency ($\eta_\gamma$) assuming $\epsilon_B=0.1$ (dark grey) and $\epsilon_B=0.01$ (light grey), taking into account the $1\sigma$ uncertainties in $\eta_\gamma$ for each burst. Median values are listed in the figure and denoted by the color-coded arrows from the bottom. Arrows are staggered for clarity.
\label{fig:energyhist}}
\end{figure*}

To quantify the distributions of circumburst densities and isotropic-equivalent kinetic energies for the entire sample, we calculate the combined probability distributions from the sum of the 1-dimensional probability distributions, $P(n)$ and $P(E_{\rm K,iso})$. We sum the individual distributions for all bursts with valid solutions at $\epsilon_B=0.1$, as well as GRB\,050724A ($\epsilon_B=10^{-4}$) and GRB\,140903A ($\epsilon_B=10^{-3}$), to create cumulative probability distributions for both density and kinetic energy, shown in Figure~\ref{fig:nhist}.

The median values for the population of 38 bursts are $\langle n \rangle \approx 2.9 \times 10^{-3}$~cm$^{-3}$ and $\langle E_{\rm K,iso} \rangle \approx 1.9 \times 10^{51}$~erg (Figure~\ref{fig:nhist} and Table~\ref{tab:stat}). The density which corresponds to $90\%$ of the cumulative distribution ($n_{90}$) is $\approx 0.4$~cm$^{-3}$, and the fraction of probability that lies at densities of $\lesssim 1$~cm$^{-3}$ ($f_{n<1\,{\rm cm^{-3}}}$) is 0.95 (Figure~\ref{fig:nhist} and Table~\ref{tab:stat}). We also calculate these statistics for the sub-sample of bursts with $\nu_c<\nu_X$, corresponding to the events in Figure~\ref{fig:wm}, that utilize all three branches of Equation~\ref{eqn:fnu} and therefore have relatively well-constrained energies and densities (compared to the broader probability distributions for events with $\nu_c>\nu_X$; see Section~\ref{sec:nmin}). We find that this distribution has a median of $\langle n\rangle\approx 3.7 \times 10^{-2}$~cm$^{-3}$, $n_{90}\approx 1.0$~cm$^{-3}$, and that $f_{n<1\,{\rm cm^{-3}}} \approx 0.90$. Furthermore, the median value for kinetic energy decreases by a factor of 2, to $E_{\rm K,iso} \approx 9.6 \times 10^{50}$~erg. The cumulative distributions in density and energy are shown in Figure~\ref{fig:nhist} and the population statistics are listed in Table~\ref{tab:stat}.

\subsection{Isotropic-equivalent $\gamma$-ray energy and Efficiency}

We compare the inferred isotropic-equivalent kinetic energy to the $\gamma$-ray energy by computing the isotropic-equivalent $\gamma$-ray energy, $E_{\gamma, {\rm iso}}$, to represent the energy range $\approx 1-10^4$~keV (to match the widest energy ranges for current GRB detection satellites),

\begin{equation}
E_{\gamma, {\rm iso}}=k_{\rm bol} \times \frac{4\pi d_L^2}{1+z} f_{\gamma} ~{\rm erg}
\end{equation} 

\noindent where $k_{\rm bol}$ is the bolometric correction factor to convert the fluence to an energy range of $\approx 1-10^4$~keV, $d_L$ is the luminosity distance in cm, and $f_{\gamma}$ is the fluence in units of erg cm$^{-2}$. For cases in which the fluence is calculated over the $15-150$~keV \swift\ energy range, we use $k_{\rm bol}=5$. If a burst is detected by other $\gamma$-ray satellites which cover a wider energy range of $\approx 10-1000$~keV (e.g., {\it Fermi}, {\it Konus-Wind}, {\it Suzaku}), we utilize the measured fluences from these satellites and $k_{\rm bol}=1$ to calculate the $\gamma$-ray energy. Individual values of $E_{\gamma, {\rm iso}}$ are shown in Figures~\ref{fig:rad}--\ref{fig:th} and listed in Table~\ref{tab:prop}, while the distribution for 38 bursts is shown in Figure~\ref{fig:energyhist}). We find that the range in isotropic-equivalent $\gamma$-ray energy is $\approx (0.04-45) \times 10^{51}$~erg, with a median value of $\langle E_{\gamma, {\rm iso}} \rangle \approx 1.8 \times 10^{51}$~erg, similar to the ranges and median values of the isotropic-equivalent kinetic energy (Figure~\ref{fig:energyhist})

We also calculate the $\gamma$-ray efficiency,

\begin{equation}
\eta_\gamma = \frac{E_{\gamma, {\rm iso}}}{E_{\gamma, {\rm iso}}+E_{K, {\rm iso}}},
\end{equation} 

\noindent as well as the $1\sigma$ uncertainties in $\eta_\gamma$ for each burst, following standard propagation of errors from the $1\sigma$ uncertainties in $E_{\rm K,iso}$. The resulting values of $\eta_\gamma$ are listed in Table~\ref{tab:prop} and the cumulative distributions, after incorporating the $1\sigma$ uncertainties, are shown in Figure~\ref{fig:energyhist}. We find a wide range in $\eta_\gamma$, $\approx 10^{-3}-1$, and note that the lower bound is set by the single outlier, GRB\,150101B (Table~\ref{tab:prop}). Excluding this burst, the lower bound is $\approx 0.03$ (GRB\,140622A).

The distribution for $\epsilon_B=0.1$ has a median of $\langle \eta_\gamma \rangle~\!\!\!=~\!\!\!0.56^{+0.36}_{-0.37}$ ($1\sigma$ uncertainties). The isotropic-equivalent kinetic energy scale for the $\epsilon_B=0.01$ case is comparatively high (c.f., Figure~\ref{fig:nhist}), and thus the median value for the $\gamma$-ray efficiency is relatively low, $\langle \eta_\gamma \rangle~\!\!\!=~\!\!\!0.40^{+0.49}_{-0.35}$ ($1\sigma$ uncertainties; Figure~\ref{fig:energyhist}).

\subsection{Alternative Cases}

We have thus far made assumptions about the values of the microphysical parameters ($\epsilon_e=\epsilon_B=0.1$), the redshifts ($z=0.5$ unless otherwise determined), the minimum allowed density ($n_{\rm min}=10^{-6}$~cm$^{-3}$), and the cooling frequency (0.3~keV for $\nu_c<\nu_X$; 10~keV for $\nu_c>\nu_X$). To explore the impact of these assumptions on our resulting distributions for the kinetic energy and circumburst density, we consider alternative values for these parameters.

\subsubsection{The Value of $\epsilon_B$}

In some cases, a valid solution using the fiducial input of $\epsilon_B=0.1$ (with fixed $\epsilon_e=0.1$) cannot be found. For instance, three of the four bursts with radio afterglows require that $\epsilon_B \lesssim 0.1$ (Figure~\ref{fig:rad} and Table~\ref{tab:prop}).  At a fixed value of $\epsilon_e=0.1$, $\epsilon_B$ is constrained to $\lesssim 10^{-4}$ for GRB\,050724A, $\lesssim 10^{-2}$ for GRB\,051221A, and $\lesssim 5 \times 10^{-3}$ for GRB\,140903A (Figure~\ref{fig:rad}). For the remaining bursts, we do not have enough information to rule out the $\epsilon_B=0.1$ scenario. Thus, we consider two additional cases for all bursts: $\epsilon_B=0.01$ and $\epsilon_B=10^{-4}$ (with fixed $\epsilon_e=0.1$). Of the 38 bursts in our sample, 33 have valid solutions for $\epsilon_B=0.01$; the median and uncertainties in circumburst density and kinetic energy for each burst are listed in Table~\ref{tab:prop}. To create cumulative probability distributions, we repeat the same exercise as described in Section~\ref{sec:results} for the 33 events with valid solutions, displayed in Figure~\ref{fig:nhist}.

Using $\epsilon_B=10^{-4}$ as the fiducial value, the constraint from the cooling frequency conflicts with the solutions allowed by the afterglow observations in 16 cases, indicating that data do not allow such a low value of $\epsilon_B$ for these bursts. The population statistics for the 22 bursts with valid solutions are listed in Table~\ref{tab:stat}.

\subsubsection{Redshift} 

For bursts with no determined spectroscopic redshift, we have assumed $z=0.5$, set by the median of the short GRB population with known redshifts \citep{ber14}. However, in three cases, GRBs\,060313, 111020A, and 121226A, we do not find a valid joint solution at $z=0.5$. We find that there are valid solutions at redshifts of $z \gtrsim 1$, and thus assume $z=1$ for these bursts.

\subsubsection{The Value of $n_{\rm min}$}
\label{sec:nmin}

In the 20 cases in which the cooling frequency lies above the X-ray band ($\nu_m<\nu_X<\nu_c$), the X-ray and optical/NIR bands occupy the same spectral regime (branch 2 of Equation~\ref{eqn:fnu}) and the resulting $E_{\rm K,iso}-n$ relations have the same slope (Figure~\ref{fig:rad} and \ref{fig:th}). Thus, the lower bound on the density is set by our minimum grid value of $n_{\rm min}=10^{-6}$~cm$^{-3}$; the density is otherwise unconstrained at the low end and results in broad probability distributions in both circumburst density and energy. To understand the impact of our choice of $n_{\rm min}$ on the resulting distributions, we repeat the individual probability analysis, employing a more stringent lower bound of $n_{\rm min}=10^{-4}$~cm$^{-3}$, at the low end of gas densities for the interstellar medium (ISM; \citealt{kbs+99,mur00,gsf+13}). Since kinetic energy and density are inversely related, the upper bound on $E_{\rm K,iso}$ for each burst is naturally set by our choice of $n_{\rm min}$. We consider this alternative value of $n_{\rm min}$ for $\epsilon_B=0.1$ and $\epsilon_B=0.01$.

\afterpage{
\tabletypesize{\small}
\begin{deluxetable*}{lccccc}
\tablecolumns{6}
\tablewidth{0pc}
\tablecaption{Circumburst Density and Kinetic Energy Population Statistics
\label{tab:stat}}
\tablehead {
\colhead {Scenario}                &
\colhead {No. of Events}           &
\colhead {$\langle n \rangle$}     &
\colhead {$n_{90}^a$}           &
\colhead {$f_{n<1\,{\rm cm^{-3}}}^b$}    &
\colhead {$\langle E_{\rm K,iso} \rangle$} \\
\colhead {}                        &
\colhead {}                        &
\colhead {(cm$^{-3}$)}               &
\colhead {(cm$^{-3}$)}               &
\colhead {}                       &
\colhead {(erg)}                                      
}
\startdata

\noalign{\smallskip}
\multicolumn{6}{c}{$\epsilon_B=0.1$}  \\
\noalign{\smallskip}
\hline
\noalign{\smallskip}

All bursts, $n_{\rm min}=10^{-6}$~cm$^{-3}$ & 38 & $2.9 \times 10^{-3}$ & 0.40 & 0.95 & $1.9 \times 10^{51}$ \\
All bursts, $n_{\rm min}=10^{-4}$~cm$^{-3}$ & 37 & $3.7 \times 10^{-3}$ & 0.49 & 0.95 & $1.2 \times 10^{51}$ \\
All bursts, $n_{\rm min}=10^{-6}$~cm$^{-3}$, $\nu_c=1.7$~keV & 38 & $2.2 \times 10^{-3}$ & 0.20 & 0.95 & $1.6 \times 10^{51}$ \\
Bursts with $\nu_c<\nu_X$ & 19 & $3.7 \times 10^{-2}$ & 1.01 & 0.90 & $9.6 \times 10^{50}$ \\

\noalign{\smallskip}
\hline
\noalign{\smallskip}
\multicolumn{6}{c}{$\epsilon_B=0.01$} \\
\noalign{\smallskip}
\hline
\noalign{\smallskip}

All bursts, $n_{\rm min}=10^{-6}$~cm$^{-3}$ & 33 & $5.2 \times 10^{-3}$ & 2.7 & 0.79 & $2.9 \times 10^{51}$ \\
All bursts, $n_{\rm min}=10^{-4}$~cm$^{-3}$ & 32 & $1.4 \times 10^{-2}$ & 3.1 & 0.78 & $1.7 \times 10^{51}$  \\
All bursts, $n_{\rm min}=10^{-6}$~cm$^{-3}$, $\nu_c=1.7$~keV & 34 & $1.6 \times 10^{-2}$ & 2.1 & 0.83 & $2.4 \times 10^{51}$ \\
Bursts with $\nu_c<\nu_X$ & 14 & $0.96$ & 23.3 & 0.51 & $6.4 \times 10^{50}$ \\

\noalign{\smallskip}
\hline
\noalign{\smallskip}
\multicolumn{6}{c}{$\epsilon_B=1 \times 10^{-4}$} \\
\noalign{\smallskip}
\hline
\noalign{\smallskip}

All bursts, $n_{\rm min}=10^{-6}$~cm$^{-3}$ & 22 & $3.0 \times 10^{-2}$ & 771 & 0.68 & $1.8 \times 10^{52}$ 

\enddata
\tablecomments{All scenarios include GRBs\,050724A and 140903A, with $\epsilon_B=10^{-4}$ and $10^{-3}$, respectively. \\
$^{a}$ This is the circumburst density which corresponds to $90\%$ of the cumulative distribution. \\
$^{b}$ This is the fraction of the circumburst density cumulative distribution below a value of $1$~cm$^{-3}$. }
\end{deluxetable*}
 }

\subsubsection{The Value of $\nu_c$}

In all cases, we have assumed that the cooling frequency is on the edge of the X-ray band (0.3~keV for $\nu_c<\nu_X$; 10~keV for $\nu_c>\nu_X$). To test whether this assumption has any impact on our results, we repeat the individual and joint probability analysis assuming that the cooling frequency is at the logarithmic mean of the $0.3-10$~keV \swift\ X-ray band, $\nu_{c,{\rm mid}}=4.1 \times 10^{17}$~Hz (1.7~keV). We consider this alternative value of the cooling frequency for $\epsilon_B=0.1$ and $\epsilon_B=0.01$.

\subsubsection{Trends}

Taking these alternative values into account, we repeat the individual and joint probability analysis for each burst for nine different sets of input parameters in total. The population medians, as well as values for $n_{90}$ and $f_{n<1\,{\rm cm^{-3}}}$ are listed in Table~\ref{tab:stat}. In addition, cumulative distributions for kinetic energy and circumburst density for varying values of $\epsilon_B$ and $n_{\rm min}$ are shown in Figure~\ref{fig:nhist}.

Overall, we find that a change in $\epsilon_B$ results in an increase in the circumburst density. For instance, assuming $n_{\rm min}=10^{-6}$~cm$^{-3}$ for all bursts, decreasing $\epsilon_B$ by a factor of 10 to $\epsilon_B=0.01$ results in an increase in the median density by a factor of $\approx 1.8$. This trend becomes more drastic for other scenarios: when assuming $n_{\rm min}=10^{-4}$~cm$^{-3}$, the median density increases by a factor of $\approx 3.8$, and when considering only the bursts with $\nu_c<\nu_X$, the factor of increase is $\approx 26$. When comparing the values of $n_{90}$ and $f_{n<1\,{\rm cm^{-3}}}$, we also find overall trends commensurate with an increase in circumburst density. In particular, the factor of ten decrease in $\epsilon_B$ causes $f_{n<1\,{\rm cm^{-3}}}$ to decrease by a larger factor, from $\approx 0.95$ to $\approx 0.8$ for all bursts. When considering only bursts with $\nu_c<\nu_X$, $f_{n<1\,{\rm cm^{-3}}}$ decreases from $\approx 0.90$ to $\approx 0.51$. The effect of $\epsilon_B$ on the median kinetic energy is less pronounced, with increases by factors of $\approx 1.5$ in all cases except when considering only bursts with $\nu_c<\nu_X$; in this case, the median undergoes a slight decrease by a factor of $\approx 1.5$ (Table~\ref{tab:stat}). Overall, we find that a decrease from $\epsilon_B=0.1$ to $0.01$ results in higher median densities by factors of $\approx 2-25$ depending on the considered scenario, and a uniform increase in the density cumulative distributions (Figure~\ref{fig:nhist}). When considering the more extreme input of $\epsilon_B=10^{-4}$ for all bursts, the median density and kinetic energy both increase by factors of $\approx 10$, compared to the $\epsilon_B=0.1$ case (Table~\ref{tab:stat}).

We next investigate the effects of the values of $n_{\rm min}$ and $\nu_c$ on the parameter distributions. We find that at constant $\epsilon_B$, the upper $\approx 30-50\%$ of the density cumulative distributions are virtually independent of our choice of $n_{\rm min}$, provided that $n_{\rm min}\lesssim 10^{-4}$~cm$^{-3}$ (Figure~\ref{fig:nhist}). Importantly, this also allows us to place robust $90\%$ upper limits on the density that are largely unaffected by our choice of $n_{\rm min}$: $n_{90}\approx 0.4-0.5$~cm$^{-3}$ for $\epsilon_B=0.1$ and $n_{90}\approx 2.7-3.1$~cm$^{-3}$ for $\epsilon_B=0.01$ (Table~\ref{tab:stat}). We find that the choice of cooling frequency within the X-ray band has a minor effect on the circumburst density, either a $\approx 1.3$ factor of decrease ($\epsilon_B=0.1$) or a factor of $\approx 3$ increase ($\epsilon_B=0.01$), while the median kinetic energy only decreases by a factor of $\approx 1.2$ in both cases (Table~\ref{tab:stat}).

When considering all cases with less extreme values of $\epsilon_B$ for all bursts, the median density range is $(3-15) \times 10^{-3}$~cm$^{-3}$, with 90\% upper limits of $n_{90}\approx 0.4-3$~cm$^{-3}$. Furthermore, $\approx~\!\!\!\!\!80-95\%$ of the probability is below $\approx 1$~cm$^{-3}$ regardless of the input parameters considered (Table~\ref{tab:stat}). The median isotropic-equivalent kinetic energy ranges from $\approx (1.2-2.9) \times 10^{51}$~erg. Including more extreme scenarios like the subset of bursts with $\nu_c<\nu_X$ and $\epsilon_B=10^{-4}$, the full median density range is $\approx (3-1000) \times 10^{-3}$~cm$^{-3}$ and the median isotropic-equivalent kinetic energy range is $\approx (0.6-20) \times 10^{52}$~erg (Table~\ref{tab:stat}).

\section{Discussion and Implications}
\label{sec:disc}

\subsection{Afterglow Properties}

The X-ray, optical, and radio afterglow catalogs (Tables~A1-A3) allow us to analyze the observational afterglow properties of short GRBs as a population. By fitting the afterglow light curves, we find a weighted mean pre-jet break decline rate of $\langle \alpha_X \rangle \approx -1.07$ at $\delta t \gtrsim 1000$~s for bursts with measured temporal indices, similar to the pre-jet break declines measured from long GRB light curves \citep{nfp09,rlb+09,kkz+10,zbm+13}, and slightly shallower than the value of $\alpha_X \approx -1.2$ found for 11 short GRBs in an earlier study \citep{nfp09}. We measure the optical decline rates and find the same weighted mean decline rate of $\langle \alpha_{\rm opt} \rangle \approx -1.07$ from 19 well-sampled bursts with measured indices.

From spectral fitting of the optical, near-IR, and X-ray data, we find 12 short GRBs which require rest-frame extinction, with measured values of $A_V^{\rm host} \approx 0.3-1.5$~mag, and two bursts with lower limits of $A_V^{\rm host} \gtrsim 2.5-6$~mag (Table~\ref{tab:prop}). We note that GRB\,080919 has the highest value of rest-frame extinction; however, the sightline to this burst is close to the Galactic plane and therefore has a highly uncertain Galactic extinction which likely affects the measurement of $A_V^{\rm host}$. Prior to this study, evidence for $\gtrsim 0.5$~mag of extinction has only been reported in three short GRBs: GRB\,070724A \citep{bcf+09,ktr+10}, GRB\,111020A \citep{fbm+12}, and GRB\,120804A \citep{bzl+13}. Afterglow modeling in this work results in the same conclusions for those three events, and includes nine additional events with rest-frame extinction. We note that most of the events with non-zero extinction and robust host associations are in star-forming host galaxies; the single exception is GRB\,150101B which has an inferred value of $A_V^{\rm host}\approx 0.5$~mag and is located on the outskirts of an early-type galaxy (Fong~et~al., in prep). In comparison, $\approx 15-20\%$ of \swift\ long GRBs have optically sub-luminous afterglows that have been attributed to dust extinction. For long GRBs, inferred values of $A_V^{\rm host} \approx 0.5$~mag are common, with a substantial fraction of events with $A_V^{\rm host} \gtrsim 1-2.5$~mag \citep{ckh+09,pcb+09,plt+13}. 

While rest-frame extinction can be explained by dust in star-forming regions in the local environments of long GRBs, substantial extinction in short GRBs cannot be easily explained in the context of the double compact object merger progenitor. It is possible that these events are ``impostors'' which in actuality have massive star progenitors; in that case, we might expect them to be distinct in their $\gamma$-ray properties with longer durations and softer $\gamma$-ray spectra. However, there does not appear to be a correlation between short GRBs with extinction and their durations as they span the full range, with $T_{90} \approx 0.02-2$~sec, and only three events with non-zero extinction have $T_{90}\gtrsim 1$~sec (GRBs\,060121, 070714B, and 121226A). Furthermore, a study conducted by \citet{bnp+13} assigned six of these events probabilities of having non-collapsar progenitors based on their $\gamma$-ray properties. According to this study, 4/6 events have $\gtrsim 60\%$ probabilities that they do {\it not} originate from collapsars. Thus, it is unlikely that the majority of these events are in fact ``impostors''. Instead, this suggests that some short GRBs may originate in a star-forming regions, or have progenitor systems that can produce dust.

 \subsection{Opening Angles}
 
\begin{figure}
\tabcolsep0.0in
\includegraphics*[width=0.495\textwidth,clip=]{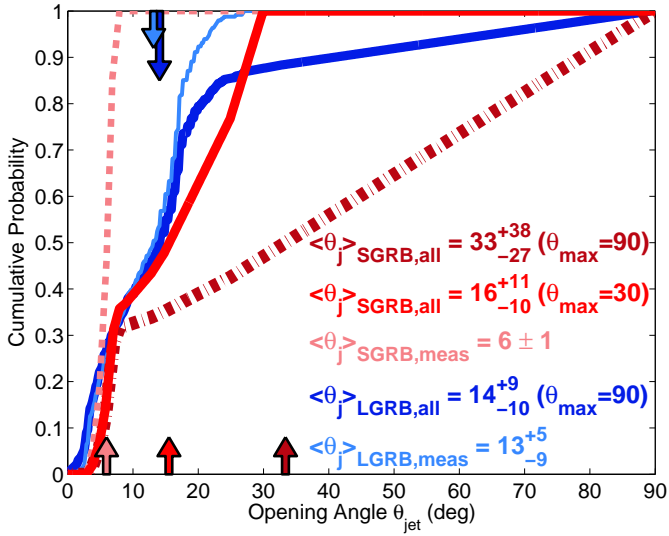}
\caption{Cumulative histograms of 11 short GRBs (dark red dash-dotted) and 265 long GRBs (dark blue) with opening angle measurements or limits, assuming a maximum opening angle of $\theta_{\rm j,max}=90^{\circ}$. Also shown is the distribution for 11 short GRBs assuming a more realistic maximum opening angle of $\theta_{\rm j,max}=30^{\circ}$ (red). The distributions for the subset of four short GRBs (light red dotted) and 248 long GRBs (light blue) with opening angle measurements are also shown. Medians are denoted by color-coded arrows, and listed with their $1\sigma$ uncertainties in units of degrees.
\label{fig:anglehist}}
\end{figure}

Most well-sampled short GRBs exhibit a single afterglow decline rate in the X-ray (at $\delta t \gtrsim 1000$~sec) and optical bands. However, there are four short GRBs (GRBs\,051221A, 090426A, 111020A, and 130603B) which have temporal steepenings on timescales of $\delta t \approx 2-5$~days, attributed to jet breaks (Table~\ref{tab:angle}; \citealt{sbk+06,nkr+11,fbm+12,fbm+14}). Jet breaks are achromatic features, and in principle can be detected in the X-ray through radio bands. The measurement of jet break time, in conjunction with the energy, density, and redshift, can be converted to a jet opening angle, $\theta_j$, using \citep{sph99,fks+01}

\begin{equation}
\theta_j=9.5\,t_{j,{\rm d}}^{3/8}(1+z)^{-3/8}E_{\rm K,iso,52}^{-1/8}n_0^{1/8} \text{ deg},
\label{eqn:jb}
\end{equation}

\noindent where $t_{j,{\rm d}}$ is in days. The opening angle measurements for the four short GRBs with jet break detections are listed in Table~\ref{tab:angle}. Using these four short GRB opening angle measurements alone, taking into account the published range of angles for individual bursts, the median is $\langle \theta_j \rangle=6 \pm 1$~deg (Figure~\ref{fig:anglehist}).

 \tabletypesize{\small}
\begin{deluxetable}{lcccc}
\tablecolumns{5}
\tablewidth{0pc}
\tablecaption{Short GRB Opening Angles
\label{tab:angle}}
\tablehead {
\colhead {GRB}                &
\colhead {Band$^{a}$} &
\colhead {$\theta_j$}          &
\colhead {$\delta t_{\rm last}^b$} & 
\colhead {Reference} \\
\colhead {}  &
\colhead {}     &
\colhead {(deg)}   &        
\colhead {(days)}  &
\colhead {}               
}
\startdata
050709 & O & $\gtrsim 15^{\circ}$ & 16.2 & 1 \\
050724A & X & $\gtrsim 25^{\circ}$ & 22.0 & 2 \\
{\bf 051221A} & X & $6-7^{\circ}$ & 26.6 & 3\\
{\bf 090426A} & O & $5-7^{\circ}$ & 2.7 & 4 \\
101219A & X & $\gtrsim 4^{\circ}$ & 3.9 & 5, This work \\
{\bf 111020A} & X & $3-8^{\circ}$ & 10.2 & 6 \\
111117A & X & $\gtrsim 3-10^{\circ}$ & 3.0 & 7, 8 \\
120804A & X & $\gtrsim 13^{\circ}$ & 45.9 & 9, This work \\
{\bf 130603B} & OR & $4-8^{\circ}$ & 6.5 & 10 \\
140903A & X & $\gtrsim 6^{\circ}$ & 3.0 & 11, This work \\
140930B & X & $\gtrsim 9^{\circ}$ & 23.1 & This work
\enddata
\tablecomments{Bursts with opening angle measurements are in bold. \\
$^a$ This indicates the band in which the jet break was detected or the lower limit was placed. X=X-ray, O=optical, R=radio. A range of angles is due to uncertainty in the redshift, kinetic energy or circumburst density. \\
$^b$ This is the time elapsed between burst detection and the last observation. \\
{\bf References:} (1) \citealt{ffp+05}; (2) \citealt{gbp+06}; (3) \citealt{sbk+06}; (4) \citealt{nkr+11}; (5) \citealt{fbc+13}; (6) \citealt{fbm+12}; (7) \citealt{mbf+12}; (8) \citealt{sta+13}; (9) \citealt{bzl+13}; (10) \citealt{fbm+14}; (11) \citealt{ebp+09}  }
\end{deluxetable}

However, the majority of short GRBs do not have detected jet breaks and instead exhibit a single power-law decline as long as they are detected. In these cases, the time of the last observation ($\delta t_{\rm last}$) can be used to place lower limits on the opening angles. The inclusion of these bursts is essential in understanding the true opening angle distribution. In most cases, \swift/XRT observes short GRBs until they fade below the detection threshhold at $\delta t \lesssim 1$~day, and enables relatively shallow lower limits of $\theta_j \gtrsim 2-5^{\circ}$ \citep{chp+12,fbm+12}. The inclusion of such limits will not have a significant effect on the opening angle distribution, as they virtually span the entire range of allowable angles.

Therefore, in order to have a more complete understanding of the opening angle distribution, we collect all existing published lower limits of $\theta_{j}\gtrsim 5^{\circ}$, and calculate lower limits for GRBs\,101219A, 120804A, 140903A, and 140903B using the observations and physical parameters presented in this work. The inferred lower limits, the band in which the jet break was detected, and the value of $\delta t_{\rm last}$ used to compute the limits, are listed in Table~\ref{tab:angle}. These seven events demonstrate that multi-wavelength afterglow observations to $\delta t_{\rm last} \approx 3-25$~days enable more meaningful lower limits on the opening angles of $\gtrsim 5-25^{\circ}$ (Table~\ref{tab:angle}).

To calculate the opening angle distribution, we give each of the 11 events equal weighting, where measurements are given Gaussian probability distributions to represent their allowed range of angles (Table~\ref{tab:angle}), while lower limits are given probability that is evenly distributed between the lower limit and the maximum possible opening angle, $\theta_{\rm j,max}=90^{\circ}$. The resulting cumulative distribution for 11 short GRBs, including measurements and lower limits, is shown in Figure~\ref{fig:anglehist}. Assuming an upper bound of $\theta_{\rm j,max}=90^{\circ}$, the short GRB population median is $\langle \theta_{j} \rangle=33^{+38}_{-27}$~deg ($1\sigma$). Motivated by simulations of post-merger black hole accretion predict jets with $\theta_j \sim 5-30^{\circ}$ \citep{rj99a,ajm+05,ros05,rgb+11}, we also calculate the cumulative distribution employing a more realistic maximum value of $\theta_{\rm j,max}=30^{\circ}$, and find a median of $\langle \theta_{j} \rangle=16 \pm 10$~deg ($1\sigma$).

To compare these distributions to those for long GRBs, we collect opening angle measurements for 265 long GRBs, including 17 events with limits \citep{fks+01,bkf03,bfk03,ggl04,fb05,rlb+09,cfh+10,cfh+11,fkg+11,gpb+11,rmz15} and calculate the cumulative distributions in the same manner (Figure~\ref{fig:anglehist}). We find a median value for the 248 long GRBs with measurements of $\langle \theta_{\rm j} \rangle=13^{+5}_{-9}$~deg. Including the 17 events with limits ($\theta_{\rm j,max}=90^{\circ}$), the median becomes $14^{+9}_{-10}$~deg.

The opening angle distribution of short GRBs impacts the true energy scale, as the true energy is lower than the isotropic-equivalent value by the beaming factor, $f_b$, where $f_b \equiv 1-{\rm cos}(\theta_j)$ and therefore $E_{\rm true}=f_bE_{\rm iso}$. To calculate the cumulative beaming factor distribution, we use the individual opening angle probability distributions for each burst to convert to individual distributions in beaming factor. We then sum the individual distributions in a cumulative sense and calculate the median and $1\sigma$ uncertainties about the median. Including all short GRBs with opening angle measurements and limits and assuming the more realistic scenario of $\theta_{\rm j,max}=30^{\circ}$, the median beaming factor is $f_b=0.04^{+0.07}_{-0.03}$. The beaming correction is less substantial if we assume $\theta_{\rm j,max}=90^{\circ}$, $f_b=0.17^{+0.52}_{-0.16}$, and is much more substantial if we only include short GRBs with measurements, $f_b = 0.005 \pm 0.002$.

We find median isotropic-equivalent $\gamma$-ray and kinetic energy scales of $E_{\gamma,{\rm iso}}\approx 2 \times 10^{51}$~erg and $E_{\rm K,iso}\approx (1-3) \times 10^{51}$~erg. Applying the beaming correction for the most realistic scenario gives median beaming-corrected $\gamma$-ray and kinetic energy scales of $\langle E_{\gamma} \rangle = 0.8^{+1.4}_{-0.6} \times 10^{50}$~erg and $\langle E_K \rangle=0.8^{+2.5}_{-0.7} \times 10^{50}$~erg, resulting in a total beaming-corrected energy release of $\langle E_{\rm tot} \rangle = 1.6^{+3.9}_{-1.3} \times 10^{50}$~erg. The inferred energy scales can be used to constrain the mechanism of energy extraction to power the relativistic jet:  the thermal energy release from $\nu\bar{\nu}$ annihilation in a baryonic outflow \citep{jar93,mhi+93} and magnetohydrodynamic (MHD) processes in the black hole's accretion remnant (e.g. \citealt{bz77,rrd03}). The general consensus is that $\nu\bar{\nu}$ annihilation can only produce beaming-corrected total energy releases of $10^{48}-10^{49}$~erg, while MHD processes can more easily produce energy releases in excess of $10^{49}$~erg \citep{rj99a,rj99b,pwf99,ros05,baj+07,lr07}. Thus, if the majority of short GRBs have wider opening angles than the four short GRBs with measurements, and thus have a smaller overall correction to the isotropic-equivalent energy scale, it will be necessary to invoke MHD processes to explain the observed energy releases.

The opening angles also impact the event rate, as the true event rate is elevated compared to the observed rate by a factor of $f_b^{-1}$, so $\Re_{\rm true}=f_b^{-1}\Re_{\rm obs}$. The current estimated observed short GRB volumetric rate is $\Re_{\rm obs}\approx 10$ Gpc$^{-3}$ yr$^{-1}$ \citep{ngf06}. Using $f_b^{-1}=27^{+158}_{-18}$, which corresponds to all short GRBs with opening angle measurements and limits ($\theta_{\rm j,max}=30^{\circ}$), we find a true event rate of $\Re_{\rm true}\approx 270^{+1580}_{-180}$~Gpc$^{-3}$ yr$^{-1}$. The observed all-sky event rate of $\approx 0.3$~yr$^{-1}$ within 200~Mpc \citep{gp05} then becomes $8^{+47}_{-5}$~yr$^{-1}$. We note that this rate is conservative compared to previously reported rates based on short GRB observations \citep{chp+12,fbm+12,fbm+14}, as it properly incorporates opening angle lower limits, with the only assumption of an upper bound on the opening angle of $30^{\circ}$. This range is fully consistent with the expected detection rates of neutron star mergers within a volume of 200~Mpc by Advanced LIGO/VIRGO of $\approx 0.2-200$~yr$^{-1}$ \citep{aaa+13}. Since there are a limited number of short GRBs with meaningful information on the opening angles, any additional measurements will greatly help to elucidate the true opening angle distribution, and therefore true energy scale and event rate.

\subsection{Connection to Galactic Environments}


\begin{figure}
\includegraphics*[width=\columnwidth,clip=]{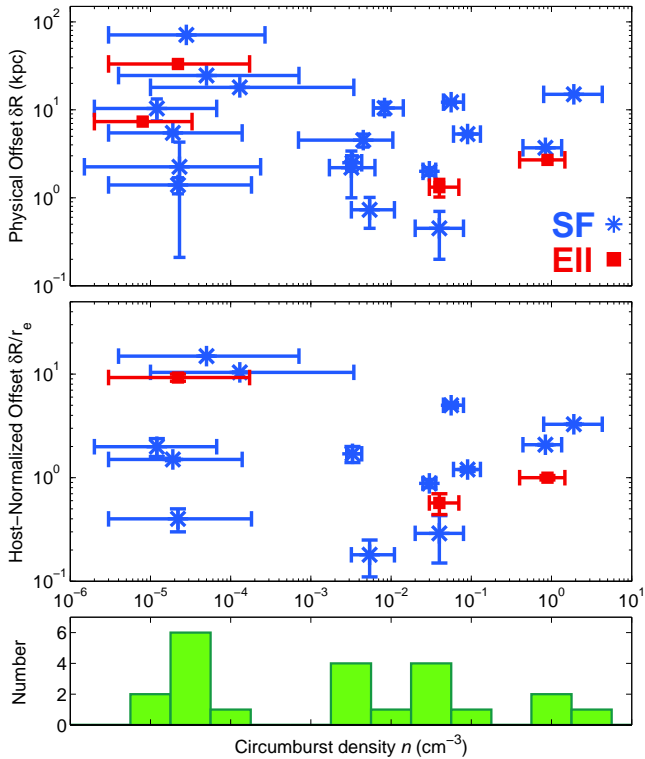} 
\caption{{\it Top:} Projected physical offset, $\delta R$, versus circumburst density for 22 short GRBs with inferred circumburst densities and sub-arcsecond localization allowing for precise offset measurements. The sample is separated by host galaxy type: star-forming hosts (blue asterisks) and elliptical hosts with no signs of star formation (red squares) according to \citet{fbc+13}. For bursts with no spectroscopic redshift, we assume $z=0.5$ to calculate the physical offset, except in two cases, GRBs\,060313 and 111020A, where we have assumed $z=1$ as suggested by the afterglow modeling. {\it Middle:} Host-normalized offset, $\delta R/r_e$, versus circumburst density for 16 bursts with measured host galaxy sizes. Physical and host-normalized offsets are from \citet{fbf10}, \citet{mbf+12}, \citet{bzl+13}, \citet{fb13}, \citet{sta+13}, and this work. Error bars correspond to $1\sigma$ confidence. {\it Bottom:} Histogram of inferred median circumburst densities for 22 short GRBs with physical offsets.
\label{fig:noffset}}
\end{figure}

We connect the afterglow properties of short GRBs to their larger-scale, galactic environments. In particular, we investigate trends between the circumburst densities, which probe the explosion environment on sub-parsec-scales, to the predicted distributions for NS-NS mergers, host galaxy type, and locations within the host galaxies.

We find a wide range of inferred circumburst densities for the 38 bursts that we have studied in detail, and can compare the properties of the sub-parsec-scale environment with the global host galaxy environment. Separating the bursts by host galaxy type according to \citet{fbc+13} and additional data collected since, we find that bursts in both star-forming and elliptical host galaxies span a wide range of densities, $10^{-6}$-5~cm$^{-3}$ (Figure~\ref{fig:noffset}). We compare the inferred densities to predictions for NS-NS mergers from population synthesis for varying Galactic potentials, which have input distributions for merger timescales and kick velocities \citep{bpb+06}. Considering only the four bursts with elliptical hosts, we find particularly good agreement with the distributions for the large elliptical galaxy model ($M_*=10^{11} M_{\odot}$, $M_{\rm halo}=10^{12} M_{\odot}$) which has probability peaks at $10^{-5}$~cm$^{-3}$ and $1$~cm$^{-3}$, and poor agreement with the small elliptical galaxy model ($M_*=10^{8} M_{\odot}$), which is dominated by very low densities of $\lesssim 10^{-6}$~cm$^{-3}$ \citep{bpb+06}. This conclusion is commensurate with the known stellar masses of short GRB elliptical hosts, which have a median of $M_*\approx 10^{11.0}~M_{\odot}$ \citep{lb10,ber14}. There are no predicted density distributions for star-forming galaxies; however, we would expect short GRBs which originate in elliptical galaxies to have lower inferred densities due to the lower average ISM densities in elliptical galaxies (e.g., \citealt{fbp+06}). Based on the small number of events, we do not find any significant difference between the inferred circumburst densities of short GRBs in star-forming versus elliptical hosts. 

We next investigate the relationship between the inferred circumburst densities and burst offsets from their host galaxies. If short GRBs trace the large-scale distribution of the ISM, we expect the inferred circumburst densities to decrease as a function of offset. In this vein, we gather all available projected physical offsets, $\delta R$, between the afterglow location and host galaxy center, derived from ground-based \citep{mbf+12,bzl+13,sta+13} and {\it Hubble Space Telescope} ({\it HST}) observations \citep{fbf10,fb13}. For bursts with no spectroscopic redshift, we calculate the physical offsets at $z=0.5$ to be consistent with this work. However, for GRBs\,060313 and 111020A, afterglow modeling implies that $z>0.5$ (c.f. Section~\ref{sec:ipd}; \citealt{fbm+12,rvp+06}) so we keep the original fiducial value of $z=1$ for these bursts for complete uniformity. The distribution of circumburst densities with respect to projected physical offset for 22 bursts is shown in Figure~\ref{fig:noffset}. We find that four bursts with $\delta R \gtrsim 10$~kpc have very low densities of $\lesssim 10^{-3}$~cm$^{-3}$, while bursts with $\delta R \lesssim 10$~kpc have a wider range of densities, spanning $\approx 6$~orders of magnitude (Figure~\ref{fig:noffset}). Overall, we find that for $\delta R \approx 1-15$~kpc, there is no obvious trend between circumburst density and projected physical offset. We also find no obvious trends when considering only bursts with relatively well-measured densities ($\nu_c<\nu_X$).

To analyze the relationship with offsets in a more uniform manner, we utilize offsets that have been normalized by the sizes of their host galaxies ($\delta R/r_e$ where $r_e$ is the galaxy half-light radius). The sample of bursts with host-normalized offsets is smaller since precise galaxy size measurements require the resolution of {\it HST}; thus the sample comprises 16 events \citep{fbf10,fb13}. The circumburst densities as a function of projected host-normalized offset is provided in Figure~\ref{fig:noffset}. Our analysis suggests that for $\lesssim 5~r_e$, the inferred densities are largely independent of host-normalized offset. We discuss a couple of possible contributing factors. First, since we can only measure projected offsets, we are not sensitive to the distance component along our line of sight, which could contribute a significant amount to the absolute distance. This may explain the case of GRB\,061006, which has a small projected offset $\approx 0.4 r_e$ but has a very low density of $\approx 2 \times 10^{-5}$~cm$^{-3}$ (Table~\ref{tab:prop} and Figure~\ref{fig:noffset}). Second, the afterglow only probes the sub-parsec circumburst environment and to a certain extent will be more sensitive to small-scale fluctuations in the ISM rather than the average ISM density on kiloparsec scales.

However, bursts that appear to have no coincident host galaxy to deep optical/NIR limits of $\gtrsim 26$~mag and are located $\approx 30-75$~kpc from the nearest most probable host galaxy (``host-less'' bursts; \citealt{ber10,fb13,tlt+13}) are expected to have low inferred densities. Indeed, the three bursts located at offsets of $\gtrsim 10 r_e$ have low densities of $\lesssim 10^{-4}$~cm$^{-3}$, as expected if these bursts occur in the IGM or outer halos of their hosts.
 
\section{Conclusions}

We present the most comprehensive catalog of short GRB afterglows to date, representing a decade of observations since the launch of \swift\ in 2004. This catalog is comprised of 103 short GRBs with prompt X-ray, optical/NIR and radio follow-up, enabled by broad-band Target-of-Opportunity programs. Applying the synchrotron afterglow model to the observations, we also place constraints on the isotropic-equivalent kinetic energies and circumburst densities for a subset of 38 events with well-sampled data sets. While a handful of short GRB afterglows have been studied in detail on an individual basis, our work presents the energy and density scales for a large population of events for the first time. We come to the following key conclusions:

\begin{itemize}

\item The afterglow observations presented in this work include 71 X-ray detections, 30 optical/NIR detections, and 4 radio detections. The detection fractions are 91\%, 40\% and 7\%, respectively, after accounting for observing constraints. We present new optical/NIR observations for 11 events, and new radio observations for 25 bursts.

\item Applying a synchrotron model to the broad-band afterglows of 38 bursts, we calculate the inferred circumburst density. Considering a range of scenarios with varying values for the microphysical parameters, cooling frequency, and minimum circumburst density, the median circumburst density is $(3-15) \times 10^{-3}$~cm$^{-3}$, with 90\% upper limits of $n_{90}\approx 0.4-3$~cm$^{-3}$. Furthermore, $\approx 80-95\%$ of the probability is below $\approx 1$~cm$^{-3}$. This indicates that overall short GRBs explode in low-density environments.

\item Depending on the set of assumptions in our analysis, we infer a median isotropic-equivalent kinetic energy of $\approx (1-3) \times 10^{51}$~erg (considering all scenarios), and an isotropic-equivalent $\gamma$-ray energy scale of $\approx 2 \times 10^{51}$~erg. We find a median $\gamma$-ray efficiency of $\approx 0.40-0.56$.

\item We find no obvious trends between circumburst density and host galaxy offset for projected offsets of $\lesssim 10$~kpc (or $\lesssim 5~r_e$), and no trend between density and host galaxy type, indicating that the circumburst density is not strongly dependent on the average ISM density. However, three bursts in our sample with offsets of $\gtrsim 10$~kpc have low densities of $\lesssim 10^{-4}$~cm$^{-3}$, as expected if these bursts explode in the IGM.

\item Using 11 short GRBs with opening angle measurements and lower limits, and assuming a maximum value on the opening angle of $30^{\circ}$, we calculate a median jet opening angle of $16 \pm 10$~deg and a median beaming factor of $0.04^{+0.07}_{-0.03}$. This results in a beaming-corrected total energy release of $\approx 1.6^{+3.9}_{-1.3} \times 10^{50}$~erg ($1\sigma$ range), which is broadly consistent with the two primary proposed mechanisms of energy extraction, $\nu\bar{\nu}$ annihilation and MHD processes. The beaming-corrected volumetric rate is $\approx 270^{+1580}_{-180}$~Gpc$^{-3}$~yr$^{-1}$ with an all-sky event rate within a volume of 200~Mpc of $8^{+47}_{-5}$~yr$^{-1}$. This range is fully consistent with the expected detection rates of gravitational wave signals from neutron star mergers by Advanced LIGO/VIRGO within the same volume.

\end{itemize}

Our study highlights the importance of broad-band observations in constraining the basic properties of short GRBs. For bursts with detected radio afterglows, we can start to constrain the microphysical parameters, which has thus far only been possible for long GRBs. While our study provides the isotropic-equivalent $\gamma$-ray and kinetic energy scales, the true energy release depends on the degree of jet collimation for short GRBs. Current knowledge of the collimation of short GRBs relies on only a handful of events with measured opening angles from their light curves, primarily due to the faintness of short GRB afterglows which prevent temporal monitoring on timescales longer than $1-2$ days. Therefore, it is imperative to use the most sensitive ground- and space-based resources to uncover additional collimated events or place meaningful lower limits on the opening angles. It is especially important to undertake these studies while \swift\ is in operation, since this satellite has the unique capability of providing multi-wavelength light curves within minutes after the bursts.

The past decade of short GRB observations has enabled significant progress in understanding the basic properties of short GRBs, namely their energetics, circumburst densities, and opening angles. Furthermore, in addition to informing the behavior of on-axis afterglows, the circumburst density and energy are key parameters which feed in to predictions for electromagnetic counterparts to compact object mergers, such as off-axis afterglows \citep{gpk+02,vzm+10} and long-lived radio flares from mildly relativistic ejecta \citep{np11}. Advanced LIGO/VIRGO will detect NS-NS mergers within a horizon distance of $\approx~\!200$~Mpc \citep{aaa+13}, making these alternative electromagnetic signatures promising for joint detection with gravitational waves. In a subsequent work, we will assess the detectability of such counterparts by using the distributions of circumburst densities and energies of on-axis short GRBs as inputs, which will help to inform search strategies in the upcoming revolutionary era of gravitational wave discovery. 

\acknowledgments
We acknowledge Matthew Bayliss, Kathy Cooksey, Francesco Di Mille, Steven Elhert, Michael Florian, Tolga Guver, Traci Johnson, Dan Kelson, Michael McDonald, Andy Monson, David Osip, Benjamin Rackham, Nathan Sanders, Anil Seth, Meghin Spencer, Brian Stalder, Tony Stark, Rik Williams, and Amanda Zangari for their assistance in Magellan Target-of-Opportunity observations. Partial support for this work was provided by NASA through Einstein Postdoctoral Fellowship grant number PF4-150121 and Chandra Award Number G04-15055X issued by the Chandra X-ray Observatory Center, which is operated by the Smithsonian Astrophysical Observatory for NASA under contract NAS8-03060. BAZ acknowledges support from NSF Grant AST-1302954, EB acknowledges partial support from NSF Grant AST-1107973, and RM acknowledges support from the James Arthur Fellowship at NYU. Additional support was provided by several NASA/Swift grants. This work made use of data supplied by the UK Swift Science Data Centre at the University of Leicester. The scientific results reported in this article are in part based on observations made by the Chandra X-ray Observatory and data obtained from the Chandra Data Archive. This paper includes data gathered with the 6.5 meter Magellan Telescopes located at Las Campanas Observatory, Chile. Observations reported here were obtained at the MMT Observatory, a joint facility of the University of Arizona and the Smithsonian Institution. Based on observations obtained at the Gemini Observatory acquired through the Gemini Science Archive and processed using the Gemini IRAF package which is operated by the Association of Universities for Research in Astronomy, Inc., under a cooperative agreement with the NSF on behalf of the Gemini partnership: the National Science Foundation (United States), the National Research Council (Canada), CONICYT (Chile), the Australian Research Council (Australia), Ministério da Ciência, Tecnologia e Inovação (Brazil) and Ministerio de Ciencia, Tecnología e Innovación Productiva (Argentina). The National Radio Astronomy Observatory is a facility of the National Science Foundation operated under cooperative agreement by Associated Universities, Inc.


\newpage

\appendix
\setcounter{table}{0}
\renewcommand{\thetable}{A\arabic{table}}
\section{Broad-band Afterglow Catalogs}

\tabletypesize{\footnotesize}
\LongTables
\begin{deluxetable*}{lccccc}
\tablecolumns{6}
\tabcolsep0.05in\scriptsize
\tablewidth{0pc}
\tablecaption{Short GRB Late-time X-ray afterglow catalog
\label{tab:xcat}}
\tablehead {
\colhead {GRB}                     &
\colhead {$\delta t$}              &
\colhead {Exposure Time}           &
\colhead {$F_{X}$}                 &
\colhead {$\sigma_{X}$}            &
\colhead {References}              \\
\colhead {}                        &
\colhead {(s)}                     &
\colhead {(s)}                     &
\colhead {($\mu$Jy)}               &
\colhead {($\mu$Jy)}               &
\colhead {}                        
}
\startdata
050509B      & 831    & 381 & 0.07 & 0.02 & \citet{gso+05} \\
             & $6.2 \times 10^{3}$ & $2.0 \times 10^{3}$ & $<0.02$ & \\
	     & $^+2.1 \times 10^{5}$ & $5.0 \times 10^{4}$ & $<8.0 \times 10^{-3}$ & \\
050709  & $1.4 \times 10^{5}$ & $1.5 \times 10^{4}$ & $1.9 \times 10^{-3}$ & $1.4 \times 10^{-4}$ & \citet{ffp+05} \\ 
	& $2.1 \times 10^{5}$ & $5.0 \times 10^{3}$ & $<1.5 \times 10^{-3}$  \\ 
	& $^{+}2.2 \times 10^{5}$ & $3.8 \times 10^{4}$ & $1.1 \times 10^{-3}$ & $2.0 \times 10^{-5}$ \\ 
	& $2.8 \times 10^{5}$ & $6.4 \times 10^{3}$ & $<7.3 \times 10^{-4}$ \\ 
	& $3.7 \times 10^{5}$ & $2.1 \times 10^{3}$ & $<5.9 \times 10^{-3}$  \\
\enddata
\tablecomments{Upper limits correspond to $3\sigma$. Unless otherwise stated, all data are taken with \swift/XRT and X-ray flux densities are at 1~keV. \\
$^{a}$ We employed a fiducial spectral index of $\beta_X=-1$. \\
$^{b}$ We employed a fiducial count rate-to-unabsorbed flux conversion factor of $10^{-11}$~erg~cm$^{-2}$~s$^{-1}$ and spectral index $\beta_X=-1$. \\
$^{c}$ Reported flux is in the energy range of 0.7-10~keV. \\
$^{d}$ Late-time re-brightening in GRB\,080503 light curve is observed in both optical and X-ray bands and is unlikely the afterglow \citep{pmg+09}. \\
$^{+}$ Chandra observation. \\
$^{++}$ XMM-Newton observation. \\
}
\end{deluxetable*}

\clearpage

\tabletypesize{\footnotesize}
\LongTables
\begin{deluxetable*}{lcccccccc}
\tablecolumns{9}
\tabcolsep0.05in\scriptsize
\tablewidth{0pc}
\tablecaption{Short GRB Optical/near-IR afterglow catalog
\label{tab:ocat}}
\tablehead {
\colhead {GRB}                     &
\colhead {$\delta t$}              &
\colhead {Telescope} 		&
\colhead {Instrument}            &
\colhead {Filter} 		& 
\colhead {Exposure Time} &
\colhead {$F_{\nu}$}                 &
\colhead {$\sigma$}            &
\colhead {References}              \\
\colhead {}                        &
\colhead {(hr)}                     &
\colhead {}                        &
\colhead {}                        &
\colhead {}                        &
\colhead {(s)}                     &
\colhead {($\mu$Jy)}               &
\colhead {($\mu$Jy)}               &
\colhead {}                        
}
\startdata
050202  & $12.6$ & Mount John & MOA    & $R$ & $900$   & $<42.89$&  & \citet{gcn3018} \\
050509B & $2.1$  & WIYN  & OPTIC       & $r$ & $600$   & $<0.75$ &  & \citet{bpp+06} \\
050709  & $34.0$ & Danish tel. & DFOSC & $R$ & $7200$  & $2.34$  & $ 0.12$ & \citet{hwf+05} \\
	& $59.1$ & VLT	& FORS2	& $V$	& 3600 & 0.64 & 0.07 & \citet{cmi+06} \\
	& 59.3 & VLT & FORS2 & $R$ & 3000 & 0.90 & 0.05 &  \\
        & $60.0$ &  Danish tel. & DFOSC  & $R$ & $10200$ & $1.17$  & $0.26$ & \citet{hwf+05} \\
	& 104.7 & VLT & FORS1 & $V$ & 3600 & $<0.36$ & & \citet{cmi+06} \\
        & $134.4$ & HST & ACS         & F814W & $6360$ & $0.34$  & $0.006$ & \citet{ffp+05} \\
\enddata
\tablecomments{Upper limits correspond to $3\sigma$. All optical and near-IR fluxes are corrected for Galactic extinction in the directions of each bursts \citep{sfd98,sf11}. Instruments, exposure times, and flux density uncertainties are provided whenever available. \\
$^{a}$ Optical observing constraint, due to delayed precise localization, sightline with high Galactic extinction, crowded field, or proximity to a bright star which contaminates the GRB position. \\
$^{b}$ Optical re-brightening observed in GRB\,080503 is unlikely the afterglow \citep{pmg+09}. \\
$^{c}$ Full simultaneous $giz$-band light curves are available in \citet{nkk+12}. \\
}
\end{deluxetable*}

\clearpage

\tabletypesize{\small}
\LongTables

\begin{deluxetable*}{lcccccc}
\tablecolumns{7}
\tabcolsep0.05in\scriptsize
\tablewidth{0pc}
\tablecaption{Short GRB Radio afterglow catalog
\label{tab:rcat}}
\tablehead {
\colhead {GRB}                     &
\colhead {$\delta t$}              &
\colhead {Facility} 		   &
\colhead {Mean Frequency} 	   & 
\colhead {$F_{\nu}$}               &
\colhead {$\sigma$}               &
\colhead {References}              \\
\colhead {}                        &
\colhead {(dy)}                     &
\colhead {}                        &
\colhead {(GHz)}                   &
\colhead {($\mu$Jy)}               &
\colhead {($\mu$Jy)}               &
\colhead {}                        
}
\startdata
050202  & $0.52$ & VLA & $4.86$ & $<120$ & & \citet{gcn3007} \\
050509B & $0.49$ & WSRT & $4.86$ & $<66$ & & \citet{gcn3405} \\
050709  & $1.6$  & VLA & $8.46$ & $<115$ & & \citet{ffp+05} \\
        & $2.5$	 & VLA & $8.46$ & $<114$ & & \\
        & $4.5$  & VLA & $8.46$ & $<74$ & & \\
        & $7.5$  & VLA & $8.46$ & $<40$ & & \\
050724A & $0.57$ & VLA & $8.46$ & $173$ & $30$ & \citet{bpc+05} \\
	& $1.68$ & VLA & $8.46$ & $465$ & $29$  \\
\enddata
\tablecomments{For bursts with multiple contemporaneous upper limits, we only display the most constraining limits. Upper limits are $3\sigma$.
}
\end{deluxetable*}

\end{document}